%% file: thesis.tex
 \documentclass[oneside,12pt]{Classes/CUEDthesisPSnPDF}

\usepackage{amsmath}
\usepackage{amsfonts}
\usepackage{amssymb}
\usepackage{verbatim}
\usepackage{times}
\usepackage{pstricks}
\usepackage{cancel}
\usepackage{eso-pic}
\usepackage{graphicx}

\include{abbreviation}

\newcommand{\be}{\begin{equation}} 
\newcommand{\ee}{\end{equation}}
\newcommand{\bea}{\begin{eqnarray}}
\newcommand{\eea}{\end{eqnarray}}
\newcommand{\bi}{\begin{itemize}}
 
\newcommand{\ei}{\end{itemize}} 
\newcommand{\bc}{\begin{column}{0.50\textwidth}}
\newcommand{\ec}{\end{column}}
\newcommand{\bcs}{\begin{columns}} 
\newcommand{\ecs}{\end{columns}}
\newcommand{\Tr}[1]{\operatorname{Tr}(#1)}

\newcommand{\meanO}{\mathcal{O}_{EE}}
\newcommand{\iniE}{\overline{E}}
\newcommand{\inie}{\overline{e}}
\newcommand{\Ot}{\mathcal{O}_t}
\def\obs{{\mathcal{O}}}

\newcommand{\ket}[1]{\left| #1\right\rangle}
\newcommand{\bra}[1]{\left\langle #1\right|}
\newcommand{\braket}[2]{\left\langle #1| #2\right\rangle}
\newcommand{\Ord}[1]{{O}\left(#1\right)}

\newcommand{\avg}[1]{\left< #1 \right>}
\newcommand{\rp}{\overline{P}}
\newcommand{\Prob}{\operatorname{Prob}}

\def\cN{\mathcal N}

\def\Tr{\mathrm{Tr}}

\def\dis{h}
\newcommand{\PR}[1]{\mathrm{PR}_{#1}}
\def\ii{i}
\def\Erf{\operatorname{Erf}}
\def\id{{\mathbb I}}

\def\cN{{\mathcal N}}
\def\cI{{\mathcal I}}
\def\cC{{\mathcal C}}
\def\Hpre{H_{\mbox{\tiny pre}}}
\def\Hpost{H_{\mbox{\tiny post}}}
\def\SN{N} 
\def\HS{{\mathcal N}} 
\def\is{{\alpha}}
\def\eig{E}
\def\eigg{E'}
\def\eigo{\theta}
\def\tL{n}
\def\NP{\mathbf{NP}}
\def\PP{\mathbf{P}}
\newcommand{\IPR}[1]{\mathrm{IPR}_{#1}}
\newcommand{\AlCentroPagina}[1]{\AddToShipoutPicture*{\AtPageCenter{\makebox(0 ,0){\includegraphics[width =0.7\paperwidth]{#1}}}}}

\usepackage{StyleFiles/watermark}

\makenomenclature
\begin{document}

\begin{titlepage}
\begin{center}
\centering
\mbox{
\begin{minipage}[l]{.20\textwidth}
\includegraphics[width=\textwidth]{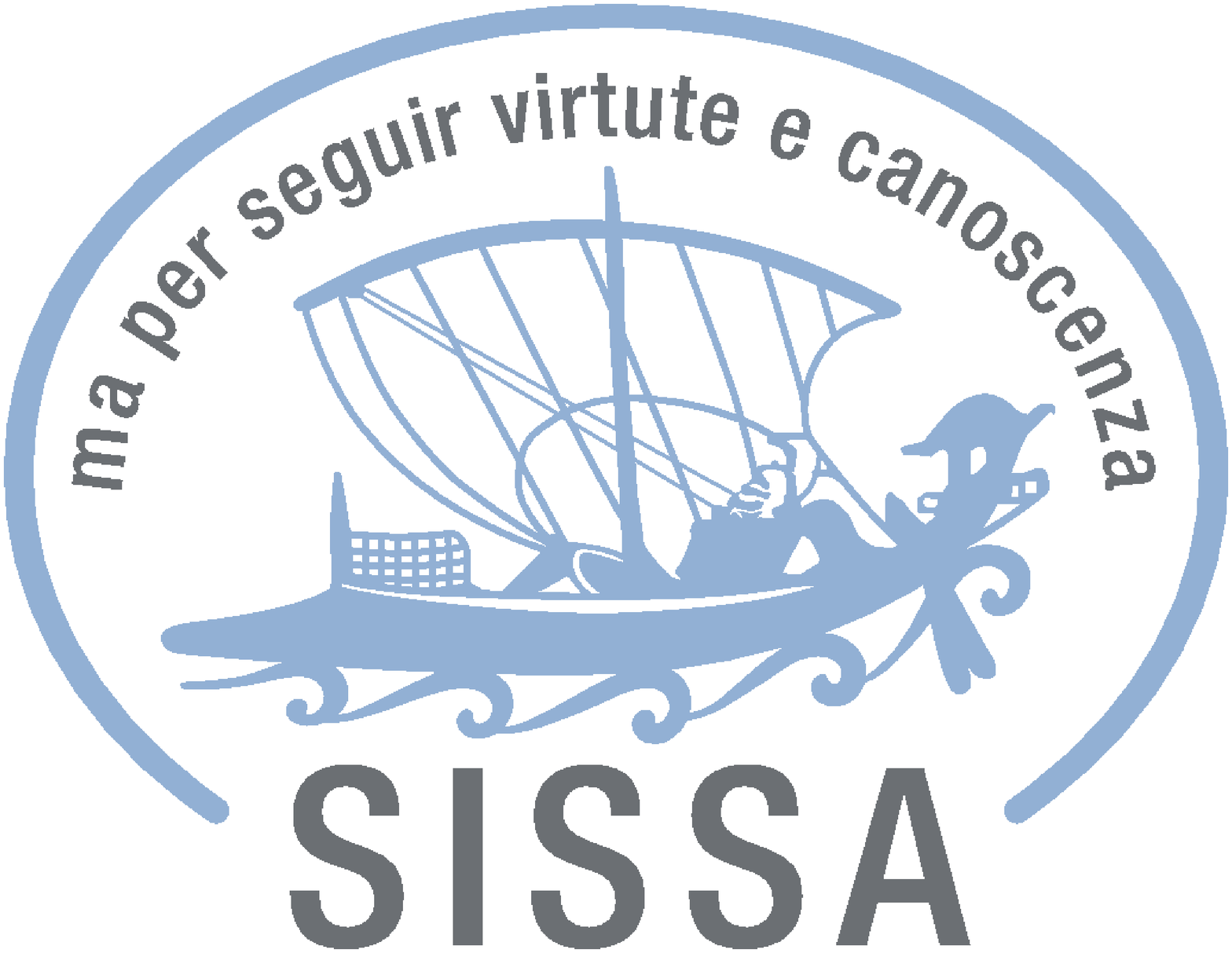}
\end{minipage}%
\quad
\begin{minipage}[c]{.8\textwidth}
\large{\textsc{International School for Advanced Studies}} \rule[0.1cm]{12cm}{0.1mm}
\rule[0.5cm]{12cm}{0.6mm}
\end{minipage}}
\\
\vspace{1cm}
\Large{\textsc{ PhD course in Statistical Physics} }\\
\vspace{0.5cm}
{\Large \textsc{Academic Year 2011/2012}}
\end{center}
\AlCentroPagina{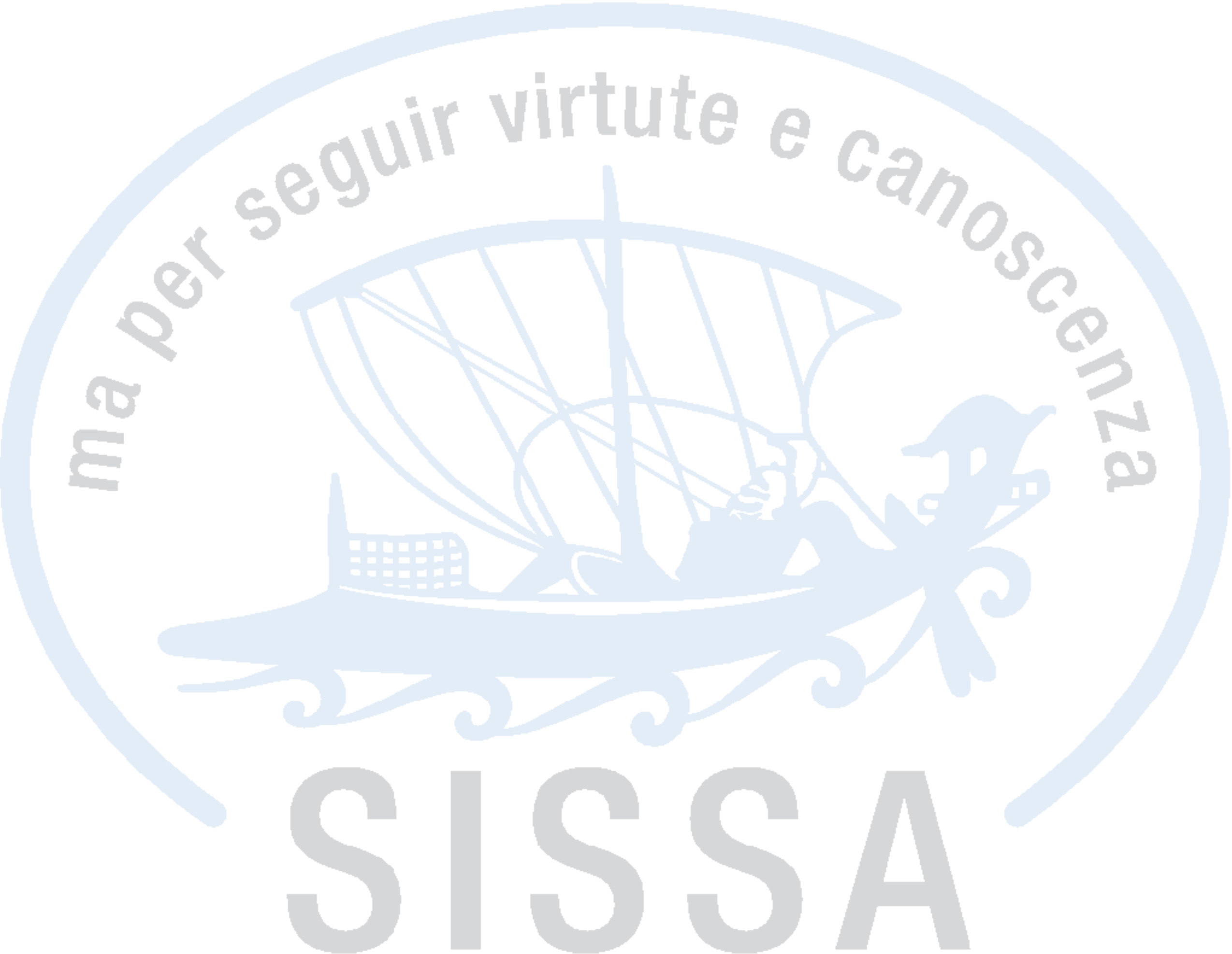}
\vspace{12mm}
\begin{center}
{\Huge{\bf A journey into}}\\
\vspace{5mm}
{\Huge{\bf localization, integrability and}}\\
\vspace{5mm}
{\Huge{\bf thermalization}}\\
\vspace{12mm} {\Large{\textsc{Thesis submitted for the degree of}}}\\
\vspace{0.5cm}
{\Large\textit{Doctor Philosophiae}}
\end{center}
\vspace{30mm}
\par
\noindent
\begin{minipage}[t]{0.5\textwidth}
{\Large{\bf Advisors:\\
Prof.
Antonello Scardicchio\\
Prof. Giuseppe Mussardo}}
\end{minipage}
\hfill
\begin{minipage}[t]{0.5\textwidth}\raggedleft
{\Large{\bf Candidate:\\
Andrea De Luca}}
\end{minipage}
\vspace{0.5cm}
\begin{center}
{\large{\textit{17th September 2012} }}
\end{center}
\end{titlepage}




\setcounter{secnumdepth}{3}
\setcounter{tocdepth}{3}

\frontmatter 
\pagenumbering{roman}
\include{Abstract/abstract}

\tableofcontents
\listoffigures
\printnomenclature 
\addcontentsline{toc}{chapter}{Nomenclature}

\mainmatter 
\include{Introduction/introduction}
\include{AndersonLocalization/andersonloc}

\include{MBL/mbl}
\include{Quench/quench}
\include{Conclusions/conclusions}

\backmatter 

\bibliographystyle{plainnat}
\renewcommand{\bibname}{References} 
\bibliography{References/references} 

\end{document}

%% file: abbreviation.tex
\def\ion#1#2{#1$\;${\small\rm\@Roman{#2}}\relax}

\def\ref@jnl#1{{\jnl@style#1}}
\def\aj{\ref@jnl{AJ}}                   
\def\araa{\ref@jnl{ARA\&A}}             
\def\apj{\ref@jnl{ApJ}}                 
\def\apjl{\ref@jnl{ApJ}}                
\def\apjs{\ref@jnl{ApJS}}               
\def\ao{\ref@jnl{Appl.~Opt.}}           
\def\apss{\ref@jnl{Ap\&SS}}             
\def\aap{\ref@jnl{A\&A}}                
\def\aapr{\ref@jnl{A\&A~Rev.}}          
\def\aaps{\ref@jnl{A\&AS}}              
\def\azh{\ref@jnl{AZh}}                 
\def\baas{\ref@jnl{BAAS}}               
\def\jrasc{\ref@jnl{JRASC}}             
\def\memras{\ref@jnl{MmRAS}}            
\def\mnras{\ref@jnl{MNRAS}}             
\def\pra{\ref@jnl{Phys.~Rev.~A}}        
\def\prb{\ref@jnl{Phys.~Rev.~B}}        
\def\prc{\ref@jnl{Phys.~Rev.~C}}        
\def\prd{\ref@jnl{Phys.~Rev.~D}}        
\def\pre{\ref@jnl{Phys.~Rev.~E}}        
\def\prl{\ref@jnl{Phys.~Rev.~Lett.}}    
\def\pasp{\ref@jnl{PASP}}               
\def\pasj{\ref@jnl{PASJ}}               
\def\qjras{\ref@jnl{QJRAS}}             
\def\skytel{\ref@jnl{S\&T}}             
\def\solphys{\ref@jnl{Sol.~Phys.}}      
\def\sovast{\ref@jnl{Soviet~Ast.}}      
\def\ssr{\ref@jnl{Space~Sci.~Rev.}}     
\def\zap{\ref@jnl{ZAp}}                 
              
\def\iaucirc{\ref@jnl{IAU~Circ.}}
\def\aplett{\ref@jnl{Astrophys.~Lett.}}
\def\apspr{\ref@jnl{Astrophys.~Space~Phys.~Res.}}
\def\bain{\ref@jnl{Bull.~Astron.~Inst.~Netherlands}}
\def\fcp{\ref@jnl{Fund.~Cosmic~Phys.}}
\def\gca{\ref@jnl{Geochim.~Cosmochim.~Acta}}
\def\grl{\ref@jnl{Geophys.~Res.~Lett.}}
\def\jcp{\ref@jnl{J.~Chem.~Phys.}}      
\def\jgr{\ref@jnl{J.~Geophys.~Res.}}    
\def\jqsrt{\ref@jnl{J.~Quant.~Spec.~Radiat.~Transf.}}
\def\memsai{\ref@jnl{Mem.~Soc.~Astron.~Italiana}}
\def\nphysa{\ref@jnl{Nucl.~Phys.~A}}
\def\physrep{\ref@jnl{Phys.~Rep.}}
\def\physscr{\ref@jnl{Phys.~Scr}}
\def\planss{\ref@jnl{Planet.~Space~Sci.}}
\def\procspie{\ref@jnl{Proc.~SPIE}}

%% file: Abstract/abstract.tex

\begin{abstracts}        

We present here the results obtained during my PhD work. Together with a broad introduction
to the Anderson localization transition, the quantum adiabatic algorithm and the quench dynamics problem,
we include our original achievements. 
We report the study of the many body localization transition in a spin chain and the breaking of
ergodicity measured in terms of return probability in a state evolution. In the many-body localized phase an initial
quantum state evolves in a much smaller fraction of the Hilbert space than would be allowed by conservation of energy only. By the anomalous scaling of the participation ratios with system size we are led to consider the eigenfunctions in configuration space by means of the distribution of the wave function coefficients, a canonical observable in modern studies of Anderson localization. We show how the delocalized phase is less ergodic than predicted by random-matrix theory and how in the localized phase some properties of such distribution are very close to those of the Anderson model on the Bethe lattice. We finally present a criterion for the identification of the many-body localization transition based on these distributions which is quite robust and perfectly suited for numerical investigations of a broad class of problems. Moreover, we introduce the Richardson model, an exactly solvable model, that turns out to be suitable for investigating the many-body localized phase. In this way, we show that this phase shares some properties common in glassy systems, such as slowing of the dynamics and fractality.
Then we turn to the analysis of the quench problem in a ensemble of random matrices. 
We analyze the thermalization properties and the validity of the Eigenstate Thermalization Hypothesis
for the typical case, where the quench parameter explicitly breaks a $Z_2$ symmetry.
 Our analysis examines both dense and sparse random matrix ensembles.
 Sparse random matrices may be associated with local quantum Hamiltonians and 
they show a different spread of the observables on the energy eigenstates with respect to the dense ones. In 
particular, the numerical data seems to support the existence of rare states, i.e. states where the observables 
take expectation values which are different compared to the typical ones sampled by the micro-canonical distribution. 
In the case of sparse random matrices we also extract the finite size behavior of two different time scales 
associated with the thermalization process. 
\end{abstracts}



%% file: Introduction/introduction.tex
\chapter*{Introduction}
\begin{figure}[htbp]
\begin{center}
\scalebox{0.30}{\includegraphics{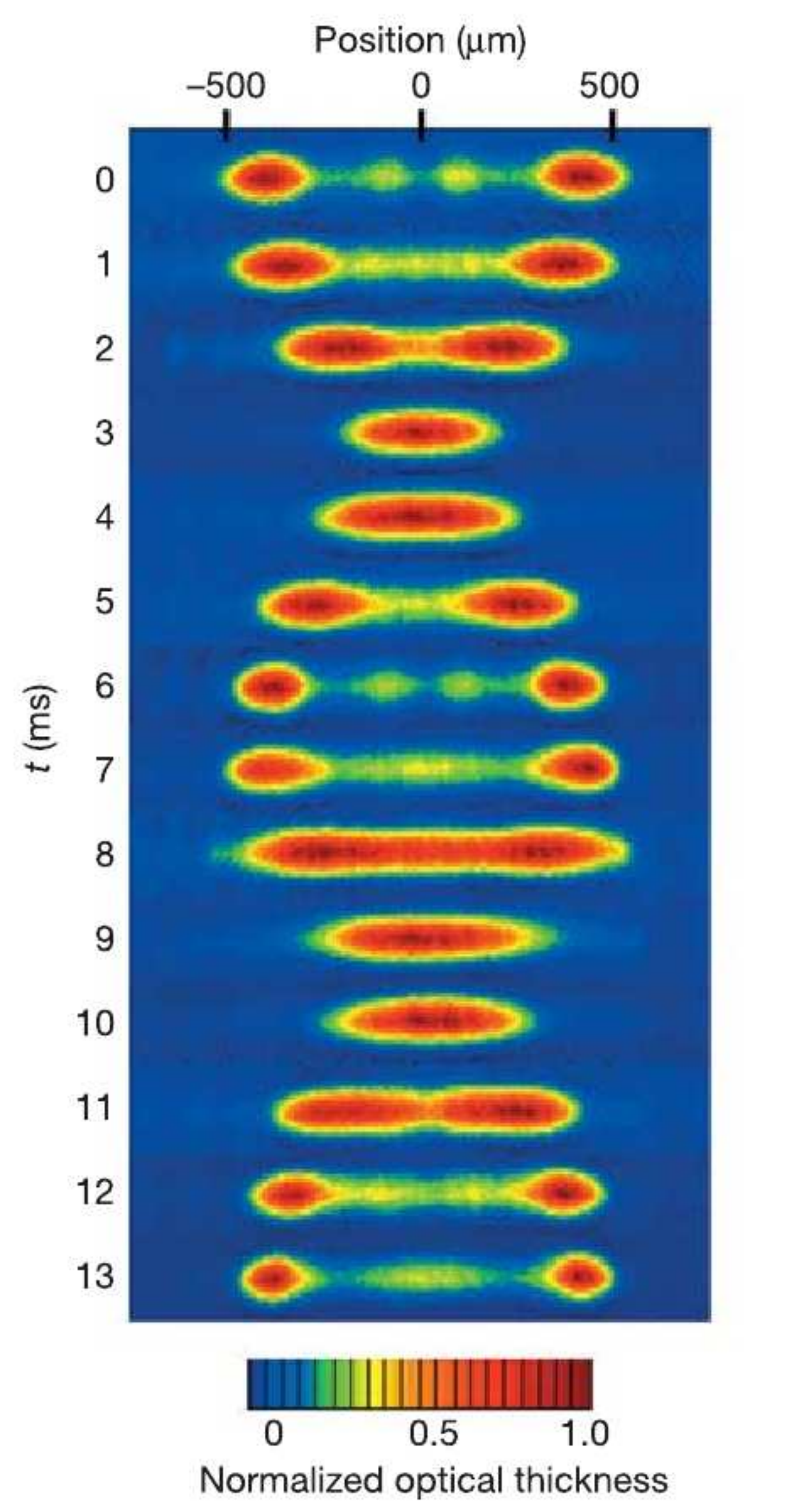}}
\caption{ From \cite{kinoshita2006quantum}, absence of thermalization in a $1d$ Bose gas with $\delta$-like interaction. An explanation of this phenomenon has been searched in the integrability of the corresponding Hamiltonian (known as Lieb-Liniger model \cite{lieb1963exact}) or in the dimensionality of the problem. But a full comprehension is still missing.
}
\label{intro:cradle}
\end{center}
\end{figure}

\begin{figure}[htbp]
\begin{center}
\scalebox{0.73}{\includegraphics{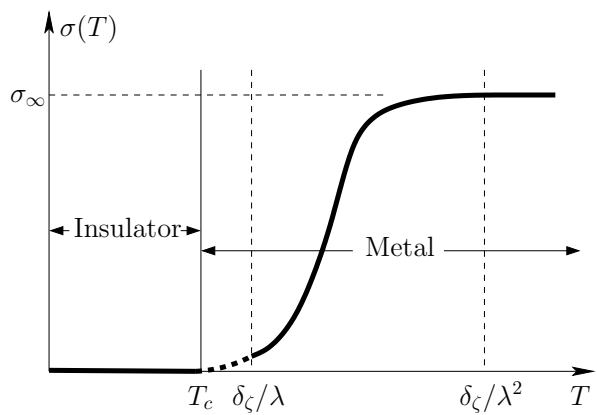}}
\caption{ From \cite{basko2007possible}, schematic temperature dependence of the dc conductivity $σ(T)$ for electrons subject to a disorder potential
localizing all single particle eigenstates, in the presence of
weak short-range electron-electron interaction $λ \delta_\zeta$ , $λ \ll 1$.
 Below the point of the many-body metal-insulator
transition, $T < T_c$, no inelastic relaxation occurs
and $\sigma(T) = 0$. At $T \gg \delta_\zeta/\lambda^2$, Drude theory is applicable.
 In the intermediate region, electron-electron interaction leads to electron transitions between
localized states, and the conductivity depends on temperature as a power-law.
}
\label{intro:baa}
\end{center}
\end{figure}

\ifpdf
    \graphicspath{{Introduction/IntroductionFigs/PNG/}{Introduction/IntroductionFigs/PDF/}{Introduction/IntroductionFigs/}}
\else
    \graphicspath{{Introduction/IntroductionFigs/EPS/}{Introduction/IntroductionFigs/}}
\fi
\markboth{\MakeUppercase{Introduction}}{Introduction}
The study of quantum system in presence of disorder or far away from standard equilibrium condition can be considered as two of the 
most important strands among the current research in condensed-matter physics. 
This is true, not only for their hardness from the theoretical point of view, but also because recent development of experimental techniques
allowed enormous progress in the control of the microscopic details of a quantum systems. So, problems that in the past could look as completely
academic, as it could be the evolution of a closed quantum systems, have nowadays become crucial also experimentally thanks, for example, to the 
cold atoms community \cite{bloch2008many} (see Fig. \ref{intro:cradle} for an important example). 
Last years represented an important development also for the quantum disordered community, with experimental works
able to test and verify the Anderson localization \cite{billy2008direct, roati2008anderson}. We can be therefore optimistic on the possibility to study in a controlled way also a quantum systems in presence of disorder and interactions. 
From the theoretical point of view, the questions regarding thermalization of closed quantum systems, especially in low dimensions, attracted an extraordinary effort, in the attempt to find general prescription and to investigate the mechanism needed to reach thermal equilibrium. While for a finite size system we expect recurrence, it is conceivable that a large system could decay towards a stationary state.
The thermodynamic characterization of this (eventual) stationary state is
one of the most intriguing puzzles of this field. 
The concept of disordered quantum system intertwines with the work of Anderson \cite{anderson1958absence}. 
The notion of localization was originally introduced for a single quantum particle in a random potential.
This concept can also be extended to many-particle systems. Statistical physics of many-body
systems is based on the microcanonical distribution, i. e., all states with a given energy are assumed to be realized
with equal probabilities. This assumption means delocalization in the space of possible states of the system. It
does not hold for non-interacting particles; however, it is commonly believed that an arbitrarily weak interaction
between the particles eventually equilibrates the system and establishes the microcanonical distribution.
However, in Fig. \ref{intro:baa}, we find the summary of the recent results from \cite{basko2006problem}, showing that interacting disordered quantum systems
can be characterized by a low-temperature phase where the system finds it hard to thermalize. This gave strength to the research of a many-body localized phase. It is interesting to notice how the low-temperature in this case, represents a new phase, where conductivity vanishes identically.
So, we see that these problems, though coming from different starting points and perspectives, end up interweaving with deep implications for the foundations itself of quantum statistical mechanics.
The intriguing part is that for both of them, standard techniques as mean-field theory or renormalization theory do not work, or have at least to be reformulated properly. Therefore, beyond the possibility of an extensive perturbative analysis, that may be fruitful in some specific case, as in \cite{basko2006metal}, one has to resort to: numerical methods or exactly solvable toy models.
In this work of thesis, we indeed tried a combination of the two things. Together with exact diagonalization and an accurate analysis of the finite-size corrections, we exploited exact results coming from integrable models, as the Richardson one, or the employment of the Bethe-lattice.
We think that the fundamental issue is coming from the complicate behavior of the typical quantum wave-function, that in presence of many-body effects and disorder may acquire complicate shape. The Anderson localization framework, though derived for a much simpler contest as the single-particle hopping problem, provided us, during the last fifty years, with a set of tools, terminology and concepts that can be applied fruitfully. Our focus was therefore on the breaking of ergodicity in the Hilbert space, an effect related both to the lack of thermalization and the lack of conductivity.

Finally, there is also a technological motivation for this research.
We have already underlined how the perfectly insulating low temperature phase of a disordered many-body system already represents a novelty.
Moreover, it is very plausible that the coherent quantum dynamics will play a major
role in future experimental set up and technologies. An example could be
provided by a quantum computer, that will definitely require the capability
of manipulate interacting system in time. Therefore, a better understanding
of out of equilibrium quantum physics could be crucial for the developing of
new technologies.

This thesis is organized as follows. In chapter \ref{chapt:andersonloc}, we present the main ingredients of the Anderson localization transition, trying to follow the historical development of same aspects. This will be useful as a background in all the rest of the thesis.
In chapter \ref{chapt:mbl}, we turn our attention to the many-body generalization of the localization transition. We will elaborate on its implication for the (quantum) complexity and we will explain our work on the subject coming from \cite{buccheri2011structure, de2012many}.
In chapter \ref{chapt:quench}, we present our work concerning non-equilibrium dynamics in closed quantum systems and in particular on the topic of quantum quenches. With a parallel use of random matrices, Anderson localization and exact diagonalization, we have been able to describe an ensemble of systems showing thermalization and characterize the mechanism behind. The results are published in \cite{brandino2011quench}.

During my PhD, I also worked on a different topic, i.e. the computation of the correction to critical entanglement entropy in a full class of $1d$ exactly solvable models, whose critical points coincide with the conformal minimal points; details can be found in \cite{de2012approaching}.


%% file: AndersonLocalization/andersonloc.tex

\chapter{The Anderson localization transition}
\markboth{\MakeUppercase{\thechapter. Anderson localization }}{\thechapter. Anderson localization}
\label{chapt:andersonloc}
\ifpdf
    \graphicspath{{AndersonLocalization/Figs/PNG/}{AndersonLocalization/Figs/PDF/}{AndersonLocalization/Figs/}}
\else
    \graphicspath{{AndersonLocalization/Figs/EPS/}{AndersonLocalization/Figs/}}
\fi

\section{Introduction}
In 1958, P. Anderson \cite{anderson1958absence} came out with a paper where
he put out the ideas at the
origin of the nowadays incredibly famous phenomenon that goes under the name of
Anderson localization (AL)%
\nomenclature{AL}{Anderson localization}%
. The motivations behind that work were related to some
experimental results from the George Feher's group at Bell Labs \cite{feher1959electron} in the
fifties that were lacking a theoretical explanation. In particular, according to
the classical Drude description of conductance in a metal, one
imagines that electrons (or more generally the charge carriers) collide with
the positive ions in the metal, thus following a diffusive motion. This theory
is at the origin of the well known Ohm's law, where clearly the conductivity
becomes proportional to the mean free path of the electron inside the metal. 
 However, experimental observations made clear that the mean free path inside a
metal is a couple of order of magnitude bigger than the lattice spacing and
only with the development of the quantum theory it was possible to provide an
explanation to this fact: in a regular lattice, electrons behave as waves and
coherently diffract on the ions. The resistivity is therefore appearing only
because of the impurities in the lattice. In this way the Drude theory becomes
again reliable, but the electron is envisaged as zigzagging between impurities.
It follows that the more are the impurities, the shorter will be the mean free
path and the conductivity. 
With this picture in mind, Anderson tried to answer the question: what happens
if the density of impurity is increased? Consistently with the known
experimental results, he was able to provide convincing arguments, showing that
the increasing of lattice disorder would not only keep decreasing the mean free
path and the conductance, but beyond a critical amount of impurity scattering the diffusive,
zigzag motion of the electron is not just reduced, but can come to a complete
halt. The electron becomes trapped and the conductivity vanishes.
Since then, the concept of Anderson localization was
shown to manifest itself in a broad variety of phenomena
in quantum physics. 
\begin{figure}[htbp]
\begin{center}
\scalebox{0.38}{\includegraphics{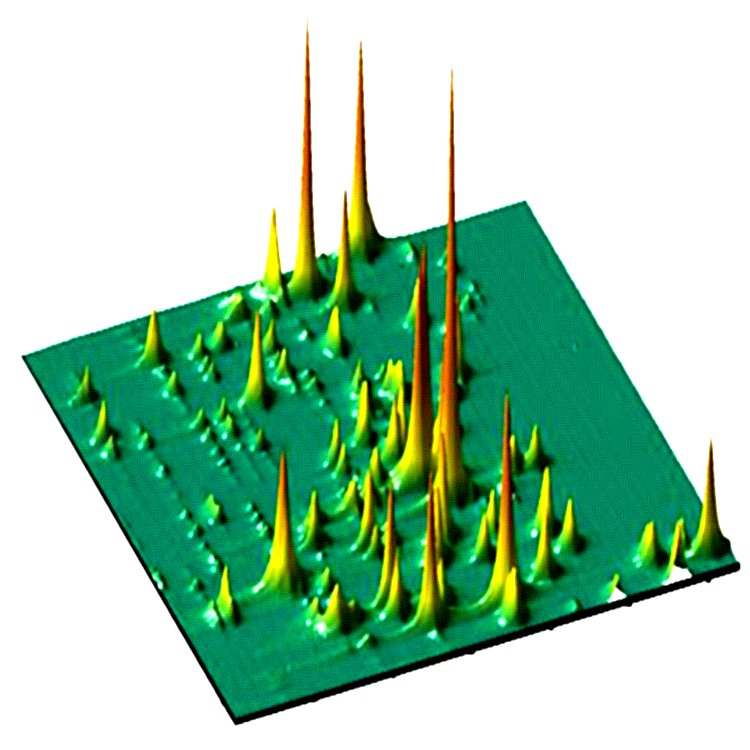}} 
\caption{From \cite{sapienza2010cavity}, Anderson localized modes of light. The high intensity peaks show the random positions where the light emitted in a disordered photonic crystal waveguide becomes strongly localized. These are signatures of Anderson localization of light.}
\label{andersonloc:light}
\end{center}
\end{figure}
Since then, the concept of Anderson localization was
shown to manifest itself in a broad variety of phenomena
in quantum physics. As an important example we report in Fig. \ref{andersonloc:light} 
the results from an experiment of localization of light.

\section{The Anderson model}
\begin{figure}[htbp]
\begin{center}
\scalebox{0.78}{\input{AndersonLocalization/Figs/contour.latex}} 
\caption{The contour integral to recover the time evolutor from the resolvent}
\label{andersonloc:contour}
\end{center}
\end{figure}
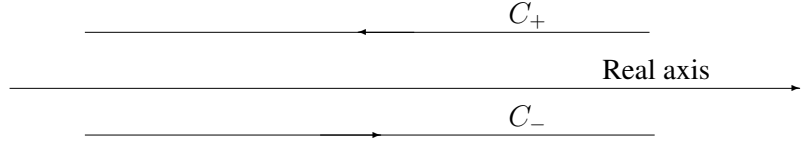

The simplest model involving (spin) transport and disorder showing this phenomenology is now known as Anderson model.
To be more specific, the Hamiltonian, in its simplest
formulation, looks like:
\begin{equation}
 \label{AndersonModel}
H = \sum_{i} \epsilon_i\ket{i}\bra{i} + V = H_0 + V
\end{equation}
here $\epsilon_i$ are independent random variables distributed according to
$P(\epsilon)$ and $V$ is the hopping term, that can be short-range, i.e. the adjacency matrix of the lattice under consideration,
or long-range with a potential of a specified form. For the moment we will consider the cases:
\begin{itemize}
 \item $P(\epsilon)= \chi_{[-\frac W 2, \frac W 2]} (\epsilon) $, where $\chi_I$
is the characteristic function of the set $I$ and $W$ can be seen as a measure of the
disorder strength;
 \item the lattice is the cubic lattice in $d$ dimensions.
\end{itemize}
The Schr\"odinger equation takes the form
\begin{equation}
 \label{schrod}
\ii \dot a_i = \epsilon_i a_i + \sum_j V_{ij} a_j
\end{equation}
Then, the question that one would like to answer regards the long-time dynamics of an initially localized
state: taking as initial condition $ a_i = \delta_{i0}$, i.e. the  wave
function is completely concentrated at the origin, the typical long
time dynamics is investigated perturbatively in the hopping. In order to formulate the perturbative expansion,
it is useful to introduce the resolvent
\begin{equation}
\label{resolventDef}
G(z) \equiv (z-H)^{-1}
\end{equation}
and the corresponding expression for $G_0(z)$ and $H_0$. It is clear that in
terms of this expression we can easily recover the time evolution operator by
contour integral
\begin{equation}
U(t) =  \int_{\mathcal{C}} G(z) e^{i t z} dz
\end{equation}
where $\mathcal{C}=C_+ \cup C_-$ is the contour depicted in the figure
\ref{andersonloc:contour}

We use the following identity holding for two invertible operators $A,B$
\begin{equation}
 A^{-1} + B^{-1} = A^{-1}(B-A)B^{-1}
\end{equation}
to derive
\begin{equation}
 \label{perturbativeResolvent}
G(z) = G_0(z) + G(z) V G_0(z) = G_0(z) + G_0(z) V
G(z) = G_0(z) + G_0(z) T(z) G_0(z) 
\end{equation}
where we introduced the $t$-matrix with its perturbative expansion as
\begin{equation}
  \label{tmatrix}
T(z) \equiv V G(z) (z - H_0) = V + V G_0(z) V + V G_0(z) V G_0(z) V + \ldots
\end{equation}
By setting
\begin{equation}
\label{Gdefinitions}
G_{ij}(z) \equiv \bra{i} G(z) \ket{j}; \quad g(z) \equiv G_{00}(z)\;.
\end{equation}
we derive the expansion, where we assume the sum over repeated indexes
\begin{equation}
\label{gexpansion}
g(z) = \frac{1}{z - \epsilon_0} + \frac{1}{z - \epsilon_0} V_{0j} \frac{1}{z -
\epsilon_j} V_{j0}\frac{1}{z - \epsilon_0} + \frac{1}{z - \epsilon_0} V_{0j} \frac{1}{z -
\epsilon_j} V_{jl} \frac{1}{z -
\epsilon_l} V_{l0}\frac{1}{z - \epsilon_0} + \ldots
\end{equation}
It is useful to introduce the self-energy $\Sigma(s)$, defined implicitly by
the equation
\begin{equation}
\label{selfenergy}
 g(z) \equiv \frac{1}{z - \epsilon_0 - \Sigma(z)}
\end{equation}
and combining \eqref{selfenergy} and \eqref{perturbativeResolvent}, one
obtains the expression useful for the perturbative expansion
\begin{equation}
\label{selfenergyexpansion0}
 \Sigma(z) = (z-\epsilon_0)\left[1-\bra{0}(1-G_0(z) V)^{-1}\ket{0}^{-1} \right]
\end{equation}
that can be written as 
\begin{equation}
\label{selfenergyexpansion}
\Sigma(z) = \sum_n \sum_{l_1,\ldots,l_n \neq 0} \frac{V_{0 l_1} V_{l_1 l_2} \ldots V_{l_n 0}}{(z-l_1)\ldots (z-l_n)} 
\end{equation}
The long-time dynamics corresponds to the behavior of $G(z)$ close to the real
axis. The self-energy is also called \textit{energy-shift
operator}\cite{cohen1998atom}, because it moves the position of the poles of
$g(z)$. In particular, if we suppose that for $z_0 = \omega - \ii \tau^{-1} $,
we have
$$z_0-\epsilon_0 - \Sigma(z_0) = 0$$
we derive by Laplace transform for large
time $t$
\begin{equation}
 \label{laplacetransform}
g(z) \simeq \frac{A}{z - \omega + \ii \tau^{-1}} \Rightarrow a_0 (t) \simeq
A e^{-\ii \omega t - \frac{t}{\tau}}
\end{equation}
It is clear that if $\tau$ is finite, the wave function will spread from the
origin, being delocalized. If instead $\tau$ is infinite, the wave function, or
at least a finite fraction of it given by the constant $A$, will remain at the
origin and no transport will take place. Two remarks are in order
\begin{itemize}
 \item being mathematically rigorous, from the definition \eqref{resolventDef}
of the resolvent, it is clear that, since $H$ is an hermitian operator,
each component of $G(z)$ can only have singularities on the real axis,
corresponding to isolated poles or branch cuts in presence of a continuous
spectrum: a finite $\tau$ would therefore be odd. However, the analytic
continuation of $g(z)$ along the branch cuts goes on an other Riemann sheet,
where the function can have new singularities \cite{cohen1998atom};
 \item since the Hamiltonian $H$ involves random variables, the question whether
$\tau$ is finite or not, becomes a probability question that, as we will see in
detail, is not simply answered considering average values. 
\end{itemize}
To be more concrete, it is easy to see that the average of the Green function does not provide a
quantity good to distinguish the localized from the delocalized phase. This was at the origin of the confusion during the first years after Anderson's one. In \cite{lloyd1969exactly}, it is considered the Anderson model, with a Cauchy distribution for the diagonal energies (Lloyd model):
\begin{equation}
 \label{cauchydistrib}
P(\epsilon) = \frac{1}{\pi}\left[\frac{\gamma}{(\epsilon-\epsilon_0)^2 + \gamma^2}\right]
\end{equation}
In this case the average of the Green function and therefore the average density of states, can be computed analitycally. It turns out that in any dimension $d$, the average of the Green function is given by the ordered one where the diagonal energies are all equal and given by $\epsilon_0 + i \gamma$. This result was used as a prove that no localization can occur in this model. However, it is clear that the long tails of the Cauchy distribution are practically irrelevant in high dimensions, where the wave function can always avoid values very far from the average. So, this result would have strong implication on the Anderson work too. However, the problem is that the average of the Green function is not a meaningful quantity for this transition. One can define an order parameter for the Anderson
transition as the long time average of the return probability, i.e. the probability
to find particle at the origin
\begin{equation}
\label{anderson:returnprobability}
 \IPR{2} \equiv \lim_{\tau \to \infty} \frac 1 \tau \int_0^\tau dt | a_0(t)|^2 dt
= \sum_e | \psi_0 (e) |^4 = \lim_{\epsilon \to 0}
\frac{\epsilon}{\pi}\int_{-\infty}^\infty g(
   p + \ii \epsilon) g(p - \ii \epsilon) dp
\end{equation}
where the sum is over the eigenstates $\psi(e)$ of $H$. Here $(\IPR 2)^{-1}$ is a measure of the portion of explored Hilbert
space during the quantum dynamics and it is usually dubbed \textit{
participation ratio} (PR). The average of this quantity, that involves the square of the Green function, can distinguish localized and extended states. Unfortunately, its analytic computation is not possible, even for the Lloyd model. We will come back on that in \ref{mbl:xxz}, where we will use a numerical approach to the problem. For now, we will use a different technique.
\section{The perturbative expansion}
\subsection{Second order}
\label{subsection:secondorder}
The use of the self-energy introduced in \eqref{selfenergy} is particularly
fruitful if one wants to consider an approximate perturbative expression.
In fact, at the lowest non-zero order in $V$ we obtain from
\eqref{selfenergyexpansion}, we get
\begin{equation}
 \label{selfenergylowest}
 \Sigma(z) = \sum_{j\neq 0}\frac{|V_{0j}|^2}{
z-\epsilon_j } + \Ord{V^3}
\end{equation}
Inserting it in \eqref{selfenergy}, we obtain
$$ g(z) = \frac{1}{z - \epsilon_0 - \sum_{j\neq 0}\frac{|V_{0j}|^2}{
z-\epsilon_j }}$$
To recover the same expression from \eqref{gexpansion}, one already needs to
re sum an infinite subset of terms. In fact, it is useful to notice that Eq.
\eqref{gexpansion} at any finite order in $V$ will still have poles
corresponding to the spectrum of $H_0$: it is only resumming an infinite subset
of them that their positions can be shifted; this procedure is automatically
encoded in the self-energy.
Now, the expression appearing in \eqref{selfenergylowest} is a random variable
due to the randomness of the energies $\epsilon_j$. We are interested in
investigating its behavior close to the unperturbed energy $\epsilon_0$ and in
presence of a small positive imaginary part: as we saw in
\eqref{laplacetransform}, the imaginary part will be related to the decaying
time $\tau$. Therefore setting $z = s + \ii x$ where $x$ is
a small real quantity, and using the Sokhotski-Plemelj theorem, we get from
the distributional limit of \eqref{selfenergylowest} 
\begin{equation}
 \lim_{x \to 0} \Sigma(s + \ii x) = \sum_{j \neq 0}
 \mathcal{P}\left(\frac{|V_{0j}|^2 }{s-\epsilon_j}\right)-\ii
\pi\sum_{j\neq 0}   |V_{0j}^2| \delta(s-\epsilon_j)
\end{equation}
When $s \to \epsilon_0$, we notice that the first term, i.e. the real part of
the self-energy, just reproduce the second order correction in $V$ to the
eigenvalue $\epsilon_0$ of $H_0$. The second term is just the Fermi's
golden rule,
directly related to the
decaying time $\tau$ and the delta functions impose the energy conservation. It
is clear that for any finite system, quantum mechanics predicts revivals, so the
amplitude of the
initial state can never really decay; indeed this term vanishes, with
probability one the delta function are never satisfied and energy conservation
forbids hopping unless one has a continuous spectrum. Here, the problem becomes
even more subtle: it is a probability question to be settle already in the
thermodynamic limit. Setting
$$ \tau_x^{-1} \equiv - \Im \Sigma(\epsilon_0 + i x) = \sum_{j \neq 0}
\frac{x |V_{0j}|^2}{x^2 + (\epsilon_0 - \epsilon_j)^2}$$ 
we can investigate this expression as a random variable, whose probability
distribution can in principle be computed. Introducing
$$ y_j =  \frac{1}{x^2 + (\epsilon_0 - \epsilon_j)^2}; \qquad Y = (x
\tau_x)^{-1} = \sum_j |V_{0j}|^2 y_j $$
In order to simplify our computations without changing the substance of the
results, we assume that energy at the origin falls in the middle of the band,
$\epsilon_0 \ll W$. Then we have for the probability distribution function (pdf)
\nomenclature{pdf}{probability distribution function}
$$ \rho(y) \equiv \left\{\begin{array}{ll}
	      \frac{1}{W y \sqrt{y - x^2 y^2}} & y \in \left[\frac{4}{4x^2 +
W^2},x^{-2} \right]\\ 0 & \mbox{otherwise}
                        \end{array}\right.
$$
The Laplace transform of the pdf can be computed for small $x$
\begin{equation}
 \label{ylaplace}
\phi(s) \equiv \int^\infty_0 dy e^{- s y} \rho(y) = 
e^{-\frac{4s}{W^2}} + \frac{2\sqrt{\pi
s}}{W}\left[\Erf\left(\frac{2\sqrt{s}}{W}\right) -
\Erf\left(\frac{\sqrt{s}}{x}\right)\right]
\end{equation}
Now, the Laplace transform of the pdf of $Y$ is obtained by product:
\begin{equation}
\label{laplaceproduct}
 \Phi(s) \equiv \int_0^\infty dY e^{- s Y} \rho(Y) = \prod_i \phi(s |V_{0i}|^2)
\end{equation}
Now since we are interested in both $s\to 0$ and $x\to 0$ limit, it is useful
to consider the expansion
\begin{equation}
 \label{phiexpansion}
 \phi(s) \stackrel{\scriptstyle{s\ll1}}{\simeq} \left\{ \begin{array}{ll}
                           e^{- \frac{2 \sqrt{\pi s}}{W}} & \sqrt{s} \gg x\\ 
			   1 - \frac{2 s}{W x} & \sqrt{s} \ll x
                          \end{array}
\right.
\end{equation}

\begin{figure}[htbp]
\begin{center}
\scalebox{0.78}{\includegraphics{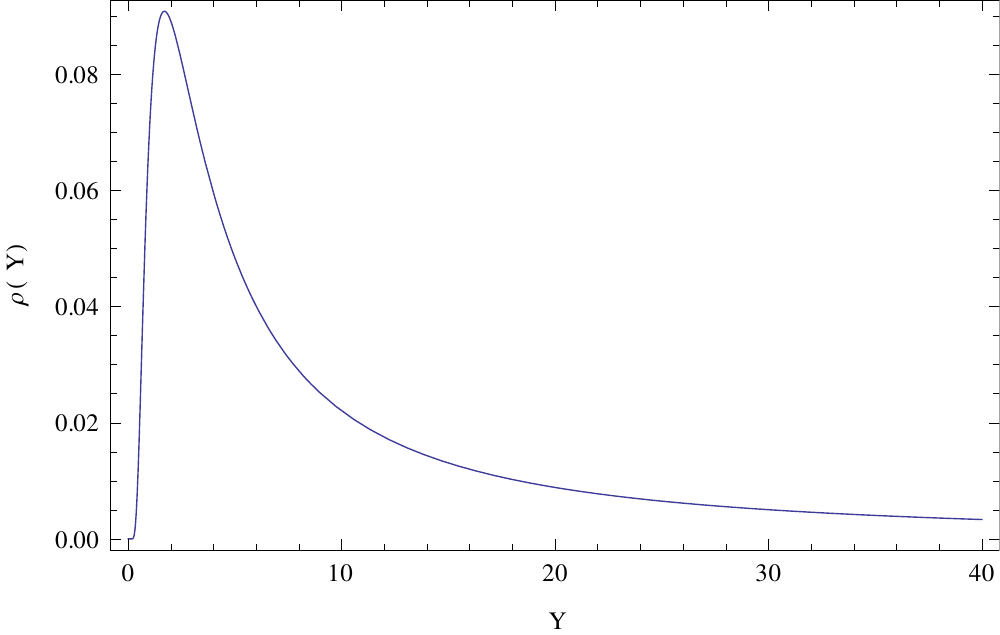}} 
\caption{Behavior of the pdf for $Y$: the finite most probable value is showed,
together with the long $3/2$ power-law tail, inducing a divergent average
value.}
\label{andersonloc:typicalvalue32tail}
\end{center}
\end{figure}

It is illustrative of the mechanism involved to consider some concrete
examples. Suppose that $V_{ij}$ is $V_0$ times the adjacency matrix of
the cubic lattice in $d$-dimensions. Being a finite product, the $x\to 0$ limit
can be taken in \eqref{laplaceproduct} without troubles. For large $d$ we can
obtain an approximate result
\begin{equation}
 \label{inverseLaplacePhi}
\Phi(s) \simeq e^{- \frac{4 d \sqrt{\pi s} V_0}{W}}  \Rightarrow \rho(Y) \simeq
\frac{e^{-\frac{4 d^2 \pi V_0^2}{t W^2}}}{t^{\frac{3}{2}}}
\end{equation}
and the resulting distribution is plotted in Fig.
\ref{andersonloc:typicalvalue32tail}.
This example contains already some of the ingredients fundamental in Anderson
localization: the derived distribution function has a long power-law tail and
so the first momentum, i.e. the average value, is divergent. However, the most
probable value is finite, and we obtain therefore
\begin{equation}
 \label{typicaltime}
 \left(\tau_x^{-1}\right)_{\mbox{\scriptsize M.P.}} \simeq 16 x d^2 \pi \left(\frac{V_0^2}{W^2}\right)
\end{equation}
\nomenclature{M.P.}{Most probable}
showing that the decaying time diverges and the state never delocalizes.
An interesting generalization is already contained in
\cite{anderson1958absence}, i.e. a potential decaying as $V(r)
= \frac{V_0}{r^{3+\epsilon}}$ for a $3$ dimensional lattice. For $\epsilon>0$,
again the $x\to 0$ limit is harmless and it is enough to replace
in \eqref{inverseLaplacePhi} and \eqref{typicaltime} 
\begin{equation}
 \label{potentialintegral}
 d \rightarrow \frac{n}{2}\int_1^\infty V(r) 4 \pi r^2 dr = \frac{4 \pi
n}{\epsilon}
\end{equation}

For the limiting case $\epsilon=0$, this integral becomes logarithmic divergent
but a natural cut-off is obtained by \eqref{phiexpansion}, i.e. $r\ll
\left(\frac{x}{V_0 \sqrt{s}}\right)^{\frac 1 3}$, from which we derive
$$ \Im \Sigma(\epsilon_0 + i x) = \left(\tau_x^{-1}\right)_{\mbox{\scriptsize M.P.}} \propto \frac{x
V_0^2 (\log x)^2}{W^2} $$
and once inserted in \eqref{selfenergy} and inverting the Laplace transform,
turns out in a slower than exponential decay ($\simeq \exp(-\alpha (\log
t)^2)$). In this example, therefore, transport takes place even though in a
particular way, where, due to the logarithmic divergence of the integral in
\eqref{potentialintegral}, long-range hopping (for large value of $r$) are
crucial. Here, we limited ourself to consider the first non-zero order in
the perturbative expansion of the self-energy, i.e. the second order in the hopping term $V$: whenever $\epsilon>0$ no delocalization, or transport occurs.
In particular, for short range hoping (e.g. nearest-neighbor), from this approach, one should conclude that no transition would ever occur. However,
as we will see in the following sections, this is due to the second order in the perturbative expansion. The AL transition, being inherently non-perturbative, requires the analysis of the full perturbative series: at any order, still finite, the decaying time will look infinite, and no delocalization, or transport, will ever occur. However, what can happen is that the full perturbative series is actually divergent, making meaningless all the finite order truncation. We will investigate therefore, the nature of the series in the following section.
\subsection{The multiple-scattering technique}
\label{anderson:multiple}
\begin{figure}[htbp]
\begin{center}
\scalebox{0.78}{\includegraphics{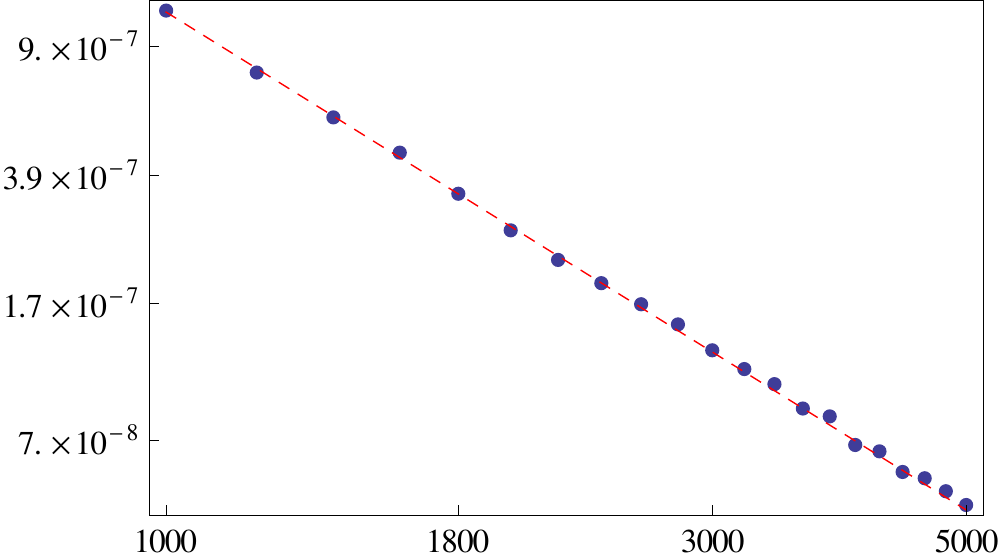}} 
\caption{The scaling of the typical minimum gap with increasing lattice size
$N$; the dashed line is the power-law $N^{-2}$}
\label{andersonloc:minimumgap}
\end{center}
\end{figure}

\begin{figure}[htbp]
\begin{center}
\scalebox{0.78}{\includegraphics{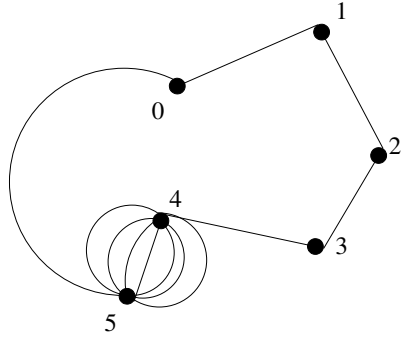}} 
\caption{An example of a path in the perturbative expansion where the repeated hopping among two sites ($4$ and $5$ in this example) can give an arbitrarily big contribution.}
\label{andersonloc:resonanceLoop}
\end{center}
\end{figure}

The perturbative series in the hopping term can be investigated considering the expansion \eqref{tmatrix} or \eqref{gexpansion}, or more specifically the self-energy expansion appearing in \eqref{selfenergyexpansion}. The main problem with this expansion is that it is affected by trivial divergences whenever
the hopping term is big enough to induce a \textbf{level crossing}. For example, for a $2$
sites system, we have
$$ H = \left(\begin{array}{cc}
            e_1 & V\\ V & e_2
            \end{array}\right)$$
and the spectrum is given by
$$ e_\pm = \frac{\bar e}{2} \pm \frac{\Delta e}{2}\sqrt{1 + \frac{4
V^2}{\Delta e^2}} $$
where $\bar e = \frac 1 2\Tr H$ and $\Delta e = e_1-e_2$; by comparison with \ref{selfenergy}, we see that the self-energy is given by the second term in this expression for the eigenvalues.
Since the square root can be expanded perturbatively for small $V$ only for 
$$ | V | \leq \frac{\Delta e}{2} $$
we can conclude that this has to be the radius of convergence for the perturbative expansion appearing in \eqref{selfenergyexpansion}.
Increasing the system size, the situation gets worse and worse, since
such bound has to be replaced by the minimum level spacing, that for $N = L^d$
random variables $e_1,\ldots, e_N$ goes down as $N^{-2}$ as showed in Fig.
\ref{andersonloc:minimumgap}.
It means the radius of convergence of the perturbative expansion is vanishingly
small, making the analysis of the full series completely meaningless. 
The strategy used by Anderson to face this issue follows the technique known as multiple-scattering \cite{watson1957multiple}.
In simple terms, the perturbation series appearing in \ref{selfenergyexpansion}, can be rewritten graphically as the sum over the paths (actually loops) starting from the origin and coming back to the origin itself, without never touching the origin but at the end. 
However, repetitions are allowed in these loops and therefore if we have a pair $j,k$ such that
$$ \left | \frac{1}{z - e_j} V_{jk} \frac{1}{z - e_k} V_{kj} \right | > 1 $$
then we can get terms arbitrarily big in the perturbative expansion just taking paths that pass more and more times between $j$ and $k$: this type of diagrams involves a ``ladder'' from $j$ to $k$ and an example is shown in Fig. \ref{andersonloc:resonanceLoop}.

One can see that all these divergent terms can be formally resummed, thus producing a perturbative expansion where the sum involves only self-avoiding loops where no repetition among the indexes are allowed.
The problem with these procedure is that all the unperturbed energies $\epsilon_k$, appearing in the denominators, have to be modified, defining as in \eqref{selfenergy}, a hierarchical set of self-energy to be fixed self-consistently. To be more explicit, we consider the expansion \eqref{gexpansion}. It is useful to rewrite it in the following form
\begin{equation} 
\label{gwithhop}
g(z) = \frac{1}{z-\epsilon_0} + \frac{1}{(z-\epsilon_0)^2} \sum_{\{k,j\} \neq 0} V_{0k} G_{k,j}(z) V_{j0} 
\end{equation}
and the last term can be written in terms of path connecting $k$ and $j$
\begin{equation}
 \label{Gkj}
 G_{k,j}(z) = \frac{\delta_{k,l}}{z-\epsilon_0} + \sum_n \sum_{l_1,\ldots,l_n} \frac{V_{k l_1} V_{l_1 l_2} \ldots V_{l_n j}}{(z-\epsilon_k)(z-\epsilon_{l_1})\ldots (z-\epsilon_{l_n})(z-\epsilon_l)} 
\end{equation}
and we will show how the terms in the sum can be rearranged. Let's fix a set of indexes representing a path connecting $k$ to $j$ with no repetitions: $g_0 = k, g_1, g_2, \ldots, g_n = j$\footnote{for simplicity we assume $k\neq j$; what has to be changed for the case $k=j$ will become clear.} and no repetitions occur.
\begin{figure}[htbp]
\begin{center}
\scalebox{0.78}{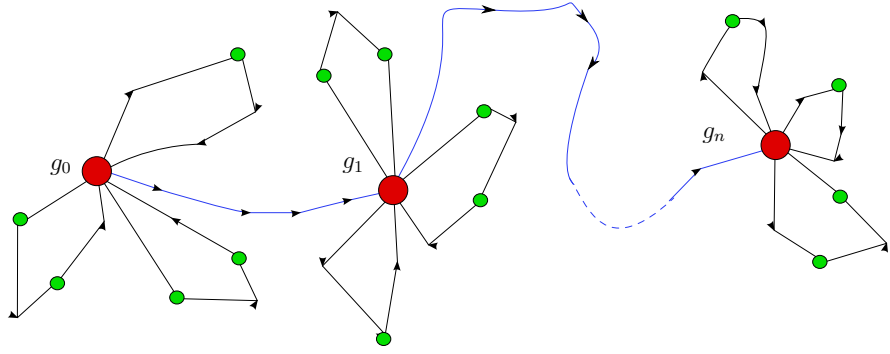} 
\caption{An example of a diagrammatic representation of one term of the perturbative series \eqref{Gkj}: repeated indexes manifest themselves as loops around $g_i$. Real processes (blue) are here indicated as opposed to virtual process (black) where the system coherently returns to the initial state and just renormalize the unperturbed energies. }
\label{andersonloc:virtualreal}
\end{center}
\end{figure}
Among all the terms in the sum of \eqref{Gkj}, as shown in Fig. \ref{andersonloc:virtualreal}, we collect the subset of them where, except for the loops corresponding to the index repetitions and that we address as \textit{virtual} processes, the system undergoes \textit{real} transition along the path $g_0 \to g_1 \to \ldots \to g_n$. The idea is to sum up step by step the virtual processes around $g_0$, then $g_1$, so on and so forth. To be more concrete, we analyze the first case $g_0 = k$. We notice that each term will start from $k$, will come back there in principle many times, and then will leave it definitely. If we put together all the terms that coincide after the last occurrence of $k$, we notice that the final part, non-containing $k$, being common among them, can be factorized. The remaining factor corresponds to the sum of all the possible loops around $g_0 = k$. Therefore we have
\begin{equation}
G_{kj}(z) = \frac{1}{z-\epsilon_k - \Sigma_k (z)} \sum_{k'\neq k} V_{k k'} G_{k'j}^{\neq k}(z)
\end{equation}
where the first term, except for the replacement $k \leftrightarrow 0$ is analogous to \eqref{selfenergy} and appears therefore as a shift to the unperturbed energy. Instead, the last term is analogous to \eqref{Gkj}, but for the fact that $k$ should not appear in any path. Therefore, repeating the previous argument, it can be rewritten as
$$ G_{k',j}^{\neq k} = \frac{1}{z-\epsilon_{k'} - \Sigma_{k'}^{\neq k} (z)} \sum_{k''\neq k'} V_{k' k''} G_{k''j}^{\neq k,k'}(z)$$
where now the self-energy in the denominator is computed using \eqref{selfenergyexpansion}, but avoiding $k$ in any path. Now it is clear how the procedure goes on taking $k' = g_1$ and iterating for the rest of the path $g_2, g_3, \ldots$: \eqref{gwithhop} has been rewritten as a sum over just the loops without repetitions, but for any of these terms, the denominators have to be shifted with the appropriate set of path-dependent self-energies. Notice that in the same way, one can arrange the expansion \eqref{selfenergyexpansion} such that it will contain only path without repetitions, and the denominators will involve all the possible $\Sigma_{k_0}^{\neq k_1 k_2 \ldots}(z) $, that have to be determined self-consistently. However, one can wonder the usefulness of this rearrangement, since the resulting set of coupled equations, in general, does not allow for any explicit solution. We will see an explicit example, i.e. the Bethe-lattice in \ref{bethelattice}, where the locally tree-like structure of the 
lattice allows for a simplification of the self-energies equations. 
For scattering problem, it happens typically that from every site, it is possible to hop in a large number of new sites. 
and the condition of no-index repetition that characterizes self-avoiding loops, becomes actually irrelevant.
Instead, for the disordered case we are considering, Anderson argued that the effect in the denominators due to the resummation is to increase the smallest of them, only slightly changing the nature of the series itself. Therefore, neglecting the self-energies in the denominators provide a larger value for the series. Since, we can estimate the position of the transition point as the value of the ration $\frac{W}{V}$ that makes the perturbative expansion divergent, it turns out that within this approximation we will be overestimating the transition point\footnote{for this reason, Anderson addresses the result obtained in (84) of \cite{anderson1958absence}, in this approximation where the denominators are just considered as the unperturbed ones, an \textit{upper limit}.}. 
\subsection{An estimation for the transition point}
\label{anderson:estimation}
In the approximation suggested by Anderson, it is possible to derive an estimation for the critical point of the Anderson transition. As we explain in the previous section, it represents an over-estimation. We will follow \cite{thouless1970anderson} in the derivation. We will assume for simplicity that the hopping matrix will just be $\mathbb{V} = V A$ where $A$ is the adjacency matrix of the cubic lattice and $V$ is a measure of the hopping strength. 
As we did in \ref{subsection:secondorder}, we will assume $\epsilon_0 \ll W$ and set $z = 0$, so that the general term at order $L$ in the expansion we derived in the last section will look like
\begin{equation}
\label{TLterm}
T_L = \frac{V^L}{\epsilon_{l_1} \epsilon_{l_2} \ldots \epsilon_{l_L} }
\end{equation} 
Here, all the indexes in the denominators have to be different, due to the multiple-scattering rearrangement we explained in the last section. Moreover, as we said, we neglect the self-energies that should shift the unperturbed energy. So the result is that the denominator will be the product of $L$ independent random variables, uniformly distributed in $\left[-\frac W 2, \frac W 2\right]$. Once set
\begin{eqnarray}
a_i &=& -\ln \left| \frac{2\epsilon_{l_i}}{W} \right|\\
A &= &\sum_i a_i 
\end{eqnarray}
we have that $|T_L| = \left(\frac{2V}{W}\right)^L e^A = \left(\frac{2V}{W}\right)^L \tau_L $. Computing the distribution functions, with the standard method of Laplace transform, we obtain
\begin{eqnarray}
 P_a(a) &=& e^{-a} \theta (a) \\
 P_A(A) &=& \frac{e^{-A} A^{L-1}}{(L-1)!} \theta(A) \\
 P_{\tau}(\tau_L) &=& \frac{(\ln \tau_L)^{L-1}}{(L-1)! \tau_L^2} \label{taudistrib}
\end{eqnarray}
Now, at each order $L$, the number of terms like $T_L$ in \eqref{TLterm} will be exponentially large in $L$, $K^L$, being given by the number of path with no loops from the origin of length $L$. 
\begin{equation}
\label{orderL}
O_L = \sum_{i \in \{\mbox{\scriptsize path}\}}^{K^L} \pm \tau_L^{(i)}
\end{equation}
We are now interested to the distribution of $|O_L|$. If we were able to assume that all the terms involved are independent random variable, then the distribution of the sum would be computable with standard methods. This is clearly false, since the number of independent variable is polynomial in $L$, while the number of path, and so of variable $T_L$ is exponential in $L$. However, assuming independence will overestimate $O_L$, therefore it is consistent with the previous approximation. 
Now, $O_L$, the sum of a large number of positive and negative random variables with a long-tailed distribution function, is dominated by the maximum of the variables. Therefore, instead that explicitly computing the distribution, for large value of $O_L$, it can be replaced with the distribution of the maximum of the $K_L$ variables $\tau_L$. The standard argument to obtain it is
$$ \Prob(|O_L|<O) \simeq \operatorname{Prob}(\max\{\tau_L^{(1)}, \ldots, \tau_L^{(K^L)}\}<O) = \left(1 - \int_O^\infty P(\tau_L) d \tau_L \right)^{K^L} $$ 
Differentiating, we get the pdf for $O_L$ for large $O$
\begin{equation}
\label{Oestimation}
P_{|O_L|}(O) \simeq K^L P_{\tau} (O) 
\end{equation}
In order to compute the probability that the series 
$$ S = \sum_L \left(\frac{2 V}{W}\right)^L O_L$$
is convergent, we compute the probability that it is dominated by a convergent geometric series, i.e. 
\begin{equation}
\label{Pconvergence}
\mathbb{P} = \Prob\left(\exists x<1, N\; |\; \forall L > N, \; |O_L| < \left(\frac{x W}{2V}\right)^L\right) = \lim_{x\to 1} \lim_{N\to \infty} \prod_{L=N}^\infty \int_0^{\left(\frac{x W}{2V}\right)^L} P_{|O_L|}(O) dO 
\end{equation}
To estimate the product of integrals, we notice that the extreme of integration will be big, therefore using \eqref{taudistrib} and \eqref{Oestimation}
$$\int_{\left(\frac{x W}{2V}\right)^L}^\infty P_{|O_L|}(O) dO = K^L \int_{\left(\frac{x W}{2V}\right)^L}^\infty  P_{\tau} (O) \simeq  \left(\frac{2V e K \ln \left(\frac{W x}{2V}\right)}{x W}\right)^L O\left(L^{-\frac 1 2}\right)$$
Inserting this approximation in \eqref{Pconvergence}, we obtain
\begin{equation}
\label{andersonUL}
  \frac{2V e K \ln \left(\frac{W}{2V}\right)}{W}<1 \Rightarrow \mathbb{P} = 1
\end{equation}
So for big values of $\frac{W}{V}$ the series is almost surely convergent and as we said, this will imply localization. It is easy to show on the same ground that in the opposite regime, the series will be almost surely divergent. 
\begin{figure}[htbp]
\begin{center}
\scalebox{0.73}{\includegraphics{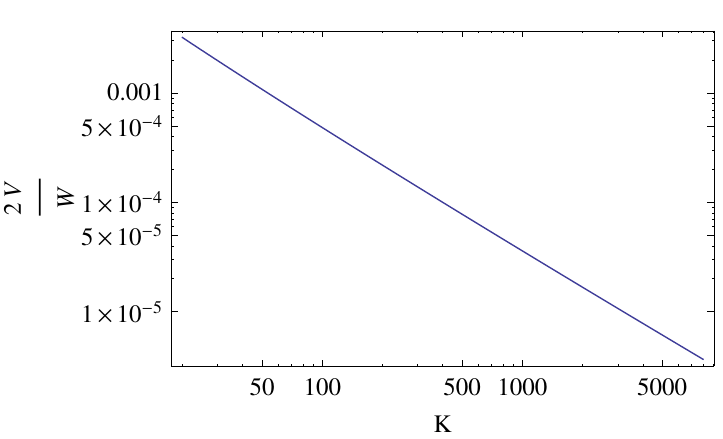}}
\caption{The estimation of the critical ratio $\left(\frac{2V}{W}\right)_c$ for different values of the graph connectivity $K$  given by the upper limit estimation \eqref{andersonUL}.}
\label{andersonloc:upperlimit}
\end{center}
\end{figure}

\section{The self-consistent approach: Bethe lattice}
\label{bethelattice}
Along the derivation in \ref{anderson:estimation}, we pointed out some of the objections, due to Thouless in \cite{thouless1970anderson}, to the derivation in Anderson original work. The analysis of the perturbative expansion is affected by a couple of weak points
\begin{itemize}
\item the complicate set of coupled equations that determine the energy shift in the denominators are neglected;
\item the $K^L$ terms of the same order obtained as different self-avoiding path are treated as probabilistically independent random variables, even though, as we already said, this assumption is clearly false.
\end{itemize} 
A substantially different approach is the one first introduced in \cite{abouchacra1973selfconsistent, abouchacra1974selfconsistent}. The fundamental idea is to study the problem on a particular kind of lattice, the Bethe-lattice \cite{bethe1935statistical}, where loops are long enough (or absent at all) to be neglected. This approach has become standard in the solution of classical statistical mechanics problems, as for example for spin glasses, where the cavity method \cite{mezard1987spin} has provided a crucial tool of investigation. A similar technique has been developed recently also in the quantum case. \cite{laumann2008cavity}. It is nice to read an anecdote from D.J. Thouless in \cite{abrahams201050}:
\begin{flushleft}
 \itshape
``At the end of a dinner in the Andersons' house in Cambridge, Anderson told me of a new approach to localization, and I left their house with a large envelope, on which a few suggestive equations were scribbled. I recognized these as being analogous to the Bethe-Peierls equations in statistical mechanics, which were known to be exact for a Bethe lattice, an infinite regular lattice with no loops. With the help of my student Ragi Abou-Chacra, we showed that these equations led to a set of nonlinear integral equations whose solutions could be found numerically, and which displayed a transition between extended and localized solutions. This result was mentioned in Anderson's Nobel lecture.''
\end{flushleft}

The fundamental equations for these results come from the expressions derived in \ref{anderson:multiple}, involving non-repeating paths: they dramatically simplify on the Bethe-lattice
\begin{equation}
 \Sigma_i (z) = \sum_{j\neq i} V_{ij} \frac{1}{z - \epsilon_j - \Sigma_j^{\neq i} (z)} V_{ji} + \cancel{\sum_{j \neq i} \sum_{k \neq i,j} V_{ij} \frac{1}{z - \epsilon_j - \Sigma_j^{\neq i}(z)} V_{jk} \frac{1}{z - \epsilon_k - \Sigma_k^{\neq i,j}(z)} V_{ki}} + \ldots
\end{equation}
All the terms, but the first, vanish, since on the Bethe-lattice, in absence of loops, longer paths, starting and ending at the origin, must contain repeated indexes. This is true also for all the nested self-energies $\Sigma_i^{\neq j,k,\ldots}(z)$ expansions.
\begin{figure}[htbp]
\begin{center}
\scalebox{0.63}{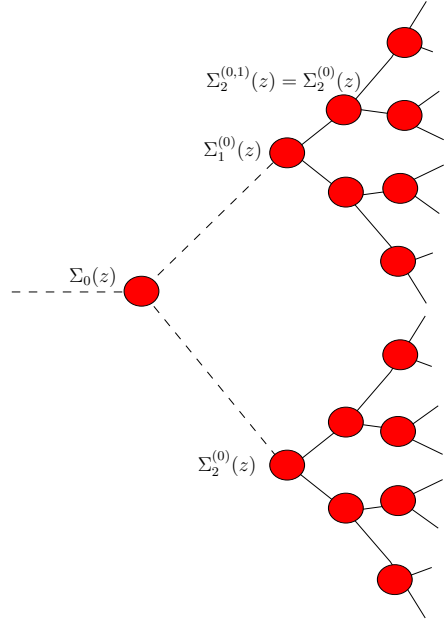}
\caption{A portion of the Bethe-lattice with connectivity $Z = K+1=3$; from the plot it becomes evident that all the descendent self-energies are on the same ground.}
\label{andersonloc:bethelatticeportion}
\end{center}
\end{figure}
Moreover, it is crucial that the equations for the descendant self-energies appearing at the denominators, are completely analogous and do not involve different variables
\begin{equation}
\label{selfenergies}
 \Sigma_j^{\neq i}(z) = \sum_{k\neq i,j} V_{jk} \frac{1}{z - \epsilon_k - \Sigma_k^{\neq j} (z)} V_{kj}
\end{equation}
where we used the fact that on-the Bethe-lattice $\Sigma_i^{\neq i_1,i_2\ldots}(z) = \Sigma_i^{\neq i_1}(z)$, since when the site $j$ has been removed, there is no link to the site $i$. This last equation is the central point for the study of the localization phenomena. It is clear that  $\Sigma_j^{\neq i}(z)$, $\Sigma_k^{\neq i}(j)$ and $\epsilon_k$ are random variables and therefore, if we assume that $\Sigma(z)$ on the lhs and rhs are equally distributed, Eq. \eqref{selfenergies} can be interpreted as an integral equation for the pdf. Unfortunately, this integral equation does not admit an explicit exact solution and one has to resort to some approximations. We will not go into the details of these computations that can be found in the original work \cite{abouchacra1973selfconsistent}. It is however instructive to mention that, going toward the middle of the band ($ \Re z \ll 1 $) and neglecting the real part of the self-energy $\Sigma(z)$, one can recover the upper limit estimation \eqref{andersonUL}. 
Even though derived with completely different methods, the coincidence of the two results shows that this estimation is quite robust and that a correct way to control the approximations made by Anderson is to go on the Bethe-lattice.
Another important concept that can be derived from this approach is the \textit{mobility edge}. We will comment more on it in the next subsection, where we will introduce a numerical method, known as \textit{population dynamics} or \textit{pool method}, to deal with equation involving random variables as \eqref{selfenergies}.
\subsection{The pool method}
A numerical procedure to solve the distributional equations \eqref{selfenergies} was already proposed
in \cite{abouchacra1973selfconsistent}, and revivified more recently in the context of finite-connectivity mean-field
disordered systems under the name of population dynamics \cite{mezard2001bethe} also called the pool
method \cite{monthus2009anderson}. The idea is as follows: suppose we have an equation involving equally distributed and independent random variables on the two sides as \eqref{selfenergies}. As we said, one can convert this equation into one for the pdf.
Instead of trying to solve the integral equation determining analytically the pdf, one tries to approximate the distribution of a random variable by the empirical distribution of a sample of a large number $M$ of representatives.
Then one starts from an arbitrary initial conditions and it is produced a sequence of samples whose empirical distributions converges to a fixed point.
Suppose we start from a sample of self-energies at a fixed value of $z$:
$$ S_0 = \{ \Sigma^{(1)}(z), \Sigma^{(2)}(z), \ldots, \Sigma^{(M)}(z) \} $$
To obtain a new sample $\mathcal{S}_{i+1}$, we replace one of its elements randomly chosen, say $\Sigma^{(j)}(z)$, with the rhs of \eqref{selfenergies}, where the $\Sigma(z)$-s are obtained selecting $K$ elements uniformly at random from the current
sample $\mathcal{S}_{i}$ and the energies $\epsilon_k$ are drawn uniformly in $[-W/2, W/2]$. Repeating these steps, if $M$ is large enough, one reaches a
sample $\mathcal{S}_{\infty}$ approximating the fixed point solution of \eqref{selfenergies}, and then the distribution, its moments and the typical values
can be obtained by computing empirically the distribution over the representatives in $\mathcal{S}_\infty$. The numerical accuracy is controlled by the size $M$ of the samples.
\begin{figure}[htbp]
\begin{center}
\scalebox{0.73}{\includegraphics{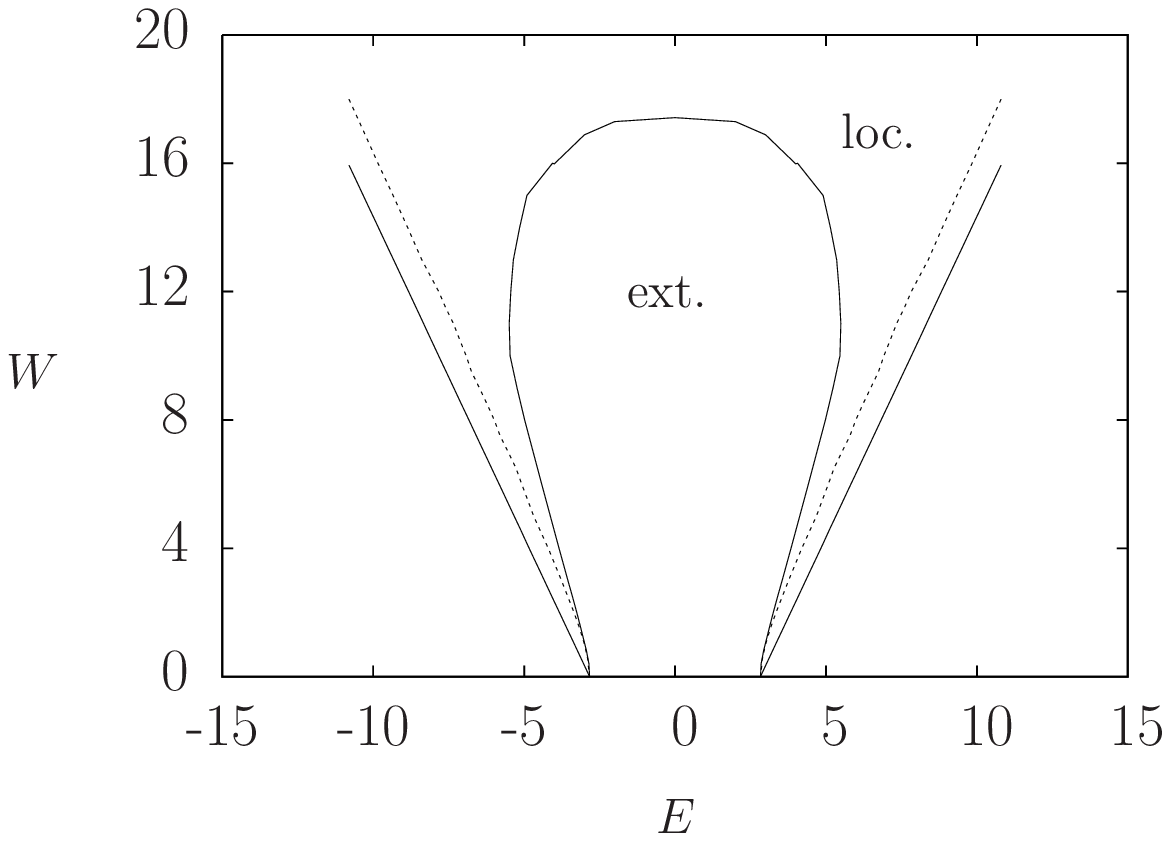}}
\caption{From \cite{biroli2010anderson}. Phase diagram for the Bethe lattice with connectivity $K+1 = 3$. The innermost solid
line indicates the mobility edge between extended and localized states, the outermost solid line
being the edge of the density of states $E = \pm(2\sqrt{k} + W/2)$. The dashed line is the numerically
estimation for the density of states.
}
\label{andersonloc:betheBiroli}
\end{center}
\end{figure}
In Fig. \ref{andersonloc:betheBiroli}, we report the result obtained with this method in \cite{biroli2010anderson}. By checking the behavior of the typical value of the imaginary part of the self-energy in the sample when $\Im z 
\to 0$ one knows whether the spectrum is localized or not. This can be done for different energies $E = \Re z$. The AL transition that we discussed corresponds to the localization of the full spectrum, but from this plot we deduce that for lower value of the disorder, the tails of the spectrum are already localized. The interface separating the middle of the band, which is delocalized, from the localized tails is known in the literature as mobility edge. 
\section{Some rigorous results}
\label{anderson:rigorous}
Being so simple to formulate, the Anderson model has attracted the interest of the mathematical community. Unfortunately, 
it must be stressed that even nowadays most of the checks are based on numerical simulations and many relevant issues still remain open, 
as one can deduce from the Anderson's Nobel lecture (1977)
\begin{flushleft}
 \itshape
``Localization [..], very few believed it at the time, and even fewer saw its importance, among those who failed to fully understand it at first was certainly its author. It has yet to receive adequate mathematical treatment, and one has to resort to the indignity of numerical simulations to settle even the simplest questions about it.''
\end{flushleft}
However, it may be useful to summarized what is already known, sometimes rigorously, sometimes not
\begin{enumerate}
 \item The RAGE theorem \cite{hunziker2000quantum} grants that under reasonable hypothesis, the spectrum of an Hamiltonian $s(H)$ can be split into two parts
$$ s(H) = s_p(H) \cup s_c (H) $$
where $s_p(H)$ corresponds to the point spectrum, i.e. bound states, while $s_c(H)$ is the continuous part, giving transport. Point spectrum states will never leave a bounded region of space.
 \item A random operator is \textit{spectrally localized} if its spectrum $H$ has with probability one an interval of pure point states:
$$ s(H) \cap [a,b] \subset s_p(H) \quad w.p. \; 1 $$
\nomenclature{w.p.}{with probability}
In the mathematical literature, the notion of Anderson localization is often identified with spectral localization. Nevertheless, counterexample exists, proving that spectral localization is not enough to prevent (sub-diffusive) transport.
\begin{figure}[htbp]
\begin{center}
\scalebox{1.3}{\includegraphics{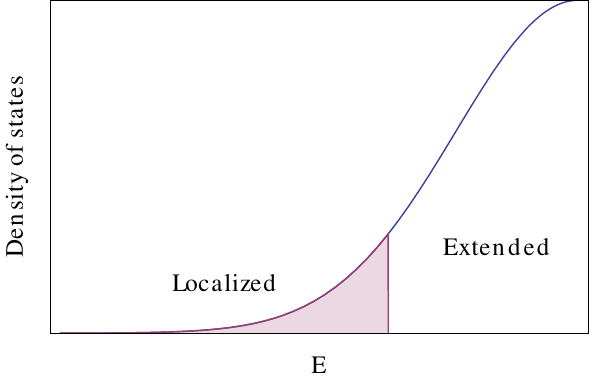}} 
\caption{A cartoon of the density of states, separated into the localized spectrum, at low energies, and extended spectrum in the middle of the band. The two regions are separated by the mobility edge.}
\label{andersonloc:mobility}
\end{center}
\end{figure}
 \item  In $d = 1$ an arbitrary amount of disorder is enough to localize the full spectrum. This result is already stated in the original Anderson paper and was proved subsequently by many authors \cite{mott1961theory}, and more rigorously in \cite{carmona1987anderson}.
 \item  In $d = 2$, in strict sense, again the spectrum is fully localized for arbitrary disorder. As shown in \cite{abrahams1979scaling}, this is the lower-critical dimensions for the Anderson transition and logarithmic scaling appears.
 \item In $d\geq 3$, for low disorder, only the tails of the spectrum (Liftshits tails) are localized; the middle of the band is delocalized and the two regions are separated by the already introduced mobility edge (see Fig. \ref{andersonloc:mobility}).
\end{enumerate}

\section{Conclusions}
In this chapter we presented some of the results coming from the original Anderson work. Many of the ideas and of the concepts will be useful in the rest of the next. In particular, the idea of distributions with long tails, whose average value is not significant, comes up often in the many-body case.
Moreover, the Bethe-lattice approach allowed us to introduce important concepts as the mobility edge. Its particular ``solvability'' makes it useful as a comparison for more complicate example. We will deepen this in the next chapters.


%% file: AndersonLocalization/Figs/contour.latex
\setlength{\unitlength}{3947sp}%
\begingroup\makeatletter\ifx\SetFigFont\undefined%
\gdef\SetFigFont#1#2#3#4#5{%
  \reset@font\fontsize{#1}{#2pt}%
  \fontfamily{#3}\fontseries{#4}\fontshape{#5}%
  \selectfont}%
\fi\endgroup%
\begin{picture}(6324,1107)(2689,-3178)
\thinlines
{\color[rgb]{0,0,0}\put(5176,-3136){\line( 1, 0){2666}}
}%
{\color[rgb]{0,0,0}\put(3301,-3136){\vector( 1, 0){2361}}
}%
{\color[rgb]{0,0,0}\put(7801,-2311){\vector(-1, 0){2325}}
}%
{\color[rgb]{0,0,0}\put(5926,-2311){\line(-1, 0){2625}}
}%
{\color[rgb]{0,0,0}\put(2701,-2761){\vector( 1, 0){6300}}
}%
\put(7426,-2686){\makebox(0,0)[lb]{\smash{{\SetFigFont{14}{16.8}{\rmdefault}{\mddefault}{\updefault}{\color[rgb]{0,0,0}Real axis}%
}}}}
\put(6676,-3061){\makebox(0,0)[lb]{\smash{{\SetFigFont{14}{16.8}{\rmdefault}{\mddefault}{\updefault}{\color[rgb]{0,0,0}$C_-$}%
}}}}
\put(6676,-2236){\makebox(0,0)[lb]{\smash{{\SetFigFont{14}{16.8}{\rmdefault}{\mddefault}{\updefault}{\color[rgb]{0,0,0}$C_+$}%
}}}}
\end{picture}%

%% file: AndersonLocalization/Figs/removingLoops.pdf_tex
\begingroup%
  \makeatletter%
  \providecommand\color[2][]{%
    \errmessage{(Inkscape) Color is used for the text in Inkscape, but the package 'color.sty' is not loaded}%
    \renewcommand\color[2][]{}%
  }%
  \providecommand\transparent[1]{%
    \errmessage{(Inkscape) Transparency is used (non-zero) for the text in Inkscape, but the package 'transparent.sty' is not loaded}%
    \renewcommand\transparent[1]{}%
  }%
  \providecommand\rotatebox[2]{#2}%
  \ifx\svgwidth\undefined%
    \setlength{\unitlength}{421.19515965bp}%
    \ifx\svgscale\undefined%
      \relax%
    \else%
      \setlength{\unitlength}{\unitlength * \real{\svgscale}}%
    \fi%
  \else%
    \setlength{\unitlength}{\svgwidth}%
  \fi%
  \global\let\svgwidth\undefined%
  \global\let\svgscale\undefined%
  \makeatother%
  \begin{picture}(1,0.440378)%
    \put(0,0){\includegraphics[width=\unitlength]{removingLoops.pdf}}%
    \put(0.04843749,0.24924418){\color[rgb]{0,0,0}\makebox(0,0)[lb]{\smash{$g_0$}}}%
    \put(0.38036225,0.25116284){\color[rgb]{0,0,0}\makebox(0,0)[lb]{\smash{$g_1$}}}%
    \put(0.78903268,0.28569835){\color[rgb]{0,0,0}\makebox(0,0)[lb]{\smash{$g_n$}}}%
  \end{picture}%
\endgroup%

%% file: AndersonLocalization/Figs/bethelattice.pdf_tex
\begingroup%
  \makeatletter%
  \providecommand\color[2][]{%
    \errmessage{(Inkscape) Color is used for the text in Inkscape, but the package 'color.sty' is not loaded}%
    \renewcommand\color[2][]{}%
  }%
  \providecommand\transparent[1]{%
    \errmessage{(Inkscape) Transparency is used (non-zero) for the text in Inkscape, but the package 'transparent.sty' is not loaded}%
    \renewcommand\transparent[1]{}%
  }%
  \providecommand\rotatebox[2]{#2}%
  \ifx\svgwidth\undefined%
    \setlength{\unitlength}{283.65109863bp}%
    \ifx\svgscale\undefined%
      \relax%
    \else%
      \setlength{\unitlength}{\unitlength * \real{\svgscale}}%
    \fi%
  \else%
    \setlength{\unitlength}{\svgwidth}%
  \fi%
  \global\let\svgwidth\undefined%
  \global\let\svgscale\undefined%
  \makeatother%
  \begin{picture}(1,1.2981724)%
    \put(0,0){\includegraphics[width=\unitlength]{bethelattice.pdf}}%
    \put(0.21276271,0.73969904){\color[rgb]{0,0,0}\makebox(0,0)[lt]{\begin{minipage}{0.42655651\unitlength}\raggedright $\Sigma_0(z)$\end{minipage}}}%
    \put(0.49294057,1.01486708){\color[rgb]{0,0,0}\makebox(0,0)[lt]{\begin{minipage}{0.58538067\unitlength}\raggedright $\Sigma_1^{(0)}(z)$\end{minipage}}}%
    \put(0.48091765,0.35310393){\color[rgb]{0,0,0}\makebox(0,0)[lt]{\begin{minipage}{0.58538067\unitlength}\raggedright $\Sigma_2^{(0)}(z)$\end{minipage}}}%
    \put(0.49878921,1.16112112){\color[rgb]{0,0,0}\makebox(0,0)[lt]{\begin{minipage}{0.59445631\unitlength}\raggedright $\Sigma_2^{(0,1)}(z) = \Sigma_2^{(0)}(z)$\end{minipage}}}%
  \end{picture}%
\endgroup%

%% file: MBL/mbl.tex
\chapter{Many-body localization}
\label{chapt:mbl}
\ifpdf
   
\graphicspath{{MBL/Figs/PNG/}{MBL/Figs/PDF/}{MBL/Figs/}}
\else
    \graphicspath{{MBL/Figs/EPS/}{MBL/Figs/}}
\fi

\section{Introduction}
\markboth{\MakeUppercase{\thechapter. Many-body localization }}{\thechapter. Many-body localization}
The result presented in the previous chapter have had a tremendous impact on the physics of condensed matter for the last fifty years \cite{lagendijk2009fifty}. We have already mentioned many applications for the AL transition and in particular, it must be stressed its fundamental role in reinterpreting the nature of metals and insulators. However, as Anderson already recognized in its original work, which was presented as a toy model, the absence of interactions makes every conclusion less robust and it is only in particular regimes that the standard AL paradigm is reliable. When interactions are negligible, the problem can be faced as a single-particle one. It means a great simplification of the problem complexity, that therefore admits, among all the other analytical treatments, a successful numerical approach. 
In the general situation, instead, it becomes necessary to take into account interactions among charge carriers and the problem must be treat as a many-body one. The Hilbert-space size is now exponentially large in the volume of the system and as usual, every numerical technique becomes rapidly inadequate. Moreover, as we already underlined in the previous chapter, when quantumness and disorder come together it is not possible to investigate at any finite perturbative order. 
For all these reasons, all these questions have remained substantially unsolved. 
One first step in this direction, is related to the study of the Anderson model in a bath of phonon at finite temperature. Even though all the single-particle eigenstates are localized, electrons can scatter with phonons and reach long distances: as a result, transport is restored and conductivity is finite, though small. This class of phenomena have been dubbed \textit{variable-range hopping} \cite{mott2012electronic, mott1969conduction}. Even in presence of a fully localized spectrum, the conductivity is small but finite and goes like
\begin{equation}
 \sigma \propto e^{-\left(\frac{A}{T}\right)^\frac{1}{d+1}}
\end{equation}
where $d$ is the space dimensionality of the problem. 
A similar question is, whether in absence of phonons, transport can be restored when the interactions between electrons are turned on. Is a sufficient amount of interactions capable of overturning AL, without the help of an external bath? An answer has come recently in the positive \cite{basko2006problem,basko2006metal}. The mechanism, which underpins this effect (dubbed many-body localization or MBL transition) %
\nomenclature{MBL}{many-body localization}%
requires the interaction to act in a substantially non-perturbative way, therefore providing an example of how disorder and strong interactions interplay in a quantum
theory. 

The natural setup to study the MBL transition is the dynamics (these were also the terms of the question posed in \cite{anderson1958absence}) and in this perspective it is a question about the foundations of statistical mechanics, namely, on the validity of the ergodic hypothesis. MBL also presents the terms in which a quantum glass can be defined and from there it is only a small leap to conjecturing that MBL is a natural ingredient for hard computational \emph{quantum} problems \cite{altshuler2009anderson,young2010first} (as Ising spin glasses are a natural scenario to discuss the Physics of hard combinatorial optimization problems \cite{hartmann2002optimization,nishimori2001statistical}).


It is useful to stress that the MBL is not only hard to treat but also difficult to define. One reason is that two complementary but somehow opposite approach are possible: 1) as we already said, it can be seen as the persistence or less of an Anderson localized phase, when interactions are turned on; 2) when the full many-body Hilbert space is taken into account, it can be seen as the standard Anderson model, but where the lattice is substituted with the complex and exotic multi-dimensional hypercube corresponding to the unperturbed many-body states. We will come back on both these points in the following sections.

MBL should be responsible, among other things, of the exact vanishing of the DC conductivity of metals below a critical temperature \cite{basko2006metal} and of the failure \cite{altshuler2010anderson, knysh2010relevance} of the simplest version (and possibly of all versions) of the quantum adiabatic algorithm (QAA)
\nomenclature{QAA}{quantum adiabatic algorithm}%
 \cite{farhi2001quantum} for the solutions of $\NP$-complete problems. 

In particular, in order to show the relevance for the quantum computation point of view, we will introduce the QAA; then we will follow \cite{altshuler2010anderson} and show how the existence of a localized phase prevents the possibility of a successful adiabatic quantum computation, inducing (at least) exponentially small gaps in the system size.

Then, we will focus on the properties of typical eigenstates of many-body disordered systems, characterizing the phase transition as a dynamical one. Our investigation will be mainly based on one dimensional systems:
they are particularly suited for studying MBL because, as we said is \ref{anderson:rigorous}, the single particle
spectrum is completely localized for arbitrarily small disorder, and
therefore any observation of delocalization must be attributed to the interaction.

Finally, we will discuss the strong-disordered regime and we will characterize some of its features with a particularly suited integrable model that we propose as toy model for the many-body localized phase: the Richardson model. It corresponds to the $XX$ model with random $z$ fields and a fully-connected hopping term. Being integrable through \textit{Algebraic-Bethe-ansatz} (ABA),
\nomenclature{ABA}{algebraic bethe ansatz}%
 it permits to access to matrix elements, correlation functions and overlaps of eigenstates in the middle of the band for quite large systems, that would be completely out of range for any exact diagonalization approach. This method allowed us to obtain quite accurate predictions for the thermodynamic limit: the integrable nature of the model completely prevents the system to delocalize even for very small (but finite) value of the disorder strength, thus defining the model as a good toy model for the localized phase.

\section{The importance for adiabatic quantum computation}
\subsection{$\NP$ vs $\PP$}
\begin{figure}[htbp]
\begin{center}
\scalebox{0.48}{\includegraphics{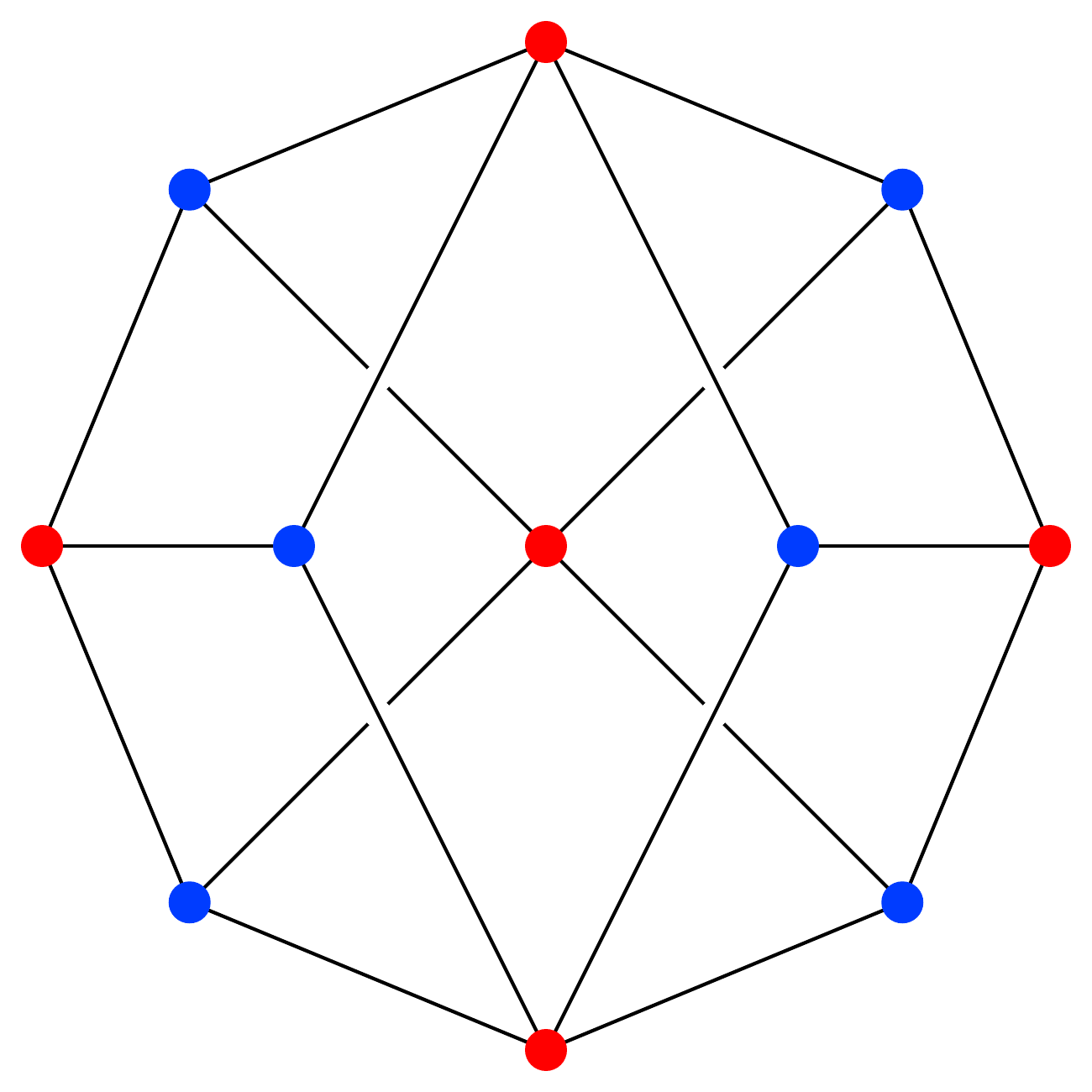}} 
\caption{The Herschel graph, the smallest possible polyhedral graph that does not have a Hamiltonian cycle.}
\label{mbl:herschel}
\end{center}
\end{figure}

The existence of an algorithm to perform a specific computational task is usually not enough to really determine its concrete applicability: a central concept becomes that of complexity, usually related to the number of operations (therefore the computational time) needed for a task of given length $\tL$. This approach, formally founded on the Turing machine itself, allowed to classify algorithms in different classes. Among them, one of the most important is for sure the $\NP$ (Nondeterministic Polynomial time) completeness.
A computational problem belongs to the class $\NP$ if its solution can be verified in a time at most polynomial in the input size $\tL$. In other words, less than $c \tL^k$ computational steps are required to check whether the given solution solves the problem or not. It is clear, already at the intuitive level, that checking if a solution is correct is much faster than really finding it.  Indeed, the class $\PP$ is defined by the problems whose solution can be found in a time at most polynomial in the input length. A classical example is the \textit{Hamiltonian path problem}, that is, given a graph $G$, determining whether it exists a path visiting all its sites once and exactly once. It is clear that in this case, given a solution, checking that it satisfies the requirement is an easy task. An example of graph that does not admit an Hamiltonian cycle, i.e. one of these paths, is shown in Fig. \ref{mbl:herschel}.
However, if the definition of the computational classes was possible and provided a set of powerful ideas for the study of computational complexity, the relations among them are very difficult to investigate: for this reason, if it is obvious that $\PP \subset \NP$, it is still unknown whether $\PP = \NP$. Even though no formal proof is known either of the equality or of its opposite, it is commonly believed that they are distinct and that inside the $\NP$ class, there are problems that do not admit any polynomial algorithm for their solution. 
An useful class goes under the name of $\NP$-complete problems: a problem is $\NP$-complete if any other $\NP$ problem can be reduced to it in a time at most polynomial in the input size. In some sense, they can be considered as the real core of $\NP$-complexity, and therefore if $\PP \neq \NP$, we can be sure that no polynomial algorithm will exist for any of them. Many examples of $\NP$-complete problems are known: the already cited Hamiltonian path, 3-sat, Travelling salesman problem, Clique problem, Graph coloring problem, Exact Cover. We will see in detail this last example. 
\subsection{The exact cover problem}
In mathematics, given a collection $\mathcal{S}$ of subsets of a set $X$, an exact cover is a sub-collection $\hat{\mathcal{S}}$ of $\mathcal{S}$ such that each element in $X$ is contained in exactly one subset in $\hat{\mathcal{S}}$.
In computer science, the exact cover problem is a decision problem to find an exact cover or else determine none exists. The exact cover problem is $\NP$-complete \cite{johnson1979computers} and is one of Karp's 21 $\NP$-complete problems \cite{karp2010reducibility}. The exact cover problem is a kind of constraint satisfaction problem.

An exact cover problem can be represented by an incidence matrix or a bipartite graph.
The Knuth algorithm X, together with its implementation known as Dancing Links (DLX), can be used to find all the solutions to an exact cover problem.
Finding Pentomino tilings and solving Sudoku are noteworthy examples of exact cover problems. A particular instance can be presented as a matrix, e.g. with $X = \{1,2,3,4,5,6,7\}$ and $\mathcal{S} = \{A,B,C,D,E,F\}$ with
\begin{eqnarray*}
A &=& \{1, 4, 7\} \\
B &=& \{1, 4\}  \\
C &=& \{4, 5, 7\} \\
D &=& \{3, 5, 6\} \\
E &=& \{2, 3, 6, 7\} \\
F &=& \{2, 7\}
\end{eqnarray*}
we have the matrix
\begin{equation}
    \begin{array}{l|ccccccc}
    & 1 & 2& 3& 4& 5& 6& 7\\ \hline
    A  & 1 & 0 & 0 & 1 & 0 & 0 & 1\\
    \color{red}B  & \color{red}1 & 0 & 0 & \color{red}1 & 0 & 0 & 0\\
    C  & 0 & 0 & 0 & 1 & 0 & 0 & 1\\
    \color{red}D  & 0 & 0 & \color{red}1 & 0 & \color{red}1 & \color{red}1 & 0\\
    E  & 0 & 1 & 1 & 0 & 0 & 1 & 1\\
    \color{red}F  & 0 & \color{red}1 & 0 & 0 & 0 & 0 & \color{red}1\\
    \end{array}
\end{equation}
An other way to formulate the problem is in terms of boolean variables. If we associate to every subset $S$ in $\mathcal{S}$ a boolean variable $x_S$, a solution corresponds to a configuration of the variables satisfying $E[x] = 0$, where
\begin{equation}
\label{energyexactcover}
 E[x] = \sum_{x \in X} \left(1 - \sum_{S \in \mathcal{S}, x \in S} x_S \right)^2 
\end{equation}
If we reinterpret this expression as an energy, the decision problem corresponds
to the determination, if any, of the configurations, giving zero energy. It is easy
to convert the boolean variables to spins by simply setting $\sigma_i^z =
1- 2 x_i$ and therefore, from now on, we will consider standard spin variables.
The solution of the computational task coincide with finding the ground states. 
From now on, we will restrict to the problem known as \textit{Exact cover - 3} (EC3), where each elements of $x$ is contained in exactly $3$ sets of $\mathcal{S}$ (in other words each column of the matrix representation contains exactly $3$ non-zero values). This restriction does not change the nature of the problem since EC3 still belongs to the $\NP$-complete class. 
\subsection{Adiabatic quantum computation}
As we saw in the previous section, any instance of EC3 can be recast in a classical spin Hamiltonian and we want to determine its ground state. Since the problem is known to be $\NP$-complete, unless $\PP = \NP$, which is hard to believe, no fast solution can be ever found for a classical computer. However, the possibility that with a quantum computer, it is possible to improve the efficiency of the algorithm is still open and under debate. We will not go into the details of the wide topic of quantum computation, usually introduced in terms of a discrete succession of unitary transformations (quantum circuit). Instead, we will focus on a different implementation known as \textit{quantum adiabatic algorithm} (QAA), that however, has been proved \cite{van2001powerful, farhi2000quantum, aharonov2004adiabatic}, to be general enough and equivalent in terms of computational cost to the quantum circuit formulation.\footnote{in \cite{aharonov2009power} it was even showed that an Hamiltonian with 11-dimensional 
qudits and 
nearest-neighbor interactions is enough to encode any quantum circuit.} 

To understand better the mechanism, we start from one of its forerunner, known as \textit{simulated annealing} \cite{kirkpatrick1983optimization}. The idea is to perform the computation through a classical statistical system, whose temperature we slowly decrease. In fact, once the energy function for each configuration of spin variables, as in \eqref{energyexactcover}, has been introduced, the Gibbs-Boltzmann probability law at inverse temperature $\beta$ reads
\begin{equation}
\label{gibbsdistrib}
 \mu(\sigma) = \frac{e^{-\beta E(\sigma)}}{Z(\beta)}
\end{equation}
where $\sigma$ represents the global configuration of all the spin variables, and $Z$ is the usual partition function, acting as normalization factor. If we perform a random walk in the configuration space, with transition probabilities respecting the detailed balance condition with the Gibbs-Boltzmann distribution, we will ensure thermal equilibrium. At small temperatures (or large $\beta$), the probability distribution in \eqref{gibbsdistrib} will be more and more concentrated around the minima of $E(\sigma)$. Therefore, if we decrease the temperature slowly enough, at every time we will be at thermal equilibrium and at the end we will end up in one of the minima of $E(\sigma)$.
In this case, the thermal fluctuation are used to overcome the barriers between local minima.
The idea behind the QAA is to use quantum fluctuations \cite{apolloni1989quantum, finnila1994quantum, kadowaki1998quantum, brooke1999quantum, farhi2001quantum} instead of the thermal ones.
For the thermal case, at high temperature, the problem trivializes. Analogously, we introduce another Hamiltonian $H_i$, whose ground-state is easy to find (or in other terms it is easy to prepare the system in it). Then we define
\begin{equation}
\label{linearinterpolation}
H[\lambda] = (1-\lambda) H_i + \lambda H_f 
\end{equation}
where the final Hamiltonian is the diagonal operator given by the energy function
\begin{equation}
\label{adiabaticfinal}
H_f = \sum_\sigma E(\sigma) \ket{\sigma}\bra{\sigma}
\end{equation}
We suppose that at the initial time the system is in the ground-state of $H_i$ and we perform a quantum evolution with a time dependent Hamiltonian $H[\lambda]$, where $\lambda = t/ \mathcal{T} \in [0,1]$ and $\mathcal{T}$ will be the simulation time. If $\mathcal{T}$ is so large that the hypothesis of the adiabatic theorem are holding \cite{landau1965quantum, messiah1962quantum}, then it is granted that the system will be at each time in the instantaneous ground-state of $H[\lambda]$ and therefore at the end it will be in a configuration corresponding to a minima of $E(\sigma)$. So a solution to the computational problem is found by simply measuring the state of each spin.

After this discussion, some remarks are in order
\begin{enumerate}
 \item many possible choices for the initial Hamiltonian $H_i$ are possible: the only requirement is that its ground state must be easy to prepare;
 \item estimating the complexity of the algorithm is now converted into the estimation of the time $\mathcal{T}$ that guarantees an adiabatic evolution.
\end{enumerate}
 The configurational space is chosen to be the product space, for each site $i$, of the common eigenvectors of the $\sigma^z_i$ 
\begin{equation}
 \{ \ket{\uparrow \uparrow \ldots \uparrow}, \ket{\uparrow \uparrow \ldots \downarrow}, \ldots\} 
\end{equation}
 One possible, and typical, choice for $H_i$ is
\begin{equation}
\label{adiabaticInitial}
 H_i = -\sum_i \sigma_i^x
\end{equation}
For a discussion of the experimental realization of the QAA, we remind to \cite{brooke1999quantum, steffen2003experimental, bian2012experimental}.
\subsection{Anticrossing and small gaps} 
The standard results of the quantum adiabatic theorem allow to relate the computational time $\mathcal{T}$ with the minimum gap between the ground-state and the first excited state
\begin{equation}
\label{adiabaticscaling}
\mathcal{T} \gg \Ord{ N \Delta_{min}^{-2}}
\end{equation}
where, given the two lowest eigenvalues $E_{gs}(\lambda), E_{es}(\lambda)$ of $H[\lambda]$, the minimum gap is defined as
$$ \Delta_{min} = \min_{\lambda \in [0,1]} E_{es}(\lambda) - E_{gs}(\lambda) $$
Therefore, the estimation of the time $\mathcal{T}$ is now related to the
computation $\Delta_{min}$. We will assume that $\Delta_{min} \neq 0$, since in
absence of particular symmetries the \textit{"No crossing rule"} should hold
\cite{von1929uber}. 
What happens typically is that the two Hamiltonians $H_i$ and $H_f$ have very
different low-energy behavior, so that there will be a critical value
$\lambda_c$ where  $H[\lambda_c]$ undergoes a phase transition. 
The common belief is that the order of the phase transition determines the
scaling of the gap with the system size:
\begin{equation}
\label{scalingordertransition}
\Delta_{min} \simeq \left\{ 
       \begin{array}{ll}
         N^{-\alpha} & \mbox{second order}\\
         e^{- \gamma N} & \mbox{first order}
       \end{array}\right.
\end{equation}
So generally we can assume that whenever a first order phase transition is
encountered in the evolution of $\lambda$, the algorithm will need a time
exponentially long in the input length. One may wonder whether it is possible to
obtain any improvement to the relation \eqref{adiabaticscaling}, if instead of
choosing the linear interpolation, we would have chosen a different
interpolation scheme. It can be shown that the bound \eqref{adiabaticscaling} is
obtained if one suppose that $\Delta(s) = \Delta_{min}$ for all $s$. If we take
into account that the gap is not always as small as its minimum value, we can
obtain a better scaling $\mathcal{T} \simeq \Delta_{min}^{-1} $, that however,
according to \eqref{scalingordertransition}, does not change the nature of the
algorithm (polynomial or exponential).  

An important remark is that the complexity time is usually estimated considering
the worst case: it means that we consider, for the given algorithm, the time
needed for the worst instance. It happens often however that the worst-case is
not the typical one, and so it is not so easy to be generated. This is
meaningful if a probability distribution among the instances has been defined.
To understand better the situation, we come back on the EC3 problem. We fix the
scaling with the system size assuming that the set $X$ will contain $M = \alpha
N$ elements, where $N$ is the number of sets in $\mathcal{S}$. Then we
investigate the large $N$ limit at fixed $\alpha$. The energy function becomes
\begin{equation}
\label{ec3costfunction}
f(x) = \sum_c (x_{i_c} + x_{j_c} + x_{k_c} - 1)^2 
\end{equation}
Each term in the sum is called a clause, that involves $3$ boolean variables and is satisfied if and only if only one among them is true.
We consider a standard distribution of random instances, where we pick the $M$
clauses independently, each clause being obtained by picking three boolean
variables with uniform distribution. 
We will give now some ideas from \cite{altshuler2010anderson}, showing that for the typical instance, the gap closes (at least) exponentially.
There are two characteristic values of $\alpha$: the clustering threshold $\alpha_{cl}$ and the satisfiability threshold $\alpha_s$ \cite{biroli2000variational}:
\begin{enumerate}
\item for $\alpha < \alpha_{cl}$ the density of the solution is high and essentially uniform
\item for $\alpha > \alpha_{cl}$ the solutions start to organize in clusters, each of them very far from the other (the distance between them measured by the Hamming distance)
\item for $\alpha > \alpha_{s}$ the probability that the problem is satisfiable vanishes when $N, M\to \infty$;
\end{enumerate}
An estimation has been provided in \cite{raymond2007phase} for the value is
$\alpha_s \simeq 0.6263 $. 
Now, the mechanism involved here is the so called \textit{avoided
crossing} or \textit{anticrossing} and it is the same mechanism involved in the
already cited ``No crossing rule''. Let's suppose that two diagonal
elements in the Hamiltonian $H[\lambda]$, for a given value of $\lambda$, are
particularly close. An arbitrary small off-diagonal element is already enough
to prevent crossing. This is true already if we limit to the $2 \times 2$
matrix
$$ H_{12} = \left(\begin{array}{cc}
             e_1(\lambda) & V_{12}\\ V_{12} & e_2(\lambda)
            \end{array}\right) $$
\begin{figure}[htbp]
\begin{center}
\scalebox{0.78}{\includegraphics{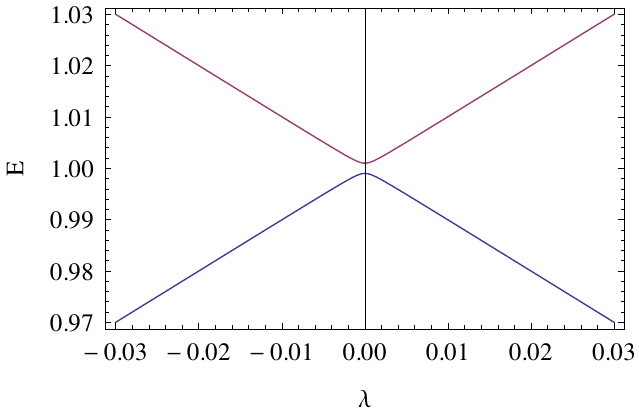}} 
\caption{A typical example of level repulsion for the eigenvalues of the matrix
depending on a parameter: an exact degeneracy would need an additional
condition and is therefore generically very rare.}
\label{mbl:levelcrossing}
\end{center}
\end{figure}

\begin{figure}[htbp]
\begin{center}
\scalebox{0.45}{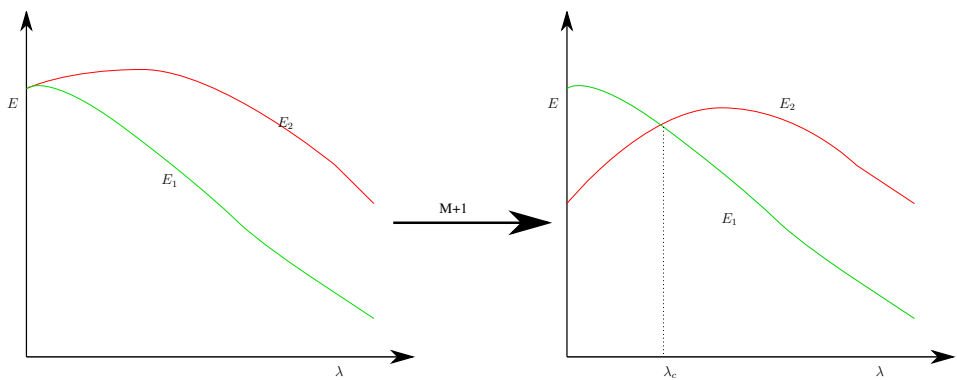} 
\caption{The behavior with $M$ clauses for the evolution with $\lambda$ of the two states $\sigma_1$ and $\sigma_2$: the degeneracy is lifted by the off-diagonal elements (left); adding one more clause, there is a finite probability that the degeneracy at $\lambda = 0$ is lifted in an opposite way with respect to large $\lambda$, thus producing an avoided crossing at $\lambda_c$.}
\label{mbl:avoidedcrossing}
\end{center}
\end{figure}

The gap between the two levels is then
\begin{equation}
\label{2by2gap}
 \Delta[\lambda] = \sqrt{(e_1 - e_2)^2 + |V_{12}|^2}  
\end{equation}
and reaches its minimum value when the two diagonal elements meet, with
$\Delta_{min} = |V_{12}|$: this is called \textit{avoided crossing}, because the two lines would cross only if $V_{12} = 0$.
The idea is to construct two states forming one of these anti-crossings, thus giving a small gap.
We set $\alpha \lesssim \alpha_s$, so that we are in
a region where solutions are few and far one from the other. Now, we
suppose $\sigma_1$ and $\sigma_2$ are two spin configurations giving a
solutions, $f (x_{\sigma_1}) = f (x_{\sigma_2}) = 0$ (where $x_\sigma$ is the configuration of the boolean variables corresponding to the spin configuration $\sigma$), and so they are degenerate
ground-states of $H[\lambda = 0]$\footnote{this does not contradict the
\textit{No crossing rule}, due to the symmetry $[H[\lambda = 0], \sigma_z^i] =
0$.}. When $\lambda \neq 0$, the two eigenvalues $E_1(\lambda), E_2(\lambda)$, corresponding at $\lambda = 0$ to $\sigma_1, \sigma_2$,
will move separately , the degeneracy will be lifted and and it can be shown
that it exists $\lambda^\ast \ll 1$ such that
\begin{equation}
\label{largegap}
\forall \lambda > \lambda^\ast \quad  |E_1(\lambda) - E_2(\lambda_2)| > 4
\end{equation}
Without losing generality we can assume $E_1(\lambda)< E_2(\lambda)$.
If we now add randomly a new clause, going to $M+1$, there is a finite
probability that $\sigma_1$ is still a solution, while $\sigma_2$ is not. It means that $f_{M+1}(x_{\sigma_2}) = 1 $ or $4$ and comparing with Eq. \eqref{largegap}, the order of the eigenstates for large $\lambda$ will remain the same. The resulting behavior is shown in Fig. \ref{mbl:avoidedcrossing} and it follows that at $\lambda = \lambda_c$ we will have an anticrossing.
It remains to estimate the gap at $\lambda_c$ and we can use Eq. \eqref{2by2gap}; therefore we need to compute the matrix element between the two unperturbed states $\sigma_1,\sigma_2$. For the given value of $\alpha$, with high probability, the Hamming
distance between them will be quite high: $d_{H} (\sigma_1, \sigma_2) \simeq \nu(\alpha) N = n$. It follows that the two states are connected only going to order $n$ perturbation theory
\begin{equation}
 \label{matrixelementPT}
V_{12} = \sum_{path}\frac{\lambda^n}{E_{p_1} \ldots E_{p_n}} \simeq \frac{n!}{[(n/2)!]^2} \lambda^n \simeq \left(\frac\lambda 2\right)^n
\end{equation}
where the sum is over all the possible path connecting $\sigma_1$ and $\sigma_2$. Here we used that for most of these paths, since solutions are very rare, every time we change a spin value, we break some clauses, therefore the energy increase linearly from the solution $\sigma_1$, reaches a maximum in the middle, and start decreasing again toward $\sigma_2$. In this way, the denominator produces a $(n/2)!^2$ that cancels out with the $n!$ coming from the sum over paths. Instead, the value of $\lambda$ can be estimated imposing the condition \eqref{largegap} at the first non-zero order in perturbation theory
\begin{equation}
|E_1(\lambda) - E_2(\lambda)| = A \sqrt{N} \lambda^4 + \Ord{\lambda^6} \;\Rightarrow\;\lambda^\star = \Ord{N^{-\frac 1 8}}
\end{equation}
where we used that at each order, the perturbative correction having zero-mean, will scale as $\sqrt{N}$. Finally we get
$$ \Delta_{min} \simeq \exp\left[-\frac{\nu(\alpha) N}{8} \log \frac{N}{N_0}\right] $$
where $N_0$ is a constant of $\Ord{1}$. This shows that the gap for this situation is exponentially small. 

Actually, we said that this situation happens with a finite probability. Is it possible that if we are lucky no anticrossing occurs and the adiabatic computation can perform without problems? Unfortunately, this is not the case, because we took into account only two degenerate states at the beginning $\sigma_1$ and $\sigma_2$. By considering all the possible states that can in principle have the same role as $\sigma_2$, it is easy to show that the probability to have at least one avoided crossing with the ground states, is practically $1$.
\subsection{The role of localization} 
From the discussion in the previous subsection, it may appear obscure what is the role of localization. However if we consider the Hamiltonian in Eq.\eqref{linearinterpolation}, it is easy to recognize that it has the same form as the Hamiltonian \eqref{AndersonModel} of the Anderson model. Here the randomness in the diagonal elements is coming from the random choice in the instance of the problem. The main difference is coming from the off-diagonal part: for Eq.\eqref{AndersonModel}, the hopping term coincided with the adjacency matrix of a finite-dimensional graph, e.g. the $d$-dimensional cubic lattice, therefore $K \simeq 2d$ and the number of sites is $L^d$. Here instead the connectivity is $N$ and the number of sites is given by the number of possible configurations $2^N$.
It is crucial in the previous derivation that perturbation theory holds. If so, as we saw, one can prove that with high probability an avoided crossing with exponentially small gap will appear. 
The perturbation term appearing in Eq. \eqref{matrixelementPT} is meaningful only if the full series is convergent. This, by turn, was the criteria used in \ref{anderson:estimation} to find the transition point. It follows that the perturbation theory is convergent in we are in the discrete region of the spectrum, or in the Anderson localized phase \ref{anderson:rigorous}. In this case, due to the exponential scaling of the Hilbert space size, we are considering the many-body version of the Anderson problem, \textit{the MBL} transition. If we n\"aively set $K \simeq N$ in Eq. \eqref{andersonUL} to estimate the critical value of $\lambda_{MBL}$ for the localization transition, we find that $\lambda^\star \ll \lambda_{MBL}$; so this procedure becomes meaningful, thanks to the existence of the many-body localized phase. 

In the following sections, we will investigate more on the structure of the eigenstates on the two sides of the MBL transition.

\section{The structure of the eigenstates in a disordered many-body problem}
We move now to the specific analysis of some features characterizing and distinguishing the two phases. 
\subsection{The XXZ spin chain}
\label{mbl:xxz}
\begin{figure}[htbp]
\begin{center}
\scalebox{0.48}{\includegraphics{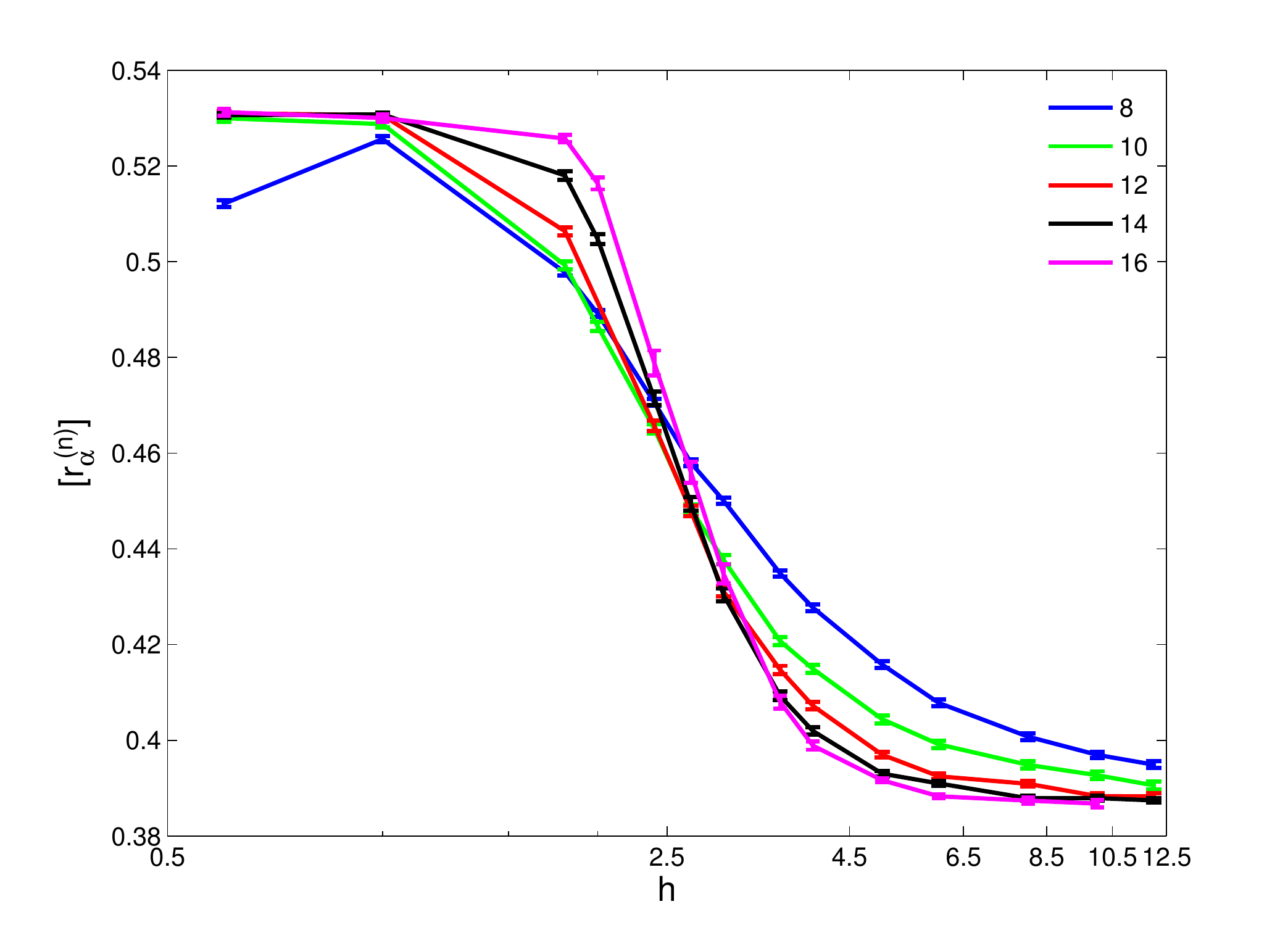}} 
\caption{From \cite{pal2010mb}, finite-size scaling of the parameter $r$ vs $h$. The crossing of the different curves seems to indicate a transition in the spectral properties at a finite $h$.}
\label{rHuse}
\end{center}
\end{figure}

We will analyze in detail the ergodicity properties of an XXZ chain with random fields. The Hamiltonian is
\begin{equation}
H=-J\sum_{i=1}^N(s^x_i s^x_{i+1}+s^y_i s^y_{i+1})-\Delta\sum_{i=1}^N s^z_i
s^z_{i+1}
-\sum_{i=1}^Nh_i s^z_i,
\label{mbl:xxzHam}
\end{equation}
with periodic boundary conditions. 
\nomenclature{$N$}{unless specified in the text, it is the number of sites in the system}
This particular example has already provided different indications of
the MBL transition for sufficiently large disorder: in \cite{pal2010mb}
correlation functions and spectral properties were studied, 
while in
\cite{znidaric2008many,Bardason2012} tDMRG was used to investigate the
different saturation properties of the entanglement entropy in the two phases.
In Fig. \ref{rHuse}, we report the finite-size scaling for the parameter $r$ defined as:
\begin{equation}
 \label{rHuseDef}
r_N = \avg{\frac{\min \left(\delta_n, \delta_{n+1}\right)}{\max\left(\delta_n, \delta_{n+1}\right)}}_{n,h}
\end{equation}
where $\delta_n = E_{n+1}-E_n$ is the spectral distance between two subsequent eigenvalues and the average is taken over $n$ and disorder realization. This parameter is known to have different limits according to the level-space statistics
\begin{equation}
 r_{N\to \infty} \simeq \left\{\begin{array}{ll}
                  0.39 & \mbox{Poisson statistics} \\
                  0.54 & \mbox{Wigner-Dyson statistics}
                 \end{array}\right.
\end{equation}
and in the figure the finite-size scaling gives strong indication of the different thermodynamic limit for this quantity in the weak and strong disorder regions.
Therefore, while the existence of a transition in the dynamics of this model is now almost
certain, its precise location, the possible existence of a critical phase and
the nature of the phases that it separates are subject of debate. This should
not be regarded as  a debate about a particular spin chain but rather as an
attempt at characterizing as much as possible the differences between MBL and
AL.

Consider the real time evolution of a state $\ket{\psi_0}$ as it is encoded into the Green's function
\begin{equation}
 \label{greenfun}
G(t) \equiv \bra{\psi_0}e^{-itH}\ket{\psi_0} \; .
\end{equation}
Generalizing what we did in \eqref{anderson:returnprobability}, we introduce 
\begin{equation}
 \label{prq}
\IPR{q} = \sum_{E} |\braket{E}{\psi_0}|^{2q}.
\end{equation}
where the sum runs over the full set of eigenstates $\ket{E}$. Therefore, we already know that the long time average of the return probability is given by
\begin{equation}
\label{mbl:returnprobability}
\rp \equiv \lim_{\tau\to\infty}\frac{1}{\tau}\int_0^\tau dt\, \left| G(t)\right|^2 =
\IPR{2}\; .
\end{equation}
As we said, $\IPR{2}^{-1}$ can be seen as a measure of the explored volume of Hilbert space during the dynamics. Higher order $\IPR{q}$'s describe finer details of the dynamics.

Let us now comment on the choice of a suitable initial state for a
\emph{gedankenexperiment} aimed at testing the breaking of ergodicity. First of
all, consider what happens if we take a random state in the Hilbert space
(therefore not an eigenstate) conditioned just to have an expectation value of
the energy $E$ and standard deviation $\delta$ (with high probability, for a
random state and a local Hamiltonian $\delta=\Ord{N^{1/2}}\ll E= \Ord{N}$). In
fact, in the phase dominated by a strong disorder there are states very close in
energy ($\Delta E=\Ord{e^{-S}}$, where $S$ is the microcanonical entropy at
energy $E$) which are macroscopically different and the expectation value of a
local operator will be the average of its values in these localized
eigenstates, concealing the effect of disorder (as expected from the ergodic
theorem \cite{vonneumann1929beweis}). If we want to observe the effect of
disorder on the dynamics, 
a reasonable prescription consists in choosing an eigenstate of the part of the Hamiltonian which dominates in the strong disorder limit. Starting the dynamics coincides then with \emph{turning on} the rest of the Hamiltonian.
In the delocalized phase, during the quantum dynamics, the motion covers a
finite fraction of the full Hilbert space (each eigenstate being individually
thermal, the so-called ``eigenstate thermalization hypothesis'' (ETH)
\cite{srednicki1994chaos, srednicki1999thermal,rigol2012thermalization}). 
Instead, in presence of strong disorder, ergodicity breaks down and the many-body wave
function motion is constrained on a small section of the full Hilbert space. 

We also believe that this point of view on MBL is what better brings forward its implications for quantum computation (or at least for the performance of the Adiabatic Algorithm \cite{farhi2001quantum}). In the localized phase the system gets trapped, the dynamics unable to efficiently explore the Hilbert space, so the algorithm is not efficient in finding the ground state \cite{farhi2001quantum,altshuler2009anderson,young2010first}. 

This view on the MBL transition will be the focus of this paper. We will show how the usual criteria for detecting AL need to be tweaked to capture the MBL transition; we will study the IPR's and will show how, although much information is contained in them, it is actually necessary to study the distribution of wave-function coefficients $\braket{\psi_0}{e}$, which is heavily tailed both in the localized and delocalized regions. 

As the Hamiltonian
commutes with the total $z$ spin $S^z=\sum_i s_i^z$, we focus on the subspace with $S^z=0$. 
The random fields are chosen from a box distribution
$h_i\in[-\dis,\dis]$. The model can be cast into
a theory of fermions ($S^z = 0$ corresponds to half-filling), with on-site
disorder $h_i$. 

The $\Delta s^z s^z$ term can be written as a two-body, point-like interaction for the fermions and for
zero temperature it can be included perturbatively or non-perturbatively
\cite{giamarchi1988anderson} leading to an interesting phase diagram. When $\Delta=0$ the fermions are free, an arbitrarily small disorder localizes the entire spectrum and therefore ergodicity is broken for any $h>h_c=0$. As $\Delta$ is increased the MBL conjecture implies that a peculiar phase transition exists (possibly even at infinite temperature) at a critical $h_c$ increasing away from zero. 
On the other hand, for $\Delta\gg J$ the disorder necessary to break ergodicity should \emph{decrease} again. In fact, for large $\Delta$ the relevant degrees of freedom are the domain walls of the classical Ising chain obtained by setting $J=0$ in (\ref{mbl:xxzHam}). Longer domain walls have smaller and smaller hopping matrix elements and therefore they are more prone to localization than the fermions at $J\gg \Delta$. Once a few of these large domain walls have frozen, ergodicity can be considered broken and this occurs for smaller $h$, since both the effective hopping and interaction are smaller (effective randomness is always $h$). Here we present results of exact diagonalization for $\Delta=J=1$, where the delocalized phase is largest.

\subsubsection{Return probability.}
According to the discussion of the previous section, we should test ergodicity by taking an initial state $\psi_0$ as one of the  $\HS=\binom{N}{N/2}$
\nomenclature{$\HS$}{unless specified in the text, it is the Hilbert space size, exponentially large in $\SN$}
 configuration of spins $\ket{a}$ polarized along the $z$ or $-z$ direction, (e.g.\ $\ket{a}=\ket{\uparrow\downarrow...}$ \nomenclature{$\ket{a}$}{it is an index for one state in the computational basis $\ket{\uparrow, \downarrow, \ldots, \uparrow}$}%
).
We need to stress a major difference in the  behavior of $\IPR{2}$ in the localized and delocalized phases between MBL and AL.
While in the latter one can distinguish the two phases by the participation ratio being
$\Ord{1}$ or not in the thermodynamic limit, this is not a
sufficient criterion in MBL. For a many-body state, even in absence of interaction, $\IPR{2}$ will be exponentially small in $N$ also in presence of strong disorder, simply because each degree of freedom will have a localization length small but finite, corresponding to an individual \textit{participation ratio} smaller than $<1$: multiplication of $\Ord{N}$ of these factors leads to an exponentially small $\IPR{2}$. We need to correct the previous criterion by requiring that the delocalized and localized phase are distinguished by whether the ratio $\IPR{2}/\HS^{-1}$ is $\Ord{1}$ or not. The other $\IPR{q}$'s, properly rescaled with powers of the Hilbert space dimension $\HS$, also represent indicators of ergodicity breaking. 

\begin{figure}
\begin{center}
 \includegraphics[width=0.8\textwidth]{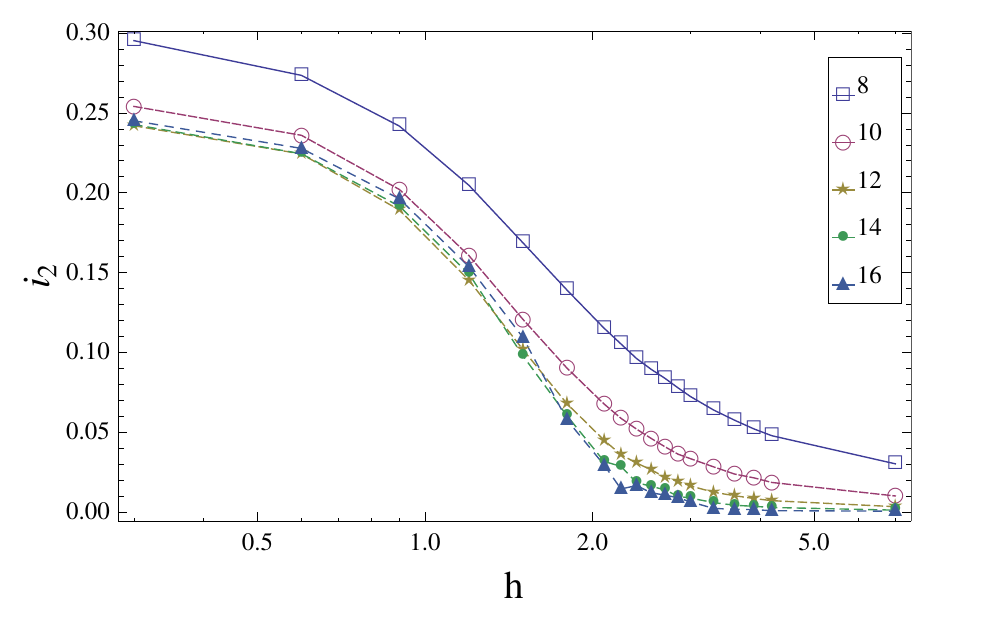}
 \caption{Average fraction of occupied Hilbert space vs h for different system
sizes $N = 8$ to $16$ using exact diagonalization. Notice how the limit for $h\to 0$ is different from $1/3$ which is the RMT prediction.
}
\label{fractionHilbert}
\end{center}
\end{figure}

\begin{figure}
\begin{center}
\includegraphics[width=0.8\textwidth]{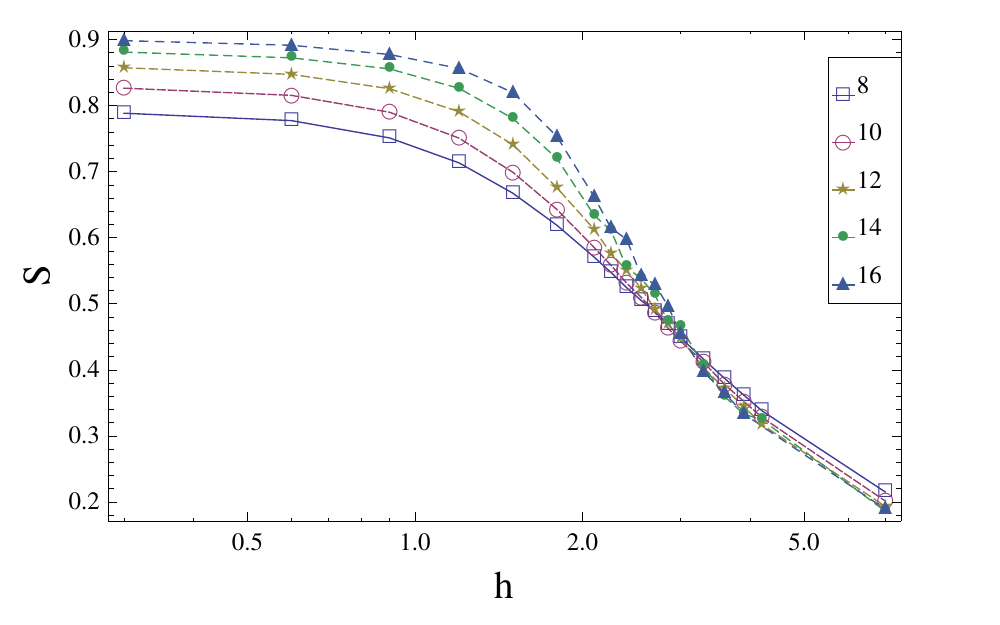}
\caption{Average diagonal entropy vs disorder strength for different sizes $N =
8,10,12,14,16$. From the $N$-dependence the transition is identified at $h_c\simeq 2.7\pm 0.3$.}
\label{entropySize}
\end{center}
\end{figure}

However, as far as averages over the initial states are involved we have found that IPR's have better finite-size behaviors (more on this later), so we considered:
\begin{equation}
 \label{intensiveIPR}
I_q^{(N)}(\dis) \equiv \left\langle \frac{\IPR{q}^{-1}}{\HS^{q-1}}
\right\rangle_{\{\dis\}, a} \; .
\end{equation}
where the subscripts in the average correspond to disorder realizations
(indicated with $h$) and initial spin configuration $\ket{a}$\footnote{The number of realizations goes from 10000 for small sizes till about 100 for the maximum size $N = 16$}.
In particular the data for $I_2$, shown in Fig. \ref{fractionHilbert}, are consistent with the $\lim_{N\to \infty}I_2^{(N)}(h)=i_2(h)$ where $i_2(h) = 0$, for $h>h_c=2.7\pm0.3$, although the finite-size corrections are strong already at $h\gtrsim1.5$. A similar information is obtained by the diagonal entropy
\begin{equation}
S^{(N)}=\lim_{q\to 1}\frac{\avg{\IPR{q-1}}}{(q-1)\ln\HS},
\label{eq:diagentr}
\end{equation}
which is plotted for varying $h$ in Figure \ref{entropySize} and also this quantity is clearly far from its thermodynamic limit of $S=1$ in the delocalized phase. If we identify the critical point as the place where the $N$ dependence sets in (for $I_2$) or drops out (for $S$) then both quantities identify a critical point consistent with $h_c=2.7\pm 0.3$ consistently with the findings of \cite{pal2010mb}.

The diagonal entropy and the IPR's testify that even in the localized region, the covered phase space is growing with the system size, even though it is an exponentially small fraction of the full Hilbert-space. This suggests
that in a many-body system, the localized phase is necessarily characterized by the breaking
of ergodicity, but not necessarily by a concrete localization ($\IPR{2} \simeq O(1)$). However, to pinpoint the transition and understand the reasons of the scalings we should analyze the full probability distribution of $|\braket{a}{e}|^2$.

\subsubsection{Distribution of wave function amplitudes.}
If one considers the various $\IPR{q}$ averaged over $\ket{a}$, one observes a peculiar scaling with $N$ of each of them, which can be considered as due to large fractal dimensions. In this scenario, the safest observable to consider is the distribution of the properly rescaled wave function coefficients.  As we are
interested in typical states (infinite temperature) we will not follow the usual route of fixing
the energy of the state but we will rather integrate over the whole spectrum. In
the thermodynamic limit this corresponds to energy density $E/N=0$.\footnote{The MBL mobility edge is not much of an issue here since the states at energy $E<E'$ for any $E'/N<0$ are an exponentially small fraction of the total spectrum, unlike in single particle AL problem.} 

We will consider therefore the average over eigenstates, initial states and disorder realizations: 
\begin{equation}
\label{distrib}
\phi(x,N)=\avg{\delta(x-\HS |\braket{a}{e}|^2)}_{a,e,\{h\}}.
\end{equation}
In the following we will drop the subscripts in the averages. This function depends both on $x$ and $N$ in general but in the delocalized phase, as $\HS$ plays the role of the space volume, we see that the dependence on $N$ drops out \cite{mirlin1993statistics, mirlin1994statistical, mirlin1994distribution}.

We can then write the various IPR's as
\begin{equation}
\label{prqdistrib}
\avg{\IPR{q}} = \HS^{1-q}\int_0^\infty dx\ x^q\phi(x).
\end{equation}

\begin{figure}[htbp]
\begin{center}
\includegraphics[width=0.95\columnwidth]{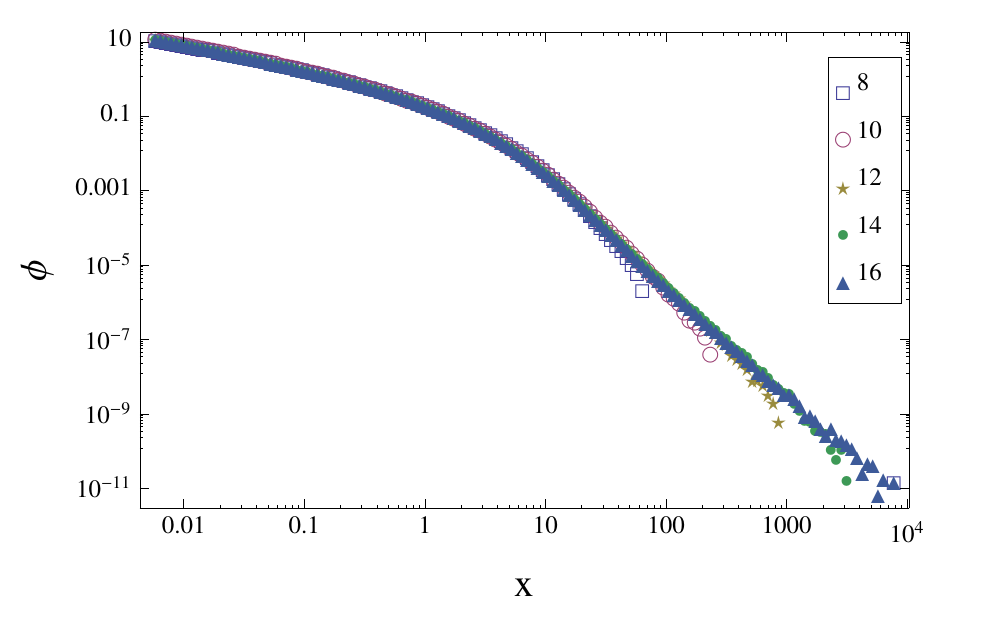}
\includegraphics[width=0.95\columnwidth]{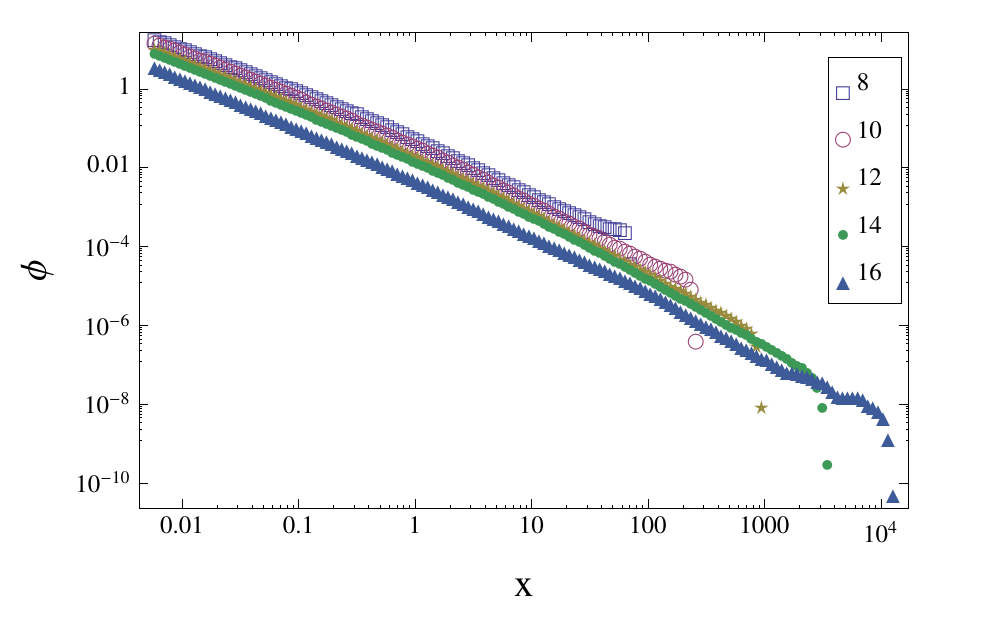}
\caption{The distribution of scaled wave function amplitudes $x=\HS |\braket{a}{e}|^2$
for different values of $h$. Upper panel: $h=1.2$ in the middle of the ergodic phase where the scaling is perfectly verified, lower panel $h=4.2$ in the many-body localized phase. In each figure the different curves correspond to different values of
$N$, from 8 to 16. Each curve is obtained by binning of not less than $3\ 10^6$
squared amplitudes.}
\label{fig:phi-h}
\end{center}
\end{figure}

Illustrative plots are shown for different regimes in Fig. \ref{fig:phi-h}.
As we said, even though in the ergodic phase, with this scaling the curves for
different sizes collapse (similarly to AL), the distribution has an elbow at $x\sim 1$ and we find
\begin{equation}
\label{delocFit}
\phi(x)\propto
\begin{cases}
x^{-\alpha} & \text{if } x \lesssim 1 \\
x^{-\beta}  & \text{if } x \gtrsim 10,
\end{cases}
\end{equation}
where $\alpha,\beta$ depend on $h$. We have $\alpha<1<\beta$ ensuring the normalization of the distribution function in the delocalized phase and their values are almost independent of $N$ for the largest sizes explored.\footnote{A residual $N$ dependence is found in the left tails, at $x\ll 10^{-3}$, that part of the distribution reaching its asymptotic form for larger $N$ ($N\geq 14$).} This is an uncommon distribution for the quantity $x$: usually $\alpha=1/2$ and the large $x$ behavior is exponential \cite{mirlin2000statistics} as in the Porter-Thomas distribution of RMT \cite{porter1965statistical}.
\nomenclature{RMT}{random matrix theory}%
As the tail is power-law, we see that the delocalized region is less so than one would expect on general grounds. The almost perfect collapse of the curves in the upper panel of Fig.\ref{fig:phi-h} allow a much better finite size scaling analysis than any of its moments or IPR's.

\begin{figure}
\includegraphics[width=0.9\columnwidth]{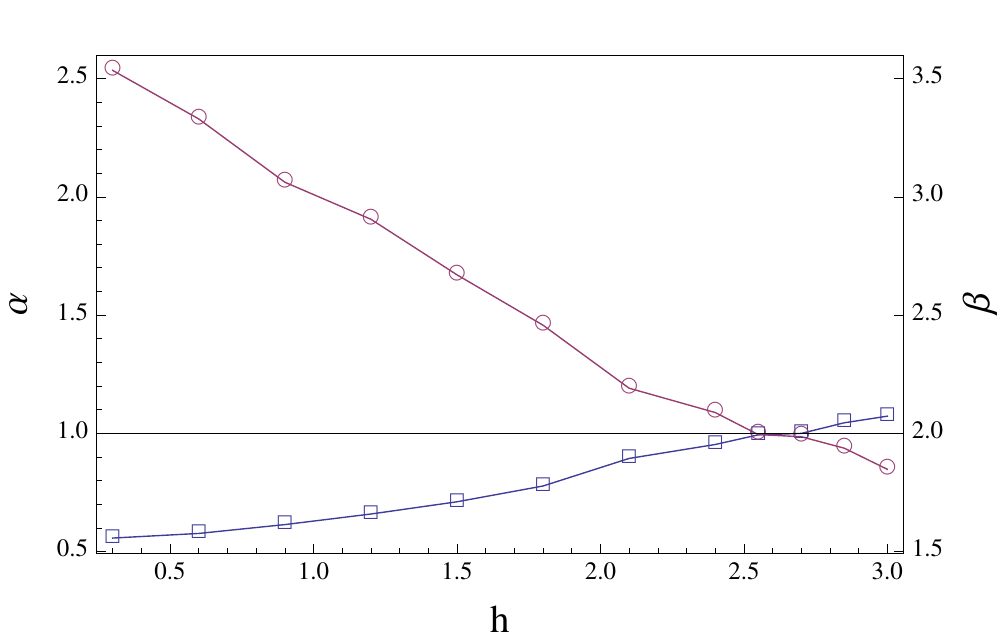}
\caption{The value of the exponent $\alpha$ (blue squares) and $\beta$ (pink circles) in Eq.\ (\ref{delocFit}) for $N=16$ (these exponents are independent of $N$ within the symbol size). The exponent $\alpha$ crosses the value $1$ required by summability, which occurs at $h\simeq 2.55\pm 0.05$, precisely where (within errors) $\beta$ crosses the value $2$, required for the existence of the first moment (normalization of the wave function).}
\label{fig:alphabeta}
\end{figure}

As $h$ approaches $h_c\simeq 2.6$ the elbow smoothens and $\alpha\to 1$ so that we can identify $h_c$ as the point at which $\alpha=1$, the distribution stops being summable and necessarily the independence on $N$ ceases.\footnote{As $\avg{x}=1$ is fixed by normalization the divergence of $\avg{1}$ implies a divergence of the first moment as well. In fact, $\beta=2$ occurs at the same value of $h_c$.} This occurs at $h_c=2.55\pm0.05$ as it can be seen in Fig.\ \ref{fig:alphabeta}. An explicit $N$-dependence of $\phi$ means that the scaling of all the IPR's and of the diagonal entropy with $N$ change abruptly and ergodicity is broken.

The exponent $\beta$ governs the scaling of the various $\IPR{q}$'s. For $0<q<\beta-1$ the integral in \eqref{prqdistrib} is finite and $\IPR{q}\sim \HS^{1-q}$. If instead $\beta-1<q$, since the integral in \eqref{prqdistrib} is divergent
the average of the participation ratio $\IPR{q}$ over the initial states $\ket{a}$ does not exist, but the typical value for a state should be found by looking of the sum $\HS$ i.i.d.\ variables $x_a^q$. One then finds the probability density for $\sum_{a\leq \HS}x_a^q\equiv Y$ (by computing and then inverting its Laplace transform) as
\begin{equation}
\label{typicalPq}
P(Y)\propto Y^{-\frac{3-\gamma}{4-2\gamma}}\exp\left(-C\left(\frac{\HS^{\frac{1}{\gamma-1}}}{Y}\right)^{\frac{\gamma-1}{2-\gamma}}\right),
\end{equation}
where $\gamma=1+(\beta-1)/q$, ($1<\gamma<2$) and $C$ is a constant of $\Ord{1}$. This distribution has a power law tail but the typical value of the sum is set by the exponential as $Y\sim \HS^{1/(\gamma-1)}\gg \HS$. This implies typical values of the $\IPR{q}$ of a state, when $q>\beta-1$:
\begin{equation}
\IPR{q}^{(N)}\sim\HS^{-q+\frac{q}{\beta-1}}.
\end{equation}
The different participation ratios define therefore different ``critical points" $h_q$ solutions of $\beta(h_q)=q+1$. The real transition, signaled by an explicit $N$-dependence of full distribution $\phi$ can then be identified by the diagonal entropy (\ref{eq:diagentr}).

As we said, the non-summable divergence of $\phi(x)$ at small $x$ signals the
beginning of the localized region. This implies an accumulation of wave-function
amplitudes towards small values typical of localized states\cite{mirlin1994distribution}.
Approaching the transition from the delocalized region, in fact, and assuming
the form $\phi(x)\propto x^{-\alpha}$ is preserved to arbitrarily small $x$, we
see that the minimum of the eigenfunctions amplitudes is the minimum of $\HS$
i.i.d.\ random variables $x_i$, which is found to scale like
$x_{\mathrm{min}}\sim \HS^{-1/(1-\alpha)}$. The values of $\alpha$ in
Fig.\ref{fig:alphabeta} predict a scaling exponent which compares well with that
obtained \emph{explicitly} from the numerics, supporting this hypothesis. And as
this exponent diverges when $h\to h_c$, we expect that the scaling becomes
faster than a power-law in $\HS$ at the transition and stays so in the whole
localized region. This observation is again verified in the numerics.

This suggests a description of the localized phase in which a typical eigenstate is described by a faster than exponential decay on ample regions of the Hilbert space, which is reminiscent of the ``small branching number" Bethe lattice picture of \cite{altshuler1997quasiparticle,basko2006problem} and of the eigenstates of a disordered but integrable model \cite{buccheri2011structure}.

This bring us again to discuss the similarities and differences with AL on the Bethe lattice (or regular random graph) \cite{abou1973selfconsistent,mirlin1994distribution}. Our case however brings three differences from this classic topic: {\it 1)} our lattice has connectivity $\Ord{N}\gg\Ord{1}$ (but still $\ll\Ord{\HS}$, the volume of the system), {\it 2)} the on-site disorder potentials of neighboring configurations $a$ and $b$ are strongly correlated ($E_a-E_b=h_{i+1}-h_i\ll E_{a},E_{b}$) and {\it 3)} our lattice is not random at all. In order to identify which of these three ingredients are necessary to preserve this phenomenology of the distribution functions we have investigated numerically a random graph with $\HS$ nodes and fixed connectivity $N/2$ and \emph{independent} random energies $\epsilon_i$ on each node. We observe the same qualitative features in the distribution of the coefficients and the same distance from the Porter-Thomas distribution, even for small $\dis$. On the contrary, for the 
Anderson model on a Bethe lattice with connectivity $\Ord{1}$ in the ergodic region we observe an exponential tail at large $x$ as predicted by the theory. Therefore we conjecture that the necessary requirement for the large $x$ power-law tail is the growing connectivity, and that one can get rid of the correlation of the energies and the specific topology of the hypercube. 

This confirms that we have the right to look at MBL as a localization phenomenon on a Bethe lattice with asymptotically large connectivity, a problem amenable of analytic treatment, beyond the locator expansion \cite{abou1973selfconsistent}.

\subsection{An integrable model for the localized phase}
We saw in the previous section which results can be obtained with the use of exact diagonalization technique. The exponential nature of the problem hardly constrains the possibility to go to big system sizes, that would be actually necessary for the estimation of the thermodynamic limit. If on one side we saw how particular quantities are less sensitive to the finite size corrections, on the other side, the other possibility is to limit ourselves to a particular class of models that are exactly solvable. These models have offered an important set of tools, with exact analytical results in many field of theoretical physics, including, in particular, low dimensional systems in condensed matter and statistical mechanics \cite{mussardo2009statistical}. They spread from lattice models, based on the Bethe-ansatz approach, to the analytic S-matrix useful for integrable field theories. In spite of their success, their range of applicability seldom had any overlap with disordered systems. In fact, the mechanism 
involved in the exact solution of integrable models, is typically based on an infinite set of symmetries constraining their dynamics and thus providing a set of equations, whose solution produces eigenstates or correlation functions. 
For disordered systems, many evident symmetries are trivially broken, such as translational and rotational invariance and the standard approach to integrable systems are therefore prohibitive. Specific examples have been built \cite{de1984families} but they usually lack of a concrete physical relevance. 

In order to find a partial compromise, we focused on the Richardson model, the most famous member of a more general class of integrable models, known as Gaudin magnets \cite{ortiz2005exactly}. It  was first introduced as a model of nuclear matter and has been studied in connection with the finite-size scaling of the BCS theory of superconductivity. 
The Richardson model \cite{richardson1963restricted, richardson1964exact, dukelsky2004colloquium} is an XX-model (i.e.\ with no $s^zs^z$ coupling) of pairwise interacting spins with arbitrary longitudinal fields
\begin{equation}\label{rich:hamiltonian}
H=-\frac{g}{N}\sum_{\alpha,\beta=1}^N s^+_\alpha s^-_\beta+\sum_{\alpha=1}^N h_\alpha s^z_\alpha,
\end{equation}
where $s^{x,y,z}$ are spin-$\frac{1}{2}$ representation of $SU(2)$ algebra. The fields $h_i$ appears as parameter in the Hamiltonian, that remains integrable for any choice of them, allowing therefore to study the model in presence of quenched disorder in the $z$ fields. The price to pay to accommodate disorder together with integrability is that the hopping term connects all the sites, in other terms, it is a fully connected graph.
As we already recall, the analogue of this model in 1-d, where the hopping $g$-term just connects nearest-neighbor sites on a chain, reduces to non-interacting fermions by Jordan-Wigner map and hence localizes for arbitrarily small disorder. So, if on one hand the fully-connectivity represents an unphysical aspects of this model, on the other hand, it is a crucial ingredient in order to make its dynamics non-trivial.
The Hamiltonian \eqref{rich:hamiltonian} is exactly solvable: it means that each eigenstate and eigenvalue can be, in principle, obtained by the solution of a set of algebraic equations. We will add more on this point in the following sections.
\subsubsection{Integrability and localization}
We have already pointed out that integrable models represent rare examples in the sea of the possible quantum Hamiltonians. It may appear therefore weird that the Hamiltonian \eqref{rich:hamiltonian} is instead integrable for arbitrary value of each parameter involved: the hopping strength $g$ and each local field $h_i$. It is interesting to investigate more on this issue. 
One of the main features of integrable Hamiltonians is the existence of local conserved quantities. It means a set of linear independent hermitian operators $Q_1, \ldots, Q_n$\footnote{There is a lot of confusion in the literature due to the rigorous definition of the conserved charges. This is due to the fact that the n\"aive application of the correspondence principle to the classical definition of integrable systems, produces a trivial quantum case. In particular, it is not even easy to fix the number of charges needed to call a finite system, integrable. We will not comment more on this point and by purpose we leave the parameter $n$ undefined here. A thorough discussion of these issues can be found in \cite{caux2010integrability}.} commuting among themselves and with the Hamiltonian
$$ \forall i \quad [Q_i, Q_j] = 0; \quad [Q_i, H] = 0 \; . $$
For any Hamiltonian, a trivial set of charges is clearly provided by the set of eigenstates
\begin{equation}
 \label{rich:projectors}
 P_E = \ket{E}\bra{E}\; .
\end{equation}
However, these operators are in most of the cases extremely complicate when expressed in terms of the local variables (e.g. the spins $s^{x,y,z}_i$ or local creation and annihilation operators for fermions or bosons); in other terms, they are non-local.
One would like to say that an exception to this general framework is indeed provided by disordered Hamiltonians in the Anderson localized phase: as we saw, every wave-function will be localized in the real space and also the projectors in \eqref{rich:projectors} will be local operators. However, attempts to explicitly write those projectors showed that, even in the localized phase, the expression would require an infinite perturbative series \cite{vosk2012many}.
The Richardson model being both disordered and integrable, somehow, provides a simple example of this. In fact a set of conserved charges can be written as
\begin{equation}\label{rich:charges}
  \tau_\alpha= s^z_\alpha-\frac{2 g}{N}\sum_{\beta\ne \alpha}\frac{1}{h_\alpha-h_\beta}\vec s_\alpha\cdot\vec s_\beta
\end{equation}
Here, we see the leading term is $\Ord{1}$ and is given by the local magnetization along the $z$ direction, with a correction due to all the other sites. 
Therefore since we have that $\tau_\alpha \simeq s^z_\alpha$ are exactly conserved, this model can provide a good toy model for the description of the many-body localized phase. 
\subsubsection{Solution of the model}
The model belongs to the class of models that are integrable through Algebraic Bethe Ansatz. It implies that all the states in the sector $S^z=(2M-N)/2$ can be found using an ansatz of the form
\begin{equation}\label{eigenstates}
\ket{E[w]}= \prod_{j=1}^M B(w_j) |\downarrow\ldots\downarrow\rangle,
\end{equation}
where the generalized raising operators are
\begin{equation}\label{B}
B(w)=\sum_{\alpha=1}^N\frac{s_\alpha^+}{w-h_\alpha} \; . 
\end{equation}
Requiring that they are eigenstates, one obtains the set of $M$ coupled Richardson equations for the roots $w_j$:
\begin{equation}\label{RE}
\forall j=1,...,M:\quad
\frac{N}{g}+\sum_{\alpha=1}^N\frac{1}{w_j-h_\alpha}-\sum_{k=1, k\neq j}^M\frac{2}{w_j-w_k}=0
\end{equation}
and the energy of the state is then given by
\begin{equation}\label{nrg}
 E[w]=\sum_{j=1}^M w_j-\sum_{\alpha=1}^N \frac{h_\alpha}{2}.
\end{equation}
We address the reader to reference \cite{links2003aba} for an extensive review of the algebraic aspects of the model and its solution.
As we said, we will focus on $S_z=0$ so $M=N/2$, which means that we have to solve $N/2$ coupled nonlinear equations, which is numerically viable only provided one has a reasonably good initial condition for root-finding algorithms. A widely used technique is that of considering that when $g\to 0$ the roots tend to some of the fields $h_\alpha$, and from (\ref{B}) it is clear that such root configuration correspond to the different choices of sets of $M$ spins which are flipped with respect to the ground state according to (\ref{eigenstates}); the choice of the set can be used to label the state at any $g$.

When one adiabatically increases $g$, by moving it of some small amount and solving (\ref{RE}) at each step, the roots start departing from their initial $h$'s values towards the negative direction. The ensuing evolution depends on the initial configuration of roots, but generally, two of them may collide and form a pair of complex conjugate solutions, then they may also recombine and return real. When $g\to\infty$, roots either diverge in the negative direction or stay trapped within a couple of levels. The number of roots that eventually diverge is equal to the total spin $S$ of the state (which is a conserved quantum number at infinite $g$). An algorithm which can follow the evolution of the roots with $g$ has to take into account these changes in the nature of the solution, where the roots become complex conjugate. These critical points, for random choices of the $h$'s can occur at particularly close values of $g$ and this can create troubles for the algorithm. \footnote{This problem is not so serious 
for the ground state and first excited states so one can go to much higher values of $N$ without losing accuracy.}

The reader may refer to \cite{faribault2009bethe} and \cite{sierra2006excitations} and references therein for further details on the solutions. Extensive study on critical points has been performed in \cite{dominguez2006critical}.

When more than a pair of roots collide in a too small interval of $g$ this change of variables may not be sufficiently accurate and one should think of something else (if one does not want to reduce the step in the increment of $g$ indefinitely). The most general change of variables which smooths out the evolution across critical points is that which goes from the roots $w_j$ to the coefficients $c_i$ of the characteristic polynomial $p(w)$ --i.e.\ the polynomial whose all and only roots are the $w_j$'s
\begin{equation}
 p(w) = \prod_{i=1}^M (w-w_i) = w^M + \sum_{i=0}^{M-1} c_i w^i
\end{equation}
This polynomial is quite interesting in itself as it satisfies a second order differential equation whose polynomial solutions have been classified by Heines and Stjielties \cite{szego1939orthogonal}
$$-h(x)p''(x)+\left(\frac{h(x)}{g}+h'(x)\right)p'(x)-V(x)p(x)=0$$ 
where 
\begin{eqnarray}
h(x)&=&\prod_{\alpha=1}^N(x-h_\alpha)\\
V(x)&=&\sum_{\alpha=1}^N\frac{h(x)A_\alpha}{x-h_\alpha}
\end{eqnarray}
Similarly to an eigenvalue problem, one has to find a set of $A_\alpha$'s such that there exists a polynomial solution of this equation. A similar approach has also been investigated in the recent work \cite{faribault2011numeric}.

Following the evolution of the coefficients $c_i(g)$ is a viable alternative to following the roots but we found out that the best strategy is a combination of both evolutions. Therefore we follow the evolution of the roots, extrapolating the coefficients and using them to correct the position of the roots at the next 
step in the evolution. In this way we do not implement any change of variables explicitly and we do not have to track the position of critical points. This algorithm\footnote{Python code is available on the webpage: \\
http://www.sissa.it/statistical/PapersCode/Richardson/} can be used on a desktop computer to find the roots of typical states with about 50 spins, although in order to collect extensive statistics we have limited ourselves to $N=40$. 

\subsubsection{A first check of the localized phase}
We have already introduced in \ref{mbl:xxz} a good indicator for the MBL problem: the return probability and the higher momenta of the probability distribution function of the wave-function defined in \eqref{mbl:returnprobability} and \eqref{prq}. Similarly, when the roots of an eigenstate have been determine, one can compute
\begin{equation}
 \label{rich:ipr}
 \PR{q}(e) = \sum_a |\braket{a}{e}|^{2q} 
\end{equation}

which is slightly different from \eqref{prq}, since the sum here involves the states of $\ket{a} = \ket{\uparrow \uparrow \downarrow \ldots}$ of the computational basis. The scalar product can be computed using
\begin{equation}
\label{rich:scalar}
 \braket{a}{e} = \frac{\det \left(\frac{1}{w_i - h_{\alpha_j}}\right)^2}{\det \left(\frac{1}{w_i - h_{\alpha_j}}\right)}
\end{equation}
where $\alpha_j$ corresponds to the indexes of the $M = N/2$ among $N$ up spins in the state $\ket a$.
However, when exact diagonalization is used, most of the computational effort goes in the diagonalization of the Hamiltonian matrix, that with the best known algorithms requires $\Ord{\HS^3}$ operations for an $\HS\times \HS$ matrix and $\HS = \binom{N}{N/2} \simeq \frac{2^N}{\sqrt N}$. In the Richardson case, instead, the determination of a single eigenstate can be done much more efficiently and the computation of the sum in \eqref{rich:ipr}, involving $\HS$ terms, becomes the hardest part, making for example $N = 40$ completely prohibitive.
To overcome this difficulty, we looked for other possible order parameters that exploit the integrable structure of the model, providing a fast way to compute them. Since the average values $\bra{E}s_\alpha^z\ket{E}$ has an expression similar to \eqref{rich:scalar}, they can be calculated efficiently (in $\Ord{N^3}$ time): therefore one is led to consider a \emph{microcanonical} version the Edwards-Anderson (EA) order parameter associated to a single eigenstate
\begin{equation}
q(E)=\frac{4}{N}\sum_{\alpha=1}^N\bra{E}s_\alpha^z\ket{E}^2,
\end{equation}
with this normalization $q\in[0,1]$. The average over eigenstates is
\begin{equation}
q=\frac{1}{2^N}\sum_E q(E).
\label{eq:avgqEA}
\end{equation}

To get the physical significance of this quantity, following \cite{pal2010mb} we start with a slightly magnetized spin $\alpha$ in an infinite temperature state:
\begin{equation}
\rho_0=(\id+\epsilon s^z_\alpha)/2^N
\end{equation}
with magnetization $\langle s^z_\alpha\rangle_0=\Tr{\rho_0 s^z_\alpha}=\epsilon/4$ (as $s_z^2=1/4$). The same magnetization at large time $t$ in the diagonal approximation reads
\begin{equation}
\langle s^z_\alpha\rangle_\infty=\lim_{t\to\infty}\Tr{e^{-iHt}\rho_0e^{iHt}s^z_\alpha}= \frac{\epsilon}{2^N}\sum_E\bra{E}s^z_\alpha\ket{E}^2.
\end{equation}
Therefore, averaging over $\alpha$ we obtain the equality with eq.\ (\ref{eq:avgqEA}):
\begin{equation}
q=\frac{1}{N}\sum_{\alpha}\frac{\langle s^z_\alpha\rangle_\infty}{\langle s^z_\alpha\rangle_0},
\end{equation}
namely the previously defined EA order parameter is the average survival fraction of the initial magnetization after very long times.

We notice two more things: \emph{one} \cite{viola2007generalized}, that $q(E)$ is related to the average purity of the state (here we use the total $S^z=0$):
\begin{equation}
q(E)=\frac{2}{N}\sum_{\alpha}\Tr{\rho_\alpha^2}-1
\end{equation}
and \emph{two}, that $q(E)$ is related to the average Hamming distance of the points in configuration space when sampled with the probability distribution $p_a$ relative to state $\ket{E}$:
\begin{eqnarray}
d(a,b)&=&\sum_{\alpha=1}^N\left(\langle a |s^z_\alpha | a \rangle - \langle b |s^z_\alpha | b \rangle \right)^2  \nonumber\\
&=&\sum_{\alpha=1}^N\frac{1-4\bra{a}s_\alpha^z\ket{a}\bra{b}s_\alpha^z\ket{b}}{2},
\end{eqnarray}
and multiplying by $p_{a},\ p_{b}$ and summing over $a,b$ we find:
\begin{equation}
\label{eq:qd}
L \equiv \avg{d}=\frac{N}{2}(1-q).
\end{equation}

So $q$ is computationally easy and it captures both some geometric properties of the covering of the configuration space by an eigenstate and the long-time correlation function for $s^z$. We averaged $q$ over the spectrum (sample over typical states) and then over realizations (the number of which depends on the size of the system but it will never be less than 100).

\begin{figure}[htbp]
\begin{center}
\includegraphics[width=7.5cm]{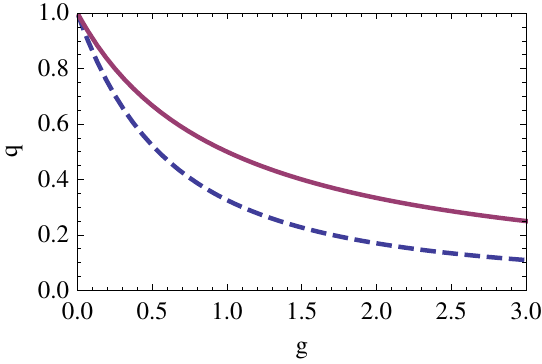}
\caption{The educated guess for $q(g)$ coming from \eqref{eq:qgpade} for $\overline{q}$ (solid line) and the lower-bound coming from the conserved charges (dashed line): the Mazur inequality is satisfied but not saturated.}
\label{fig:comparisonMazur}
\end{center}
\end{figure}

\begin{figure}[htbp]
\centering
\includegraphics[width=8cm]{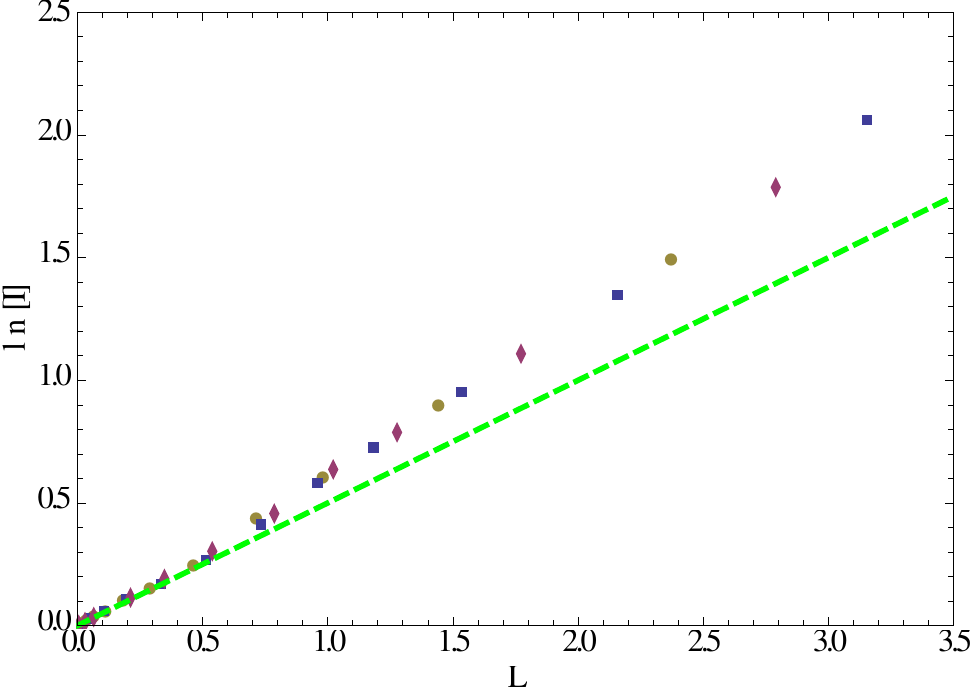}
\caption{$\ln \cI$ as a function of the average distance $L$. The points are (square, diamond, circle) $N = 28,30,32$ averaged over 100 realizations: the dashed straight line is the second order perturbation theory approximation Eq.\ (\ref{ipdrel}).}
\label{fig:IPRL}
\end{figure}

We found this average $\avg{q}$ as a function of $g$ for $g\in[0,40]$ and $N=16,...,38$ and studied the point-wise finite-size scaling (in the form $q_N(g)=q(g)+c_1(g)/N+c_3(g)/N^3$) to obtain the thermodynamic limit of $q$ (see Figure \ref{fig:comparisonMazur}). We fit the data using a ratio of polynomials with the condition that $q(0)=1$ and we found that averaging over the state and the realization of disorder
\begin{equation}
\label{eq:qgpade}
\avg{q}=\frac{1+3\times 10^{-8}g}{1+1.003g+0.009g^2}\simeq\frac{1}{1+g}
\end{equation}
works in the whole range of data to an error of at most $0.5\%$. We therefore conjecture this to be the correct functional form of the EA order parameter at infinite temperature. Since $q(g)>0$, no transport occurs in this model for arbitrary value of the hopping term $g$, as we expected being it integrable. Moreover, it is possible to obtain a lower limit for $q(g)$ employing the expression of the conserved charges \eqref{rich:charges} and the Mazur's inequalities \cite{caux2010integrability, suzuki1971ergodicity, mazur1969nonergodicity}. The two curves are shown in Fig. \ref{fig:comparisonMazur}.
One can wonder whether this parameter $q$ is related to the participation ratio. From perturbation theory for small $g$ one gets
\begin{equation}
\label{ipdrel}
\ln \cI \simeq \frac{L}{2}.
\end{equation}
where $L$ is defined in \eqref{eq:qd}. The relation is linear for small $g$ and for higher value, the relation is plotted numerically in Fig. \ref{fig:IPRL}, showing that a relation exists.
\subsubsection{Montecarlo dynamics inside a quantum state}
Using the components of each eigenstate on the computational basis, that are easily computed, we devised a Montecarlo algorithm to explore the structure of the eigenstates. Fixed an eigenstate $\ket{E}$, define the probabilities $p_a=|\braket{a}{E}|^2$ where $a\in\cC$ stands for one of the $\binom{N}{N/2}$ allowed classical configurations of spins which constitute the configuration space $\cC$. We perform a random walk with the probabilities $p_a$'s, namely start from a random configuration $a$. The neighboring configurations are those living within the same subspace $S^z=0$ and differing from $a$ by the exchange of a pair of opposite spins. We move to a random one of the $(N/2)^2$ neighboring states, say $b$, by accepting the move with probability $\min(1,p_b/p_a)$. The random walk proceeds in this way, generating a history of configurations $a$. The resulting dynamics can be compared with that of random percolation on the hypercube, which has been proposed as a model of relaxation in a glassy system \cite{
campbell1987random}. We will find that in both cases, a stretched exponential is the best fit and that the exponent depends on the coupling constant $g$. This, we believe, is a remarkable similarity.

\begin{figure}[htbp]
\centering
\includegraphics[width=8cm]{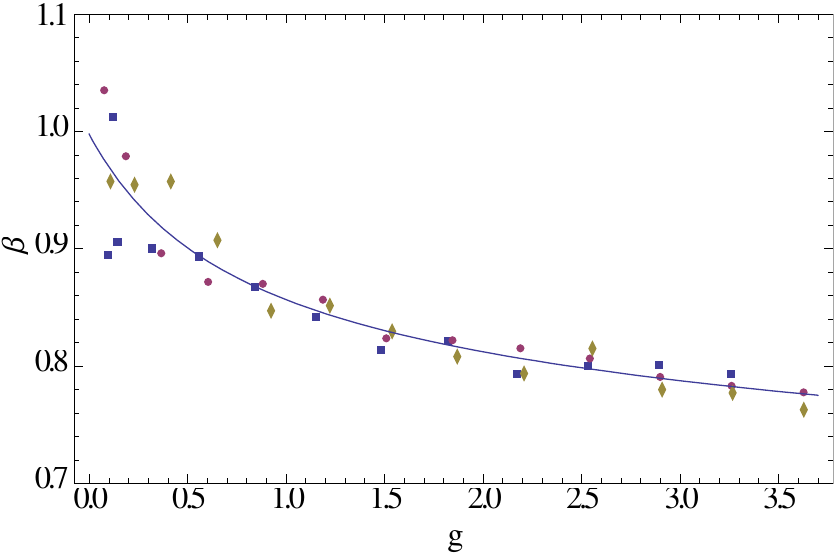}\\
\caption{The stretched exponential exponent $\beta$ data as a function of $g$ for $N=28,32,36$ (square, circle, diamond) together with a fit of the form $(1+a_1 g)/(b_0+b_1 g+b_2 g^2)$. }
\label{fig:betag}
\end{figure}

An important quantity in this sense is the time dependence of the average distance from the starting point. Consider the Hamming distance $H(t)$ from the starting point $H(t)\equiv d(a(t),a(0))$, where $a(0)$ represents a classical configuration of spins and $a(t)$ the one reached after $t$ Montecarlo steps. For $t \gg 1$, after averaging over many starting points $a(0)$, $H(t)$ is fit quite accurately by a stretched exponential ansatz of the form:
\begin{equation}
H(t) = L \left(1 - e^{-\left(\frac{t}{\tau}\right)^\beta} \right),
\end{equation}
where $L$ is the average distance introduced before and $\beta$ is a new characteristic exponent.
Let us consider the behavior of the exponent $\beta$ with respect to $g$, as plotted in Fig.\ \ref{fig:betag}. Even if the results become quite noisy for small $g$, we can still see that starting from $1$ for small values of $g$, $\beta$ decreases as $g$ increases, although quite slowly.
The small time behavior of $H(t)$ can be used to obtain some information about the local structure of the state. In particular we can set
\begin{equation} 
k \equiv \frac{H(1)}{2} = \frac{4}{N^2} \sum_{\langle a, b \rangle} \min (p_a, p_b) 
\end{equation}
where the last equality follows from the Montecarlo rate and the sum is over nearest-neighbor states. This connectivity fraction $k$ can be considered as a measure of the average fraction of active links. 
\begin{figure}[htbp]
\begin{center}
\includegraphics[width=8cm]{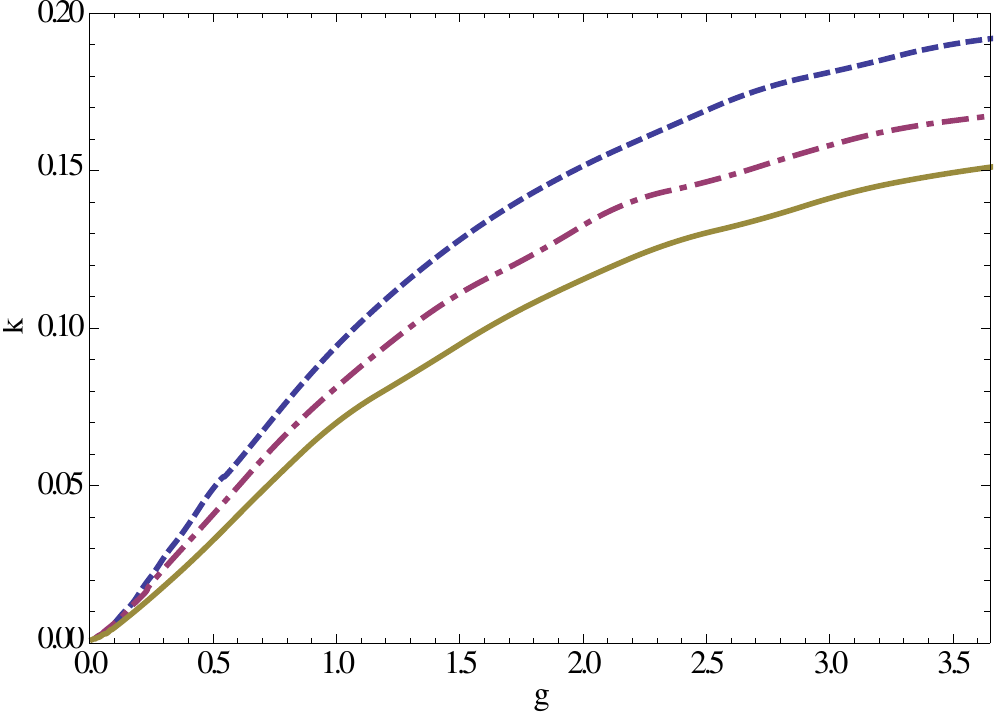}
\caption{Connectivity fraction as a function of $g$. Different lines corresponds to $N = 18 $ (dashed), $24 $ (dotted  and dashed), $30$ (solid).}
\label{fig:conng}
\end{center}
\end{figure}
From Fig.\ \ref{fig:conng}, we may deduce two things: one is that $k$ stays well below $1$ even for large $g$, confirming, as we claimed before, that the typical state is never uniformly spread over the hypercube; the second is that the connectivity scales with $N$ as $N^{-1}$ for small $g$ and with $N^{-1/2}$ for large $g$ (a fit $k=A/N^\alpha$ shows a continuously decreasing $\alpha$ from 1 to $1/2$). 
This second property seems to be related to the specific details of the Richardson model and in particular of its large $g$ limit. Instead, we conjecture that the first is a typical fingerprint of the localized phase in the MBL transition.
\section{Conclusions.}
In this chapter, we presented the MBL transition. We showed its relevance for both the study of transport in disordered metals and for quantum computation. We reported an argument for the failure of the QAA due to the existence of a many-body localized phase. Then we focused on a specific example, the XXZ spin chain in disordered $z$-fields; we have investigated the behavior of the return (or survival) probability as a possible detector of the MBL transition. We have shown how this question leads to the necessity of a thorough study of the distribution of the wave-function amplitudes of the eigenstates. We then identified the major changes which occur to said distribution at the MBL transition point. The delocalized, ergodic phase is more ``localized" than the corresponding single-particle AL and RMT does not seem to be a good approximation for the eigenstates, not even deep in the delocalized region. The localized region seems very akin to the case of single particle AL on the Bethe lattice with 
connectivity $\Ord{1}$, in particular the distribution functions of the amplitudes show a small-$x$ accumulation which points towards localized wave function on configuration space. 
To further investigate this phase, we presented the results coming from the Richardson model, an exactly solvable model for arbitrary value of disordered fields. This gave us indication that the localized phase looks like a spin glass. 
The similarities and differences with the Bethe-lattice case suggested what are the necessary ingredients for a viable analytical study of MBL.



%% file: MBL/Figs/crossingEC3.pdf_tex
\begingroup%
  \makeatletter%
  \providecommand\color[2][]{%
    \errmessage{(Inkscape) Color is used for the text in Inkscape, but the package 'color.sty' is not loaded}%
    \renewcommand\color[2][]{}%
  }%
  \providecommand\transparent[1]{%
    \errmessage{(Inkscape) Transparency is used (non-zero) for the text in Inkscape, but the package 'transparent.sty' is not loaded}%
    \renewcommand\transparent[1]{}%
  }%
  \providecommand\rotatebox[2]{#2}%
  \ifx\svgwidth\undefined%
    \setlength{\unitlength}{786.46875bp}%
    \ifx\svgscale\undefined%
      \relax%
    \else%
      \setlength{\unitlength}{\unitlength * \real{\svgscale}}%
    \fi%
  \else%
    \setlength{\unitlength}{\svgwidth}%
  \fi%
  \global\let\svgwidth\undefined%
  \global\let\svgscale\undefined%
  \makeatother%
  \begin{picture}(1,0.391037)%
    \put(0,0){\includegraphics[width=\unitlength]{crossingEC3.pdf}}%
    \put(0.28346644,0.26831169){\color[rgb]{0,0,0}\makebox(0,0)[lb]{\smash{$E_2$}}}%
    \put(0.16140184,0.20727939){\color[rgb]{0,0,0}\makebox(0,0)[lb]{\smash{$E_1$}}}%
    \put(0.34449875,0.00383837){\color[rgb]{0,0,0}\makebox(0,0)[lb]{\smash{$\lambda$}}}%
    \put(-0.00135098,0.28865579){\color[rgb]{0,0,0}\makebox(0,0)[lb]{\smash{$E$}}}%
    \put(0.81241308,0.28865579){\color[rgb]{0,0,0}\makebox(0,0)[lb]{\smash{$E_2$}}}%
    \put(0.75138078,0.16659118){\color[rgb]{0,0,0}\makebox(0,0)[lb]{\smash{$E_1$}}}%
    \put(0.91413359,0.00383837){\color[rgb]{0,0,0}\makebox(0,0)[lb]{\smash{$\lambda$}}}%
    \put(0.56828386,0.28865579){\color[rgb]{0,0,0}\makebox(0,0)[lb]{\smash{$E$}}}%
    \put(0.69034847,0.00383837){\color[rgb]{0,0,0}\makebox(0,0)[lb]{\smash{$\lambda_c$}}}%
    \put(0.4543569,0.17676324){\color[rgb]{0,0,0}\makebox(0,0)[lb]{\smash{M+1}}}%
  \end{picture}%
\endgroup%

%% file: Quench/quench.tex
\chapter{Thermalization in closed quantum systems}
\markboth{\MakeUppercase{\thechapter. Quantum quenches}}{\thechapter. Quantum quenches}
\label{chapt:quench}
\ifpdf
    \graphicspath{{Quench/Figs/PNG/}{Quench/Figs/PDF/}{Quench/Figs/}}
\else
    \graphicspath{{Quench/Figs/EPS/}{Quench/Figs/}}
\fi

\section{Introduction}
Largely triggered by recent experiments on cold atoms \cite{kinoshita2006quantum,hofferberth2007non,weiler2008spontaneous,greiner2002quantum,sadler2006spontaneous}, there has been in 
the past few years intense theoretical activity aimed at understanding the non-equilibrium dynamics in closed and isolated quantum systems following a change in one of the system parameters.
The simplest example is provided by the limit known as {\em quantum quench}: namely, the system is prepared in an 
energy eigenstate $\ket{\psi_0}$ of an initial pre-quench Hamiltonian, $\Hpre$, and then is allowed to evolve 
according to a new post-quench Hamiltonian, $\Hpost$, which differs from $\Hpre$ by some variation of a parameter. Sure enough, there will be some transient effect, but after that does the system reach a stationary state? Given that such a time evolution is purely unitary, it is clear that for finite dimensional system, quantum recurrence will always occur. However for very large systems, in the thermodynamic limit, we can expect that focusing on finite portions of the system, the remaining (infinite) part of it will act as a thermal bath. It is not difficult to prove, even rigorously, that under these conditions a stationary states will be reached \cite{venuti2010universality}. The question becomes how to characterize it and more specifically if thermalization occurs, i.e. if the system can be described with the standard approach of statistical mechanics in terms of Gibbs ensembles.

Recent progress in understanding thermalization of an extended quantum system
following a quench has involved both analytical and numerical studies. 
To be more concrete, imagine, as it typically happens, the pair $\Hpre, \Hpost$ is such that
the initial state will have an almost definite energy, meaning that it can be written as a linear superposition of $\ket{\eig}$, eigenstates of $\Hpost$, all in a shell of energies,
$|\eig-\iniE| < \Delta$, centered at $\iniE$:
\begin{equation}
| \psi \rangle \,=\,\sum_{|\eig-\iniE| < \Delta} c_{\eig} | \eig \rangle \,\,\,.
\end{equation}
The time average of the density matrix based on this state, given by
\begin{equation}
\label{quench:diag}
\rho_{diag}(\iniE)\,=\,\overline {| \psi_t \rangle \langle \psi_t |}
\,=\,\sum_{|\eig-\iniE| < \Delta}
| c_{\eig} |^2 | \eig \rangle \langle \eig | \,\,\,,
\end{equation}
defines the so-called {\em diagonal ensemble} which is, in general, different from the micro-canonical density matrix defined by
\begin{equation}
\label{quench:mc}
\rho_{mc}(\iniE) \,=\, \frac{1}{{\cN_{\iniE}}} \sum_{|\eig-\iniE| < \Delta}
 | \eig \rangle \langle \eig | \,\,\,.
 \end{equation}
where $\cN_{\iniE}$ is the number of eigenstates inside the shell. Now,
ergodicity in its classical sense, means that the time averages coincide with
the phase space averages. So, unless it happens that $| c_{\eig} |^2 =
1/{\cN_{\iniE}}$, a quantum analogous of the classical notion of ergodicity
does not hold in most of the cases.
In the attempt to define it, Von Neumann \cite{neumann1929beweis} (see also
\cite{goldstein2010long}) gave rise to a slightly different notion, still known as
quantum ergodicity or, to avoid confusions \textit{normal typicality}. The ideas
of his work are as follows. Suppose we take a quantum system. Basic quantum
mechanics tells us that, due to the possible non-commuting nature of the quantum
observables, it is not always possible to fully characterize its state in terms
of measurements. However, even though microscopic quantities are affected by
quantum effects and the uncertainty principle, there must exist coarse-grained
versions, that he dubbed \textit{macroscopic observables}, that are commuting.
Then he was able to prove that, under reasonable hypothesis for the Hamiltonian
$\Hpost$, for all possible initial states in a given energy window, for most of
the possible choices of the macroscopic observables $O_{\mbox{\tiny macro}}$ and
for most of the times $t$, thermalization occurs i.e.
\begin{equation}
 \label{vonneumanntheo}
 \bra {\psi_t}O_{\mbox{\tiny macro}} \ket{\psi_t} \simeq \Tr\left(\rho_{mc}(\iniE) O_{\mbox{\tiny macro}}\right)
\end{equation}
At first glance, this result could look as much more than we required, since no
time average is involved. However, the requirement for the observable of being
macroscopic is catchy and hard to check for a given operator. Therefore, the
question whether for a given observable $\obs$, long-time averages of expectation
values coincide with the microcanonical average remains open. Defining $\langle
{\obs} \rangle_{diag} = \rm{Tr} ({\obs}
\rho_{diag}(\iniE))$ and $\langle {\obs} \rangle_{mc} = \rm{Tr} ({\mathcal
O} \rho_{mc}(\iniE))$,
it may be true that the identity 
 \begin{equation}
\overline{ \bra{\psi_t} {\obs} \ket{\psi_t}} \,=\,\avg{\obs}_{diag}
= \langle {\obs} \rangle_{mc} \,\,\, ,
\label{identityensembles}
\end{equation}
indeed holds. 
Eq. \eqref{vonneumanntheo} may appear particularly weird if one takes into account that it holds for every initial state $\ket{\psi_0}$ including, for example, an eigenstate of $\Hpost$. In this case, no dynamics goes on and the statement can be interpret as asking that the expectation values $\meanO =\langle \eig | {\obs} | \eig\rangle$ of the (macroscopic) observable
do not fluctuate between the Hamiltonian eigenstates which are close in
energy. This is, in a nutshell, 
the scenario known in the literature as the {\em Eigenstate Thermalization Hypothesis}
(ETH)%
\nomenclature{ETH}{eigenstate thermalization hypothesis}%
) which was put forward by Deutsch and Srednicki \cite{deutsch1991quantum,srednicki1994chaos,srednicki1999approach}, based on previous work by Berry \cite{berry1977regular}, 
and which has been recently advocated by Rigol et al. \cite{rigol2008thermalization} as the mechanism behind the thermalization
 processes in quantum extended systems. In fact, for such an observable, the identity (\ref{identityensembles}) holds for all those initial states which are sufficiently narrow in energy.

Recently this hypothesis has been put under intense scrutiny by different groups. 
The main emphasis heretofore has been given to the numerical analysis of specific 
models\footnote{Analytic results for quantum quenches have been obtained only for a restricted class of exactly 
solvable lattice models, such as the XY chain, the Ising model or the XXZ quantum spin chain 
\cite{barouch1970statistical,sengupta2004quench,rossini2009effective,rossini2010long,calabrese2011quantum,mossel2010relaxation}. Analytic results have been also obtained for systems nearby the 
critical point \cite{de2010quench} or for continuous exactly solvable systems, especially in the regime of 
conformal symmetry \cite{calabrese2006time,cazalilla2006effect,iucci2009quantum}. However it has been argued that the relaxation phenomena of 
these models, ruled by an infinite number of conserved quantities, may be different from the thermalization 
of a generic model and may require the introduction of a generalized Gibbs ensemble, as proposed in 
\cite{rigol2007relaxation} (see also \cite{fioretto2010quantum} for a derivation in integrable field theories). 
 In this chapter, however, we will not deal with such systems, but rather address these issues in a separate publication.}, 
such as hard-core bosons \cite{rigol2008thermalization,rigol2009breakdown}, the Bose-Hubbard model 
\cite{kollath2007quench}, strongly correlated interacting fermions \cite{manmana2007strongly}, the Hubbard model \cite{eckstein2009thermalization,eckstein2008nonthermal,kollar2008relaxation}, etc. 
In this chapter, instead of analyzing a 
particular system, we take a different approach. 
We will not try to answer the question whether a specific observable for a given model and quench protocol will look thermal. Instead, we will study what happens \textit{typically} and if thermalization occurs, what is the involved mechanism. \textit{Typically}, here, means that Hamiltonian and observables will be drawn at random from an ensemble of random matrices \cite{mehta1991random}, which in the following will 
parameterize both the Hamiltonians and the observables\footnote{For simplicity we consider hereafter real symmetric matrices.}. 
In particular we have chosen to study the quantum quenches and the relative thermalization in a class of Hamiltonians given by 
\begin{equation}
H(h) = H_0 + h V \,\,\,,
\label{HamiltoniansH}
\end{equation}
where the quench parameter $h$ is meant to explicitly break a $Z_2$ symmetry of the \textit{unperturbed} Hamiltonian $H_0$. 
Such Hamiltonians, which are arguably among the simplest examples of quantum systems, may model spin chains in the 
presence of an external magnetic field but, as we shall see later, they may also encode the familiar quantum Ising 
chain in a transverse magnetic field.
Given the relative simplicity of this class of Hamiltonians, studying their quench dynamics may be a useful path to 
extract interesting information on generic properties of non-equilibrium systems, thus disregarding, in doing so, 
all additional complications coming from a richer structure of states of a specific model.   

The natural choice when dealing with (real) random matrix is the \textit{Gaussian orthogonal ensemble} (GOE). 
\nomenclature{GOE}{gaussian orthogonal ensemble}%
However, even upon adopting the abstract language of random matrices, an important issue governing
thermalization properties and of which to be mindful
is the locality. In fact, as we saw comparing \eqref{quench:diag} and \eqref{quench:mc}, in general we do not expect that all the possible observables, meaning with that all the possible hermitian operator, will look thermal. But we have to focus on a subclass of them, that can be considered local, and for which we can expect, as we said, the mechanism works. Nevertheless, it is not easy to generate \textit{local} random operators, even because the notion of locality is intuitive but not always rigorous. We decided therefore to focus on the structure of the hypercube corresponding to the Hilbert space: as we saw for \eqref{mbl:xxzHam} and \eqref{rich:hamiltonian}, the connectivity scales as a power of the real-space volume $\SN^k \ll \cN$, the Hilbert space size, i.e. the number of points in the graph. This will result,when written in the computational basis, in a sparse matrix ensemble (SME), i.e. matrices with a small proportion of non-zero entries. 
\nomenclature{SME}{sparse matrix ensemble defined in \ref{chapt:quench}}
The two kinds of matrices, GOE and SME,
have two different properties, going from the densities of states to the localization of the wave-functions. For these reasons one observes a different behavior under a quench of the parameter $h$.

Important features of quantum quench processes in local systems were discussed in a paper by Biroli et al. \cite{biroli2009effect}, 
in particular the role played by rare fluctuations in the thermalization of local observables. These authors considered 
the existence of certain rare eigenstates -- rare compared to the typical ones sampled by the micro-canonical 
distribution -- but which may be responsible, if properly weighted, for the absence of thermalization observed 
in certain systems. As discussed in more detail later, the presence of such states can be detected by studying 
the spread of the expectation values of the observables on the energy eigenstates, in particular by 
the finite size dependence of the distribution of expectation values. The numerical 
analysis that we have performed seems indeed to indicate the existence of these rare states in the case of 
sparse random matrices, while they are absent in the case of dense random matrices. However, in our numerics, 
thermalization is observed nonetheless in SME, simply because our averaging procedure on the 
different sampling of observables and Hamiltonians does not place a natural exponentially large weight upon the 
rare states, thus enabling them to break thermalization.

It should be underlined that the existence of rare states in the thermodynamic limit has been debated in the literature 
and in particular in a series of papers by Santos and Rigol \cite{santos2010onset,rigol2010quantum,santos2010localization}.  They have argued that in a portion of the phase
diagram of an extended t-J model with next-nearest-neighbor interactions, rare states are absent.  We will come back to this
conclusion in our presentation of results.

\section{Locality}
\label{Sec_locality}
In this section we discuss the nature of Hamiltonian matrices associated with {\em local} models. In order to 
consider finite-size matrices, we will focus on lattice models, keeping in mind that continuous ones can always be discretized adding proper cut-offs.
The main idea of this section is the following: in most of the basis of the Hilbert space, the matrix representation of 
a local Hamiltonian corresponds to a {\em dense} matrix, i.e. a matrix which has all entries different from zero 
(an explicit example will be given below). However, if the theory is local, there will exist a basis 
(in the following called the {\em local basis}), in which the Hamiltonian will be represented by a {\em sparse} matrix, 
i.e. a matrix where the great majority of its entries are zero. We have already seen that the computational basis does the job in the previous chapters. An other example is the 1d quantum Ising model in a longitudinal field (generically
a non-integrable model).
In this case the quantum Hamiltonian for $\SN$ sites is given in terms of Pauli matrices and takes the form:
\begin{equation}
\label{Ising}
 H \equiv \sum_{i=1}^\SN \sigma^z_i \sigma^z_{i+1} + h \sigma^x_i + \sigma^z_i = \sum_{i=1}^\SN H_i.
\end{equation}
The last equality makes evident the local nature of this model: the Hamiltonian
has been written as a sum of operators involving only two lattice sites. So an operator is local if it can be written as a sum over the volume of operators involving only few body terms.
In the computational basis, given as usual by the common eigenstates of the $\sigma^z_\is$ operators, the matrix elements are
\be
 H_i^{a,b} = \bra{a_1 \ldots a_N} H_i \ket {b_1 \ldots b_N} =
((a_i a_{i+1}+a_i)\delta_{a_i, b_i} + h\delta_{a_i, -b_i})\prod_{k\neq i}
\delta_{a_k, b_k'},
\label{nonzeroIsing}
\ee
where $a,b$ are labels for the computational basis and each $a_i,b_i \in \{\uparrow,\downarrow\}$ corresponds to the two
possible eigenstates of $\sigma^z_i$.
From this expression it is easy to deduce that on each row of the matrix there
are $\SN+1$ non-zero entries and therefore the total number of non-zero elements of
the $\HS\times \HS$ matrix $H$  is $ n_{nz} =  (\SN+1) \HS$, where $\HS = 2^\SN$ is the Hilbert space size. Since the total
number of matrix elements is $\HS^2$, the density of non-zero elements is given
by 
\be 
\rho \,=\,\frac{{n_{nz}} }{\HS^2} \propto \frac{\ln \HS}{\HS}\,\,\,,
\label{densityones}
\ee
In general the density of zeros will look like
\be
\rho_0 \,=\,1-\rho\,\simeq\,1-  \frac{k (\ln \HS)^\epsilon}{\HS}.
\label{densityzeros}
\ee
The constant $k$ and $\epsilon$ are related to specific properties of the model, such as dimensionality and conserved quantities. 
The Richardson Hamiltonian in \eqref{rich:hamiltonian}, being fully-connected, can not be written as sum over the volume of density operators. Nevertheless, Eq. \eqref{densityzeros} holds with $\epsilon = 2$. Therefore for large values of $\HS$, the Hamiltonian matrix $H_N$ 
is a {\em sparse} matrix, i.e. a matrix with a large number of zeros and few non-zero entries. 
This statement holds in general for any quantum Hamiltonian involving few body terms and can be shown to be true also for the discretization of quantum field theories.

\section{Quantum quenches, thermalization, and the ETH}
\label{ETH}
Let us consider an initial state $|\psi_0\rangle$ which is an eigenstate of an initial
Hamiltonian, $H(h^<)$, governed by the parameter $h^<$.  At $t=0$ we abruptly change the value 
of the parameter to $h^>$. The evolution of the initial state will be then governed 
by the dynamics given by $H(h^>)$. Our interest is in the long time behavior of expectation values of 
some one-point observable, $\langle \psi_0(t)|{\obs}| \psi_0(t)\rangle$. An observable has a thermal 
behavior if its long time expectation values
coincides with the micro-canonical prediction, i.e. 
\be
\langle \psi_0(t) |{\obs} |\psi_0(t)\rangle \xrightarrow{t \to \infty} \Tr{{\obs}
\rho_\text{mc}}\,=\,\langle {\obs} \rangle_\text{mc} \,\,\,.
\label{ltq}
\ee
Dealing with finite-size matrices, it is natural to take a notion of convergence, similar to classical ergodicity,
where time-averages are meant
\be
\overline{\langle \psi_0(t) | {\obs} |\psi_0(t)\rangle}
\equiv \frac{1}{T}\int^T_0 \langle \psi_0(t) | {\obs} |\psi_0(t)\rangle = \sum_{\eig} |c_{\eig}|^2
{\obs } =  \langle {\obs} \rangle_\text{mc} \,\,\,,
\label{ltq2}
\ee	
where $c_{\eig}=\langle \psi_0|\eig\rangle$ are the overlap of the initial state on
the eigenstate $\ket{\eig}$ of $H(h^>)$, and 
$\meanO=\langle \eig|{\obs}|\eig\rangle$ are the expectation values of the
observable, $\obs$, on the post-quench eigenstates. Eq. (\ref{ltq2}) defines the diagonal ensemble prediction, with the corresponding density matrix defined as 
\be
\rho_\text{diag}= \overline{|\psi_0(t)\rangle \langle\psi_0(t)|}=\sum_{\eig} |c_{\eig}
|^2|\eig\rangle \langle \eig| \,\,\,,
\ee
supposing the eigenstates of $H(h^>)$ are non-degenerate.


A possible mechanism for the thermal behavior of physical observables 
is based on the so called {\it Eigenstate Thermalization Hypothesis} (ETH)
\cite{deutsch1991quantum,srednicki1994chaos,srednicki1999approach}. It states that the expectation value of a physical observable,
$\meanO =\langle \eig |{\obs}|\eig\rangle$, on an
eigenstate, $\ket{\eig}$, of the Hamiltonian is a smooth function of its energy, $\eig$, with its value 
essentially constant on each micro-canonical energy shell. In such a scenario, thermalization in the 
asymptotic limit follows for every initial condition sufficiently narrow in energy.
ETH implies that thermalization can occur in a closed quantum system, different from the classical
case where thermalization occurs through the interactions with a bath.
As pointed out by Biroli et al. \cite{biroli2009effect} there are two possible interpretation of ETH: a weak one, which can be shown 
to be verified even for integrable models, which states that the fraction of non-thermal states vanishes
in the thermodynamic limit, and a strong one which states that non-thermal states completely disappear 
in the thermodynamic limit.  In the weak version of the ETH, not every initial condition will thermalize.

We briefly remind the reader of the origin of these two interpretations as it will be salient
later. Firstly, for thermalization to occur one 
needs a distribution of the overlaps peaked around the energy $E=\langle\psi_0|H|\psi_0\rangle$. 
As shown in Ref. \cite{rigol2008thermalization}, 
the energy density $e$ has vanishing fluctuations in the thermodynamic limit
\be
\Delta e =\frac{\sqrt{\langle E^2\rangle_\text{diag}-\langle
E\rangle_\text{diag}^2}}{\SN} \propto \frac{1}{\SN^{1-\sigma/2}} \to ~0 \text{~for~} \SN\to\infty
\label{enfluc}
\ee
where $\SN$ is the system size and $\sigma$ is the dimension of the space over which the
coupling $h$ is adjusted in the quench. In our case $\sigma = 1$ for SME.
We will, however, see that $\sigma$ is effectively larger when we consider quenches in dense matrices,
and consequently $\Delta e$ does not vanish in the thermodynamic limit.
Correspondingly this property means that the distribution of
intensive eigenenergies (eigenenergies scaled by $1/\SN$) with weights $|c_{\eig}|^2$ 
is peaked for large system
sizes. If the ETH is true, an immediate consequence of property Eq. (\ref{enfluc}) would
be that averages in the diagonal ensemble coincide with averages in the
micro-canonical ensemble. However, for a finite system, there will always be
finite fluctuations of $\meanO$. To characterize the ETH mechanism we need then
to have some control on the evolution of the distribution of $\meanO$ in 
approaching the thermodynamic limit.  As shown in \cite{biroli2009effect}
the width of the distribution $\meanO$ of an
intensive local observable\footnote{${\obs}$ is an intensive local
observables if it can be written as
$\frac{1}{\SN}\sum_\alpha {\obs}_\alpha$ where ${\obs}_\alpha$ are finite ranged
observables and the sum is over a local spatial region.} vanishes in the thermodynamic limit
\be
(\Delta {\obs}_e)^2\,=\,\frac{\sum_e \meanO^2}{\cN_e} - \left(\frac{\sum_e
\meanO}{\cN_e}\right)^2  \to 0 \text{ for } \SN \to \infty
\label{Ofluc}
\ee
where $e$ is the intensive energy defining a micro-canonical shell including
$|\eig\rangle$ such that $\eig/\SN\in [e-\epsilon,e+\epsilon]$ and
$\cN_e$ is the number of states in the microcanonical shell.
Eq.\, (\ref{Ofluc}) implies that the fraction of states characterized by a value of
 $\meanO$ different from the micro-canonical average vanishes in the
thermodynamic limit. Nevertheless, states with different values 
of $\meanO$ may exist. These states live in the tails of the shrinking $\meanO$ distribution and are expected to be small in number. This is why they
are called "rare".
These states, however, under proper conditions, can be relevant to the issue of thermalization. 
Indeed, if in the $|c_{\eig}|^2$ distribution they are weighted heavily,
the diagonal ensemble average will be different from the micro-canonical and
the system keeps a memory of the initial state. As emphasized in
\cite{biroli2009effect}, it is clear that the weak interpretation of ETH does not imply
thermalization in the thermodynamic limit for every initial condition, while,
with the proviso that Eq. \eqref{enfluc} holds, the strong interpretation does.

\section{ $\mathbb{Z}_2$ symmetry breaking quench protocol}
\label{Z2H}

The class of Hamiltonians we choose to study can be thought as akin to the quantum Ising
model in the presence of an additional longitudinal field.  The quench protocol involves, for the sake
of specificity, taking $h$ to $-h$.
This quench reflects that the Ising Hamiltonian is not invariant under the $Z_2$ operator, ${\cal P}=e^{i\pi( \SN/2 + S_z)}$:
$$
{\cal P}H(h){\cal P}^\dagger = H(-h).
$$
In order to mimic this in the context of random matrices, we suppose we have divided the canonical basis of 
Ising states into two groupings, even and odd under ${\cal P}$, and to then have sorted them by ordering all even states 
before any odd states.  In Ising, the transverse field term couples states with different parity, such that the 
dependence of the Hamiltonian on the external field is seen in the off-diagonal blocks, i.e.
\begin{equation}
\label{hamiltonianForm}
H(h)=\left(\begin{array}{cc} A &h B\\ h B^T & C \end{array}\right) \,\,\,. 
\end{equation}
It is this form then that we take for our random matrices.

Observables of the systems associated to the Hamiltonian (\ref{hamiltonianForm}) can be split into even and 
odd $\mathbb{Z}_2$ classes. This classification is again motivated by the case of the Ising-spin chain 
in a transverse magnetic field, where the natural observables $\sigma_z$ and $\sigma_x$ are respectively odd and 
even w.r.t. to the action of ${\cal P}$. The even observables have non-zero elements in the diagonal blocks alone, 
while the odd observables are non-zero only in the off-diagonal blocks:
\be
\label{evenodd}
 E = \frac 1 {\SN} \left(\begin{array}{cc} A & 0\\ 0& C \end{array} \right), \qquad 
O = \frac 1 {\SN} \left(\begin{array}{cc} 0 & B\\ B^T& 0 \end{array} \right) \,\,\,, 
\ee
where the volume factor $\SN$ has been added to make these quantities intensive. Since we expect the Hilbert space to be exponentially large in the volume of the system we fix the system size corresponding to an $\HS\times \HS$ random matrix via
$$ \SN = \ln \HS \,\,\,.$$
After defining the Hamiltonian as above, we will analyze the quench dynamics under the quench $h\rightarrow -h$.
We will study the long-time behavior of both odd and even observables. In our numerical analysis, we have examined 
different values of the initial and final value of the parameter $h$, and found that in the limits 
$h \ll 1$ and $h \gg 1$, the quench dynamics are essentially trivial because the initial and final Hamiltonian 
share the same eigenvectors.  For this reason, we will discuss only the intermediate case
\be
\label{quenchProtocol}
 h^{\rm pre-quench} = -1 \quad \rightarrow \quad h^{\rm post-quench} = 1 \,\,\,. 
\ee
Given these constraints we will still however consider two cases:
\begin{itemize}
\item In one case we will look at ensembles of sparse random matrices (SME), motivated by the previous 
considerations about the relationship between locality and sparseness. It should be stressed however that 
while a local observable will necessarily be sparse the converse is not necessarily true.
Nevertheless, the study of SME may provide some reliable insights into some of the 
questions of the thermalization in local Hamiltonians.
\item In the second case we will look at matrices which are dense and follow the GOE.
\end{itemize}
In both cases the (non-zero) entries of the random matrices will be generated according to the 
normal distribution. We first consider quenches involving dense matrices.
\section{Thermalization in Dense Random Matrices Ensemble}
\label{densematrices}
To define the Hamiltonian in the dense case, we generate three $\HS/2 \times \HS/2$ matrices $A,B,C$ 
and then assemble them according to Eq. (\ref{hamiltonianForm}). The matrices $A,C$ are symmetric and 
chosen according to the measure, $\mu(M)$, of a properly normalized GOE ensemble:
$$
\mu(M) \equiv \exp\left(-\frac{\HS \Tr{M^2}}{4 \SN^2} \right)\,\,\,, 
$$
while the matrix $B$ has all of its entries distributed according to a normal distribution 
with 0 mean and variance equal to 
$\frac{2\SN^2}{\HS}$. 
For  $h = \pm 1$  the Hamiltonian itself will also be distributed according to the GOE ensemble and 
therefore the eigenvalues obey the semicircle law:
\be
\label{semicircle}
\rho(E)=\frac{1}{2\pi \SN}\sqrt{4\SN^2-E^2} \,\,\,.
\ee
The spectrum thus falls in the range $[-2\SN,2\SN]$ and is therefore extensive as required. 
The observables are obtained with an analogous procedure, generating new matrices $A,B,C$ 
and then using the expressions Eq. (\ref{evenodd}).
The numerical results reported below are calculated according to the following procedure: five instances of the Hamiltonian are generated according to the prescriptions above  and for each instance of the Hamiltonian forty instances of the observable are generated. The relevant quantities are calculated for each instance of the observables, then the results are averaged.

\subsection{Numerical results}
One of the prerequisites for the ETH to operate is that given the initial state $|\psi_0\rangle$ with energy 
$$
\inie \equiv \frac 1 \SN \langle\psi_0|H_\text{post}|\psi_0\rangle \,, 
$$ 
the structure of its overlaps $|c_{\eig}|^2=|\langle\psi_0|\eig\rangle|^2$ with the post-quench eigenstates, 
as a function of the intensive energy $e = \eig/\SN$, is peaked around $\inie$. 

\begin{figure}[t]
\centering
$\begin{array}{cc}
\includegraphics[width=0.4\textwidth]{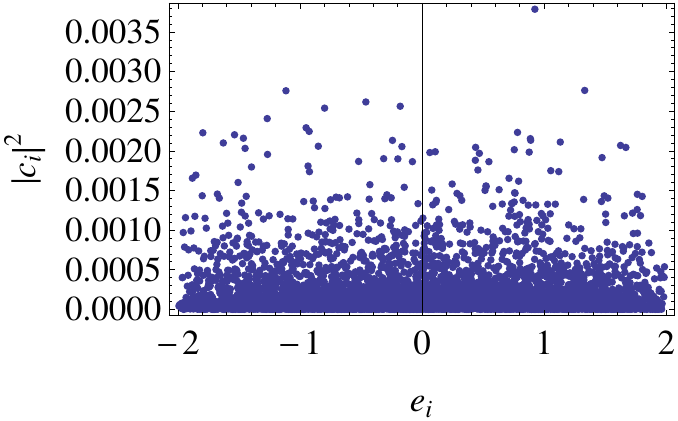} &
\includegraphics[width=0.4\textwidth]{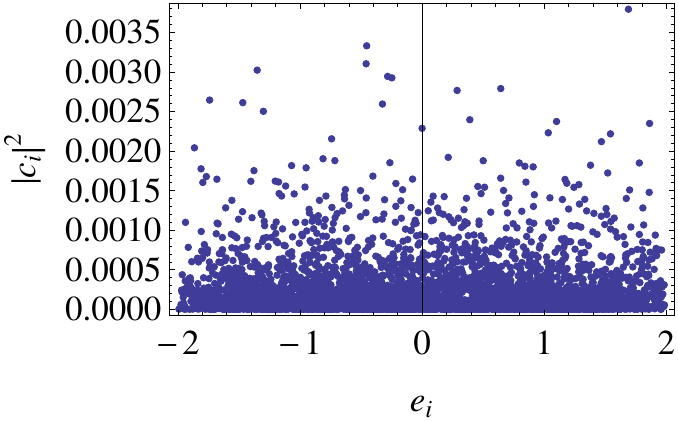}\\
\end{array}$
\caption{Dense random matrices (with  $\HS=4000$). Overlaps for the quench process. Left: 
$|c_{\eig}|^2$ for the 'ground' state.  Right: $|c_{\eig}|^2$ 
for the 2000$^\text{th}$ state, in the middle of the energy band.}
\label{sampleoverdense}
\end{figure}

\begin{figure}[b]
\centering
\includegraphics[width=0.4\textwidth]{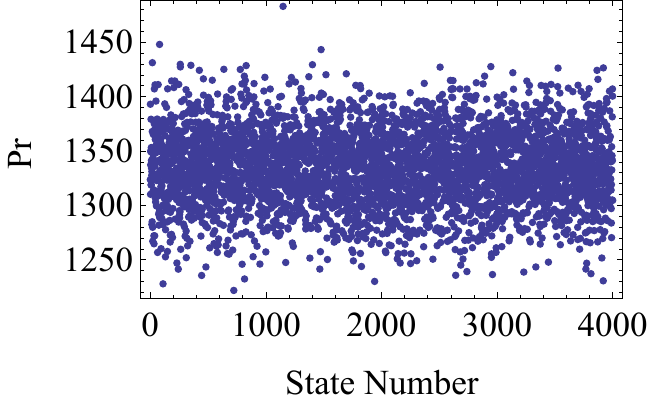} 
\caption{Dense random matrices (with $\HS=4000$). Typical $\PR{2}$ of the initial states. } 
\label{sampleiprdense}
\end{figure}

\begin{figure}[t]
\centering
$\begin{array}{cc}
\includegraphics[width=0.4\textwidth]{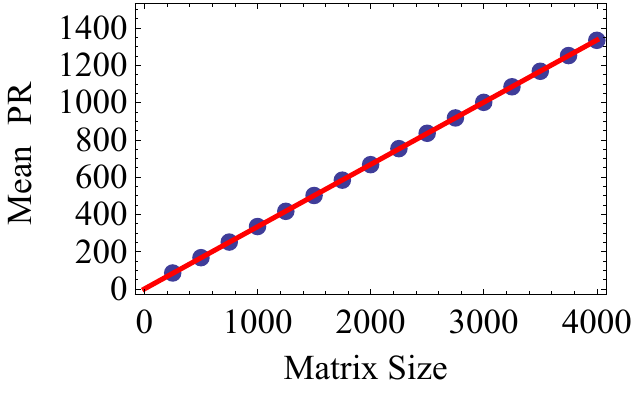} &
\includegraphics[width=0.4\textwidth]{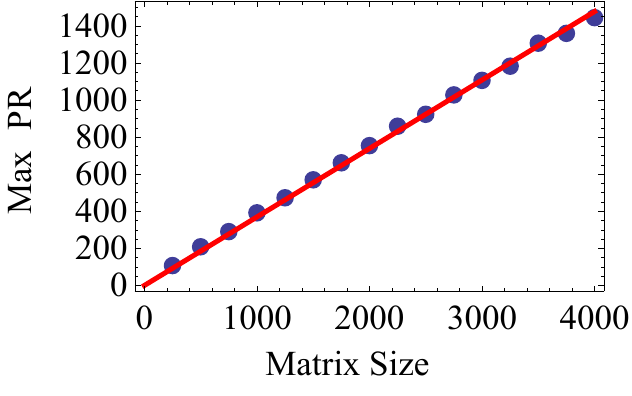}\\
\end{array}$
\caption{Dense random matrices. $\PR{2}$ vs. matrix size. The red lines are linear fits $y=ax$. 
Left panel: average $\PR{2}$ on all initial states ($a=0.3336$). Right panel: maximum $\PR{2}$ ($a=0.3764$). }
\label{iprscalingdense}
\end{figure}

We find this to be not the case for dense matrices, as can be explicitly seen by the two sample states 
in Fig. \ref{sampleoverdense} drawn from the bottom and middle of the
spectrum.
Moreover, calculating the standard deviation of the energy on the initial state, we find that 
it is always large (around 1/4 of the range of the total spectrum) for all initial states, showing that 
the relation Eq. (\ref{enfluc}) does not hold, i.e. the effective dimension $\sigma$ satisfies $\sigma>2$.
The broad distribution of overlaps is confirmed by the analysis of the already introduced Participation Ratio ($\PR{2}$), defined as 
\be
\PR{2} \,=\,\frac{1}{\sum_{\eig} c_{\eig}^4 } \,\,\,.
\ee
We show in Fig. \ref{sampleiprdense} the $\PR{2}$ for the eigenstates of a single realization of a dense matrix.
The $\PR{2}$ in this case is sharply distributed around $\HS/3$. This finding can be understood
through a simple model of random vectors on a $\HS$-sphere of unit radius (Porter-Thomas distribution). By a simple integration one finds \cite{haake2001quantum}
\be
\langle c_{\eig}^4\rangle \,=\, 3/\HS^2 \,\,\,,
\ee  
and therefore the $\PR{2}$ scales as 
\be
\frac{1}{\sum_{\eig} c_{\eig}^4 } \simeq \HS/3 \,\,\,. 
\ee
This scaling is confirmed by our data, as shown in Fig. \ref{iprscalingdense}.
Moreover, the fact that the mean $\PR{2}$ and the maximum $\PR{2}$ almost coincide is 
confirmation that all initial states are equivalent. This means that the pre-quench and the post-quench bases 
of the energy eigenvectors are completely random with respect one another.  Eigenstates therefore 
have no reason to be localized in energy. 

\begin{figure}[b]
\centering
$\begin{array}{cc}
\includegraphics[width=0.4\textwidth]{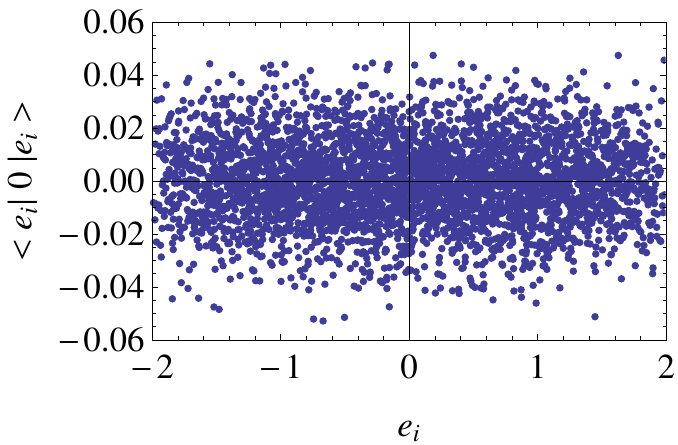} &
\includegraphics[width=0.4\textwidth]{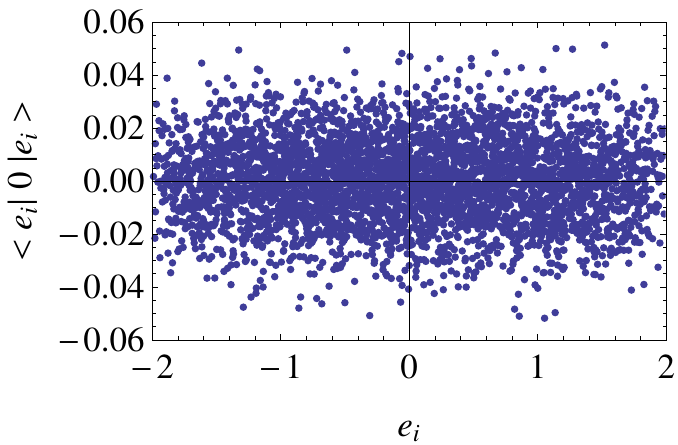}\\
\end{array}$
\caption{Dense random matrices. EEV $\langle \eig|{\obs}|\eig\rangle$ vs
$\eig$. Left: even observable. Right: odd observable.}
\label{EEVsampledense}
\end{figure}

Let us now turn our attention to the expectation values of observables since the main content of the ETH 
concerns the distribution of the eigenstate expectation values (EEVs), $\meanO$, and their behavior 
when the system size is increased. We first report two sample EEV distributions, given in Fig. \ref{EEVsampledense}, 
which show no energy dependence.
We can argue (and we have checked numerically) that the distribution of an intensive observable over the whole energy 
spectrum shrinks to zero for increasing system size. 
Moreover, it is not only the variance but even the support of the distribution of the observables that goes to zero
inasmuch as the difference of the EEV maximum and minimum is going to zero as $\HS \rightarrow \infty$ (see
Fig. \ref{semicircleintensive}).

More precisely, we have:
\be
\label{observableexpansion}
\meanO \equiv \langle \eig | {\obs} | \eig \rangle = \sum_{\eigo}
A_{\eig,\eigo} {\obs}_{\eigo} \,\,\,, 
\ee
where $\eigo$ indexes the eigenstates, $\ket{\eigo}$, of the observable
$\obs$, while ${\obs}_{\eigo}$ is the corresponding eigenvalue, and $A_{\eig,\eigo} = |\langle \eig |
\eigo \rangle|^2$. To estimate the r.h.s. we argue for an equivalence of observables and hold that
the $\PR{2}$ of an eigenvector $|\eig\rangle$ of the post-quench Hamiltonian relative to the basis
of eigenvectors $\ket{\eigo}$ equals the $\PR{2}$ of the initial state $|\psi_0\rangle$ in the basis $|\eig\rangle$.
So we can suppose that $\meanO$ can be expanded in terms of a set of $\frac{\HS}{3}$
states, each of which is given by
\begin{figure}[t]
\centering
$\begin{array}{cc}
\includegraphics[width=0.4\textwidth]{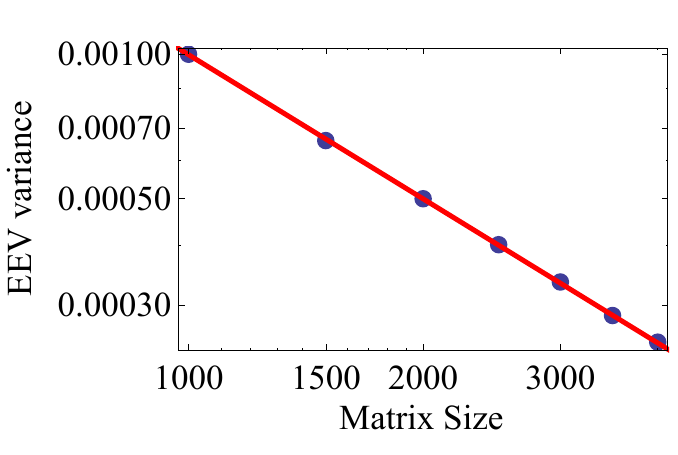} &
\end{array}$
\caption{Dense random matrices. EEV variance (averaged over the entire spectrum) vs. matrix size. The
continuous line is the fit $a/x^b$ with $ b = 0.(9)$. The data points for the even and
odd observables exactly overlap in this plot.}
\label{EEVvariancedense}
\end{figure}
\be
\label{overlapIPR}
 A_{\eig,\eigo} \simeq \frac{1}{\PR{2}(H \to {\obs})} \simeq \frac{3}{\HS} \,\,\,. 
\ee
Then, if we assume $A_{\eig,\eigo}$ and $A_{\eig,\eigo'}$ are independent and note that $\meanO$ has zero mean, we obtain
$$
 \overline{|\meanO|^2} \simeq \frac{\HS}{3} \left(\frac{3 \sigma_{
obs}}{\HS} \right)^2 
 \simeq \frac{3 \sigma_{\obs}^2}{\HS} ,
$$
where $\sigma_{\obs}^2$ is the variance of the spectrum of the observable 
${\obs}$.  In Fig. \ref{EEVvariancedense}, the numerical results are
plotted together with a power-law fit and, 
as expected, the exponent is indeed close to one.

\begin{figure}[b]
\centering
$\begin{array}{cc}
\includegraphics[width=0.4\textwidth]{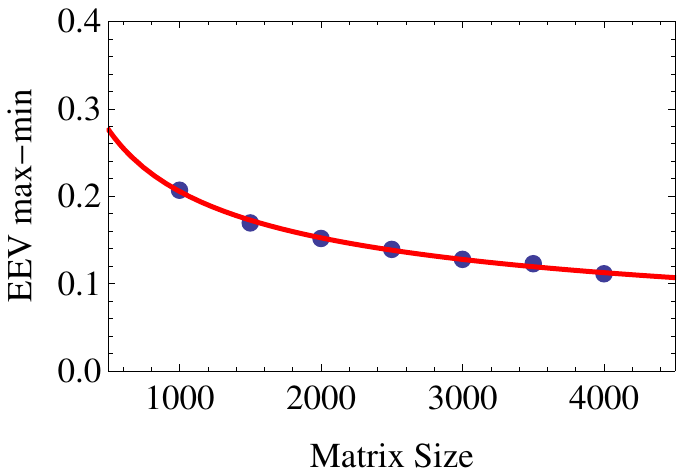} &
\end{array}$
\caption{Dense random matrices. Max-min EEV vs. matrix size for the range of even observables.  
The continuous line is the fit $a\sqrt{\frac{\ln{\HS}}{\HS}}$.}
\label{minmaxdense}
\end{figure}

Now let us consider the full support of the distribution of the EEVs where we define
$\delta_{\obs}$ as the difference of the maximum and the minimum of the EEVs 
among all the energy eigenstates $|\eig\rangle$. Since the distribution is symmetric about zero, we have:
\be
\label{maxminspread}
\delta_{\obs} \,=\, 2\max_{\eig} \{ \meanO \} \,. 
\ee
To estimate the scaling of this quantity, we again approximate all the overlaps $A_{\eig,\eigo}$ 
as in Eq. (\ref{overlapIPR}). Therefore we are led to estimate the maximum of the quantity
$$ 
\meanO \equiv \frac{3}{\HS} \sum_{\eigo}' {\obs}_{\eigo} ,
$$
where the prime on the sum indicates that only $1/3$ of the total $\eigo$'s are being summed over.
\begin{figure}[t]
\centering
$\begin{array}{cc}
\includegraphics[width=0.4\textwidth]{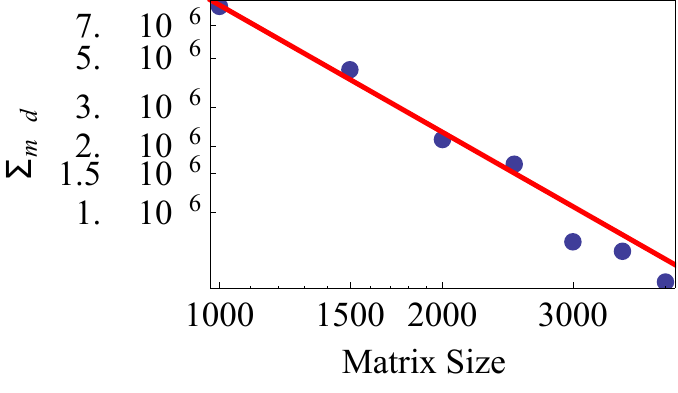} &
\includegraphics[width=0.4\textwidth]{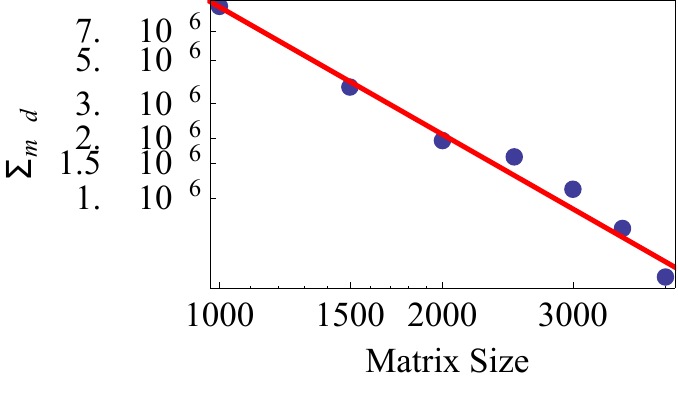}\\
\end{array}$
\caption{Dense random matrices. $\sigma=\sqrt{({\obs}_\text{micro} -
{\obs}_\text{diag})^2}$ vs. matrix size for initial states laying in the central part of the 
spectrum $\overline{e}\approx 0$. The continuous line is the fit $a/x^b$. Left: even observable $b=1.(9)$; 
right: odd observable $b=2.(1)$}
\label{centerscalingdense}
\end{figure}
As before we suppose that the random variables ${\obs}_{\eigo}$ are independently distributed 
according to the intensive semicircle law (${\obs}_{\eigo} \in [-2,2]$):
\be
\label{semicircleintensive}
\rho(x) \equiv \operatorname{Prob}({\obs}_{\eigo} = x) \,=\, 
\frac{1}{2\pi} \sqrt{4 - x^2} \,\,\,. 
\ee
In this case, from the central limit theorem it follows
\be
\label{largedeviation}
 \Prob(\meanO > x) = \frac{1}{2}\operatorname{Erfc}\left(\sqrt\frac{\HS}{6} x\right) \simeq e^{- \frac{\HS x^2}{6}} \,\,\,, 
\ee
Now the probability that $\delta_{\obs}$ is 
less than a value $M$ is given by 
$$ 
\Prob(\delta_{\obs} < M) = \prod_{E = 1}^\HS \Prob( \meanO < M) 
\simeq \left(1 - e^{- \frac{\HS M^2}{6}} \right)^\HS \simeq 1 - \HS e^{- \frac{\HS M^2}{6}}  \,\,\,. 
$$
We can find the scaling of the typical value of the maximum by requiring that the probability 
is large enough 
\be
\label{maxminscaling}
\Prob(\delta_{\obs} < M) \simeq \text{const.}  \quad \Rightarrow \quad \delta_{\obs} \,\,\,
 \simeq \sqrt\frac{\ln \HS}{\HS} \,\,\,. 
\ee
In Fig. \ref{minmaxdense}, one sees the numerical agreement with our heuristic
argument.

Finally let's consider the behavior of the difference between the diagonal and the microcanonical ensembles
with increasing matrix size.  To this end, we analyzed the difference
$\sigma = \sqrt{({\obs}_\text{micro} - {\obs}_\text{diag})^2}$ of the two ensembles for
each initial state $|\psi_0\rangle$.
Due to the broad energy distribution of the overlap of the initial states, their
intensive energy $\overline{e}=\langle \psi_0|H(-h)|\psi_0\rangle$ lies in a 
region, $[-0.5,0.5]$, smaller than the range of the post-quench energies $-2<e_{\eig}<2$.  $\sigma$
show the same behavior independent of the particular initial state $|\psi_0\rangle$
being considered, as can be argued by the constant $\PR{2}$ combined with a structure-less 
EEV distribution.
As all initial states are equivalent, we focus our attention 
on those initial states belonging to a small energy window around $e=0$: 
the result is shown in Fig. \ref{centerscalingdense}. As can be seen, 
the difference between micro-canonical and diagonal ensemble rapidly goes to zero as a function of $\HS$.

In conclusion, quenches in dense random matrices are characterized by initial states with
large PRs, EEV distributions of the post-quench eigenbasis with no
energy dependence and whose variance goes to 
zero exponentially with increasing system size. In this sense, their thermalization is trivial, 
as the spread of the micro-diagonal ensemble is governed by a distribution whose
support is increasingly localized near zero as system size grows.

\section{Thermalization in Sparse Random Matrices}
\label{sparsematrices}

\begin{figure}[t]
\centering
$\begin{array}{cc}
\includegraphics[width=0.4\textwidth]{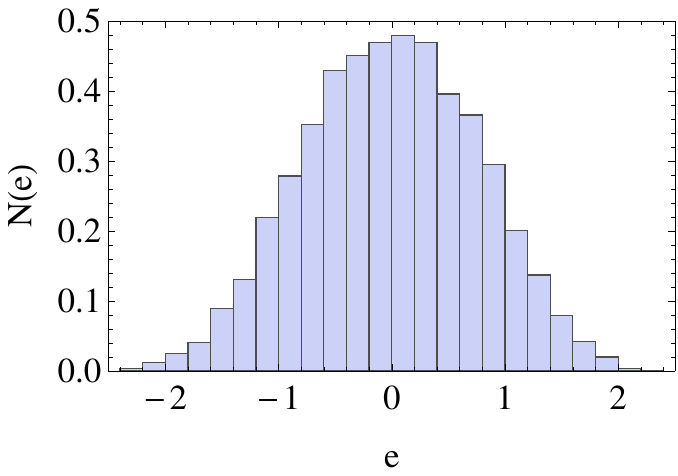} &
\includegraphics[width=0.4\textwidth]{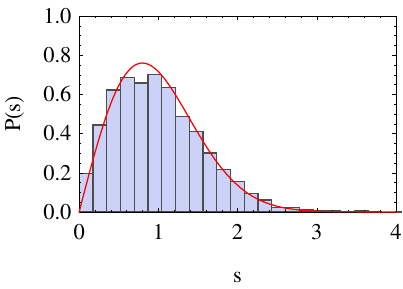}\\
\end{array}$
\caption{Sparse random matrices (with $\HS$=4000). Left: density of states, right:
level spacing statistics, the continuous line is the Wigner surmise for GOE
matrices.}
\label{dosspacingsparse}
\end{figure}

\begin{figure}[t]
\centering
$\begin{array}{cc}
\includegraphics[width=0.4\textwidth]{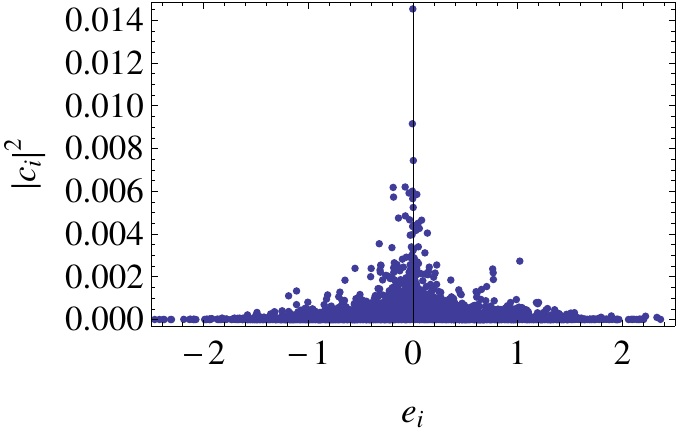}&
\includegraphics[width=0.4\textwidth]{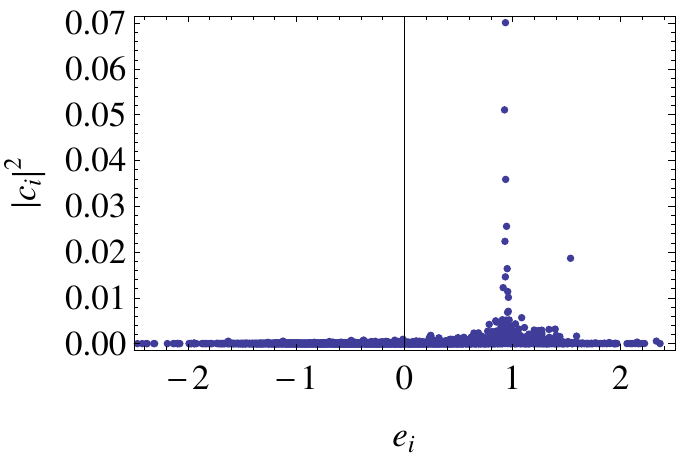}\\
\end{array}$
\caption{Sparse random matrices. Behavior of the overlaps $|c_{\eig}|^2$ for a state mid-spectrum (left) and for one 
in the upper portion of the spectrum (right).}
\label{sampleoversparse}
\end{figure}

\begin{figure}[b]
\centering
\includegraphics[width=0.4\textwidth]{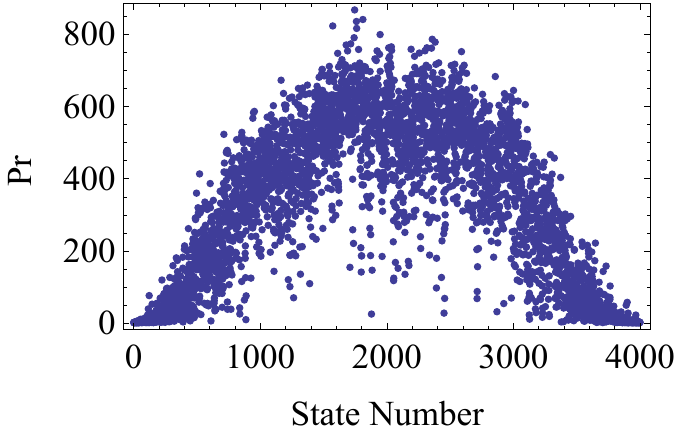} 
\caption{Sparse random matrices (with $\HS=4000$). The $\PR{2}$ for a specific realization.
The ordering of the initial states in this plot is according to their energy relative to the
post-quench Hamiltonian, $\langle \psi_0|H^>|\psi_0\rangle$.} 
\label{sampleiprsparse}
\end{figure}
\begin{figure}[t]
\centering
$\begin{array}{cc}
\includegraphics[width=0.4\textwidth]{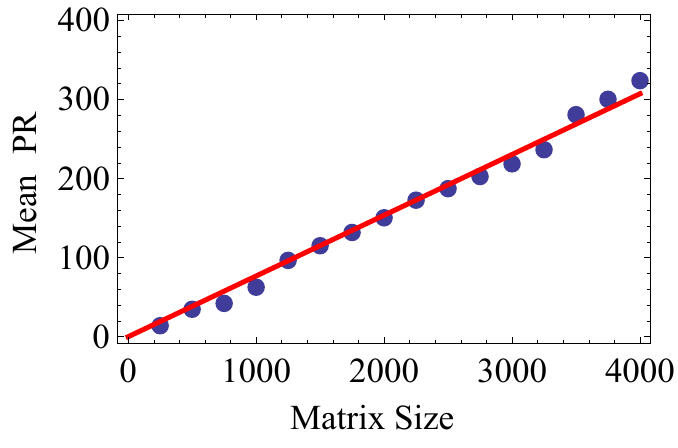} &
\includegraphics[width=0.4\textwidth]{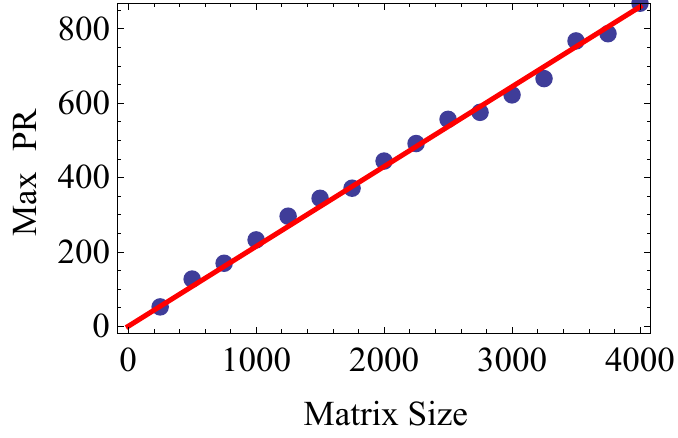}\\
\end{array}$
\caption{Sparse random matrices. $\PR{2}$ vs. matrix size. The continuous line is the $y=ax$ fit. Left, average $\PR{2}$ on the whole spectrum ($a=0.0738$). Right, maximum $\PR{2}$ ($a=0.2096$). }
\label{iprscalingsparse}
\end{figure}

We now turn to the more interesting case of thermalization in SME.  As we have indicated
such matrices describe the Hamiltonians of systems with local interactions.
In order to define the ensemble of these matrices, we employed a symmetric mask matrix, $\mathcal{M}$.
This matrix has $1$'s on the diagonal and in each of its rows 
we allow it to have on average $\ln \HS$ off-diagonal entries equal to $1$.  All
remaining entries 
of the mask matrix are 
equal to zero.  
The upper triangular part of the (symmetric) Hamiltonian is then obtained as:
\be
\label{hamiltonianScaling}
H(h)_{i<j} = 
\left\{ \begin{array}{ll}
d_i ~{\rm with~}d_{i}{\rm~drawn~from~}\mathcal{\HS}(0,\ln \HS) & {\rm if}~i=j;\\
o_{ij}\times\mathcal{M}_{ij} ~{\rm with~}o_{ij}{\rm~drawn~from~}\mathcal{\HS}(0,1) & {\rm if}~i<j,
\end{array}\right.
\ee
where $\mathcal{\HS}(\mu,\sigma^2 )$ is a normal distribution with $\mu$ mean and $\sigma^2$ variance.
Then the coefficients in the off-diagonal blocks are multiplied times $h$ to reproduce the structure in Eq. (\ref{hamiltonianForm}).
The different choice for the variances of the diagonal $d_i$ and off-diagonal elements $o_{ij}$ 
is motivated by the requirement that the spectrum be extensive. 
\begin{figure}[b]
\centering
\includegraphics[width=0.4\textwidth]{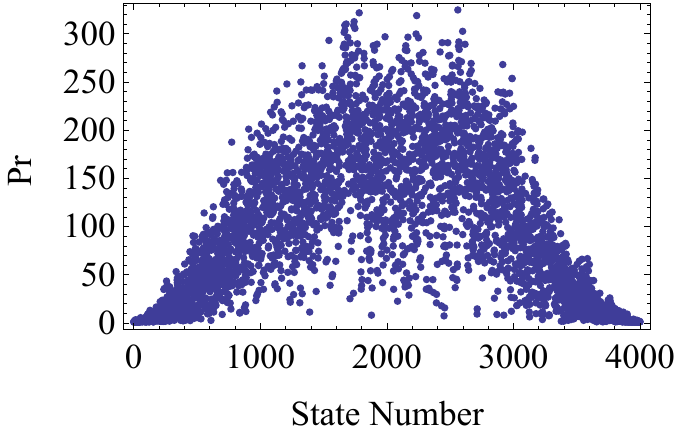} 
\caption{Sparse random matrices (with $\HS=4000$). The $\PR{2}$  of the post-quench eigenstates w.r.t. the local basis for a specific realization.
The ordering of the initial states in this plot is according to their energy eigenvalue.} 
\label{sampleiprsparsewrtlocal}
\end{figure}
\begin{figure}[t]
\centering
$\begin{array}{cc}
\includegraphics[width=0.4\textwidth]{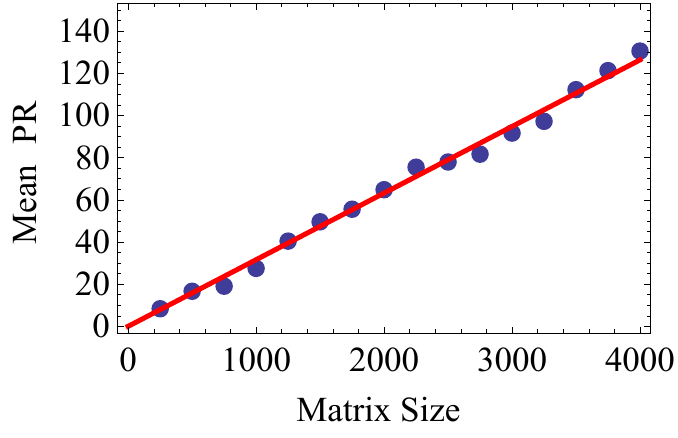} &
\includegraphics[width=0.4\textwidth]{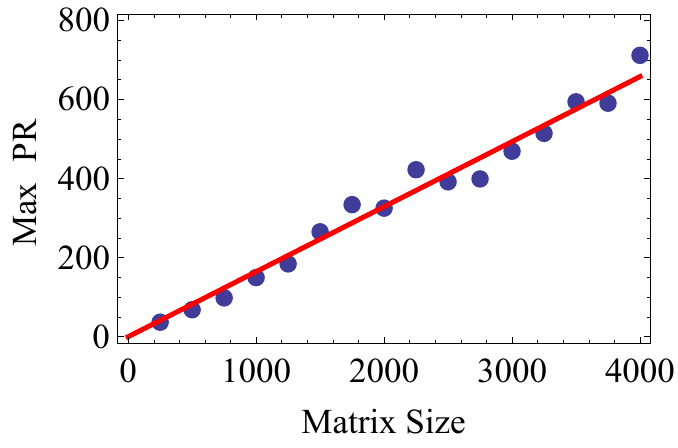}\\
\end{array}$
\caption{Sparse random matrices. $\PR{2}$ w.r.t. local basis vs. matrix size. The continuous line is the $y=ax$ fit. Left, average $\PR{2}$ on the whole spectrum ($a=0.0321$). Right, maximum $\PR{2}$ ($a=0.1632$). }
\label{iprscalingsparsewrtlocal}
\end{figure}
In fact we can
compute the variance of the spectrum:
\be
\label{sparsevariance}
\sigma_H^2 \equiv \frac{\Tr {H^2}}{\HS} \simeq 2 \ln \HS .
\ee
In the approximation in which the eigenvalues $\eig$ are considered independent
and normally distributed, one can relate the
minimum of the spectrum with the variance, obtaining for the ground-state energy
the estimate
$$
E_\text{gs} = \min_{\eig} \{\eig \} \simeq -\sigma_H \sqrt{2 \ln \HS} \simeq  - 2 \ln \HS .
$$
The eigenvalues mid-spectrum will be at most a few standard deviations $\sigma_H$ from the average $0$. 
Thus a typical eigenstate will be such that
\be
\eig^{(0)}-E_\text{gs}^{(0)} \propto \ln{\HS} \,\,\,.
\ee
We see then that our choices satisfy the requirement that energy is an extensive quantity.

We compare these estimates with numerics in 
Fig. \ref{dosspacingsparse} where we plot the density of states and the level
spacing distribution
for one realization of a sparse matrix. 
The density of states is no longer a semicircle, looking rather more like a bell-shaped distribution. 
Moreover, 
notice that, unlike the GOE case where the intensive quantities have a finite distribution 
in the large $\HS$ limit, i.e.Eq. (\ref{semicircleintensive}), 
in this sparse case the (intensive) standard deviation
behaves as $\frac{1}{\sqrt{\ln \HS }}$, while 
the support of the spectrum remains approximately $[-2,2]$.
We also see from the right side of Fig. \ref{dosspacingsparse} that the
level-spacing distribution obeys the Wigner form for 
a non-integrable model.

To generate the observables we follow a similar procedure, i.e. we employ the same mask $\mathcal{M}$. 
The idea behind this choice is that the matrix $\mathcal{M}$ is responsible for the local structure 
in the Hilbert space and so we keep it for all the physical observables (including the Hamiltonian itself). 
So we have for the upper triangular part of a symmetric observable 
$$ 
\mathcal{O}_{i<j} = \left\{ \begin{array}{ll}
                                    d_i {\rm ~with~} d_i {\rm~drawn~from~}
\mathcal{\HS}(0,1/\ln \HS) & i = j;\\
                                    o_{ij}\times M_{ij} {\rm ~with~} o_{ij}
{\rm~drawn~from~} \mathcal{\HS}(0,1/(\ln \HS)^2) & i\neq j.
                                   \end{array}\right.
$$
The even and odd parts are then obtained as before 
by splitting $\mathcal{O}$ into diagonal and off-diagonal blocks.

\begin{figure}[b]
\centering
$\begin{array}{cc}
\includegraphics[width=0.4\textwidth]{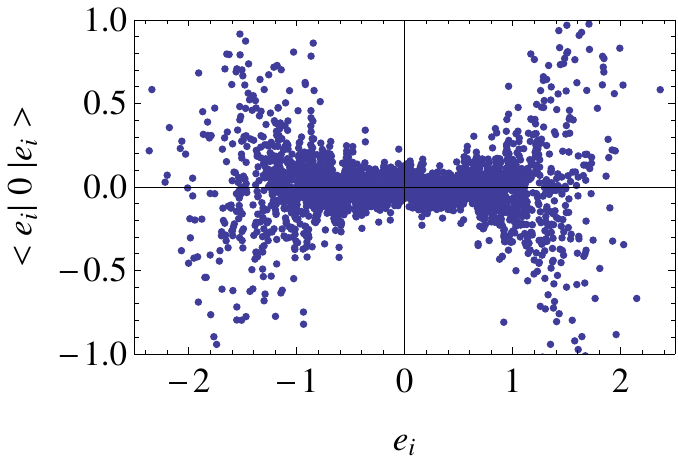} &
\includegraphics[width=0.4\textwidth]{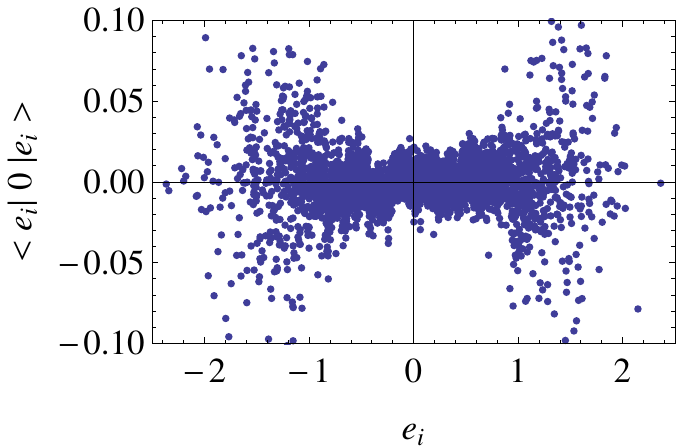}\\
\end{array}$
\caption{Sparse random matrices. EEV $\langle \eig|{\obs}|\eig\rangle$ vs
$e_{\eig}$. Left: even observable. Right: odd observable}
\label{EEVsamplesparse}
\end{figure}
\subsection{Numerical results}
Unlike the dense case, in sparse random matrices the overlap distributions are peaked around the 
initial state post-quench intensive energy $\overline{e}$, as can be seen in two examples
shown in Fig. \ref{sampleoversparse}.

The $\PR{2}$ is no longer constant (as it was for the random dense matrices), but shows behavior 
dependent on the energy of the initial state,  as demonstrated in
Fig. \ref{sampleiprsparse}.
However there is still scaling of the $\PR{2}$ with the matrix size, as can be seen by
studying the behavior of the maximum $\PR{2}$ vs the matrix size, plotted in 
Fig. \ref{iprscalingsparse}.  In this case the mean and the maximum $\PR{2}$ are
rather different, 
due to the presence of states with very small $\PR{2}$.

We see a similar phenomena when we study the $\PR{2}$ of the pre-quench and
post-quench 
(they are statistically equivalent) eigenbasis relative
to the local basis (that is, the basis of states in which the Hamiltonian matrices, $H(h^<)$ and $H(h^>)$
are expressed).  We see in Fig.\ref{iprscalingsparsewrtlocal} that the mean and
max $\PR{2}$'s here as a
function of matrix
size, $\HS$, are similar to those in Fig. \ref{iprscalingsparse}. 

\begin{figure}[t]
\centering
$\begin{array}{cc}
\includegraphics[width=0.4\textwidth]{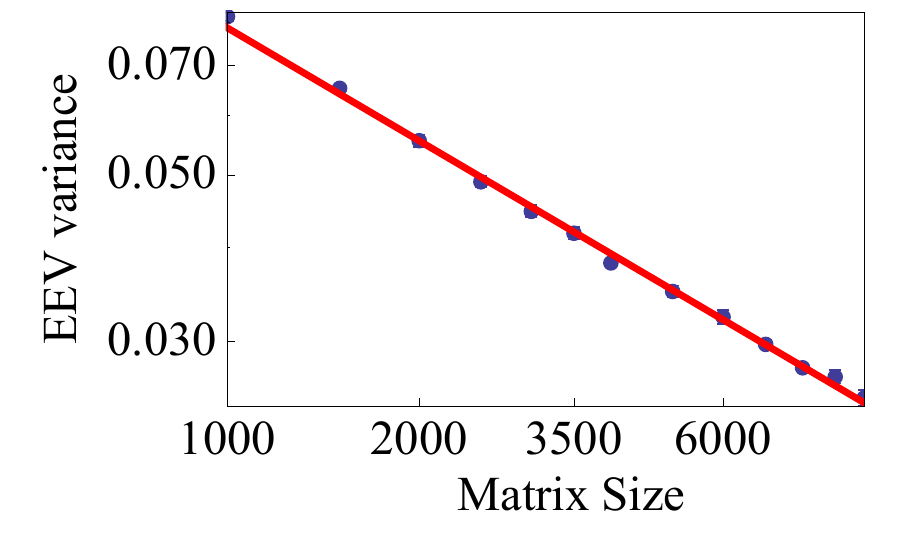} &
\includegraphics[width=0.4\textwidth]{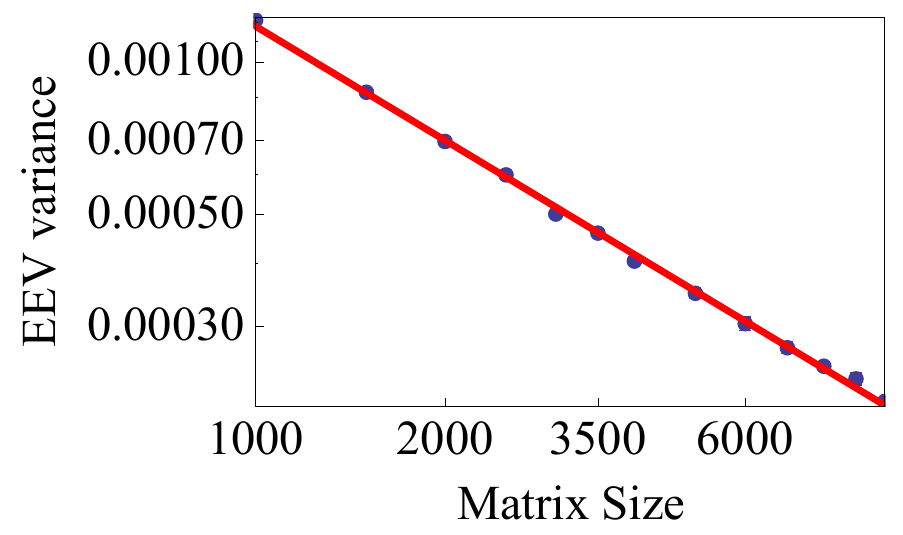}\\
\end{array}$
\caption{Sparse random matrices. EEV variance vs. matrix size for the full spectrum. The continuous lines are
the fit $a/x^b$. Left: even observable $b=0.5(0)$ Right: odd observable $b=0.7(5)$. }
\label{EEVvariancesparse}
\end{figure}
\begin{figure}[b]
\centering
$\begin{array}{cc}
\includegraphics[width=0.4\textwidth]{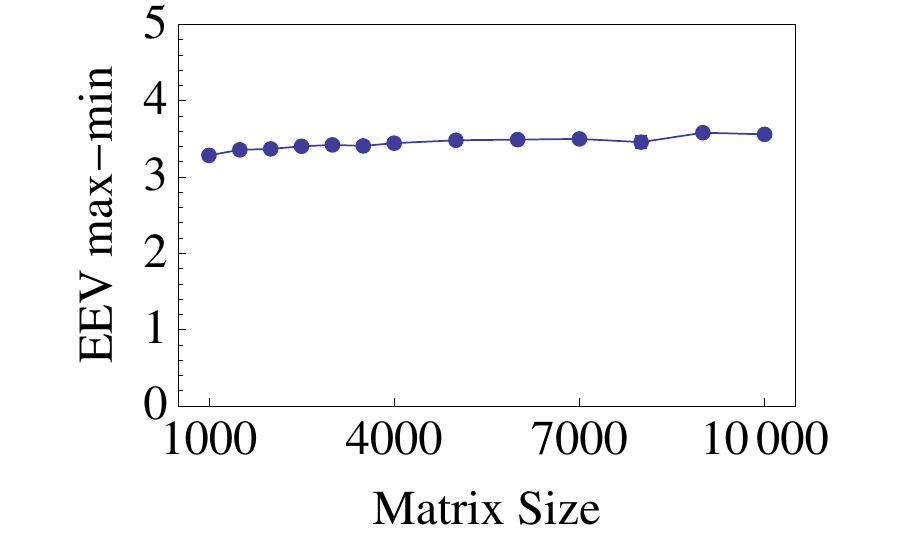} &
\includegraphics[width=0.4\textwidth]{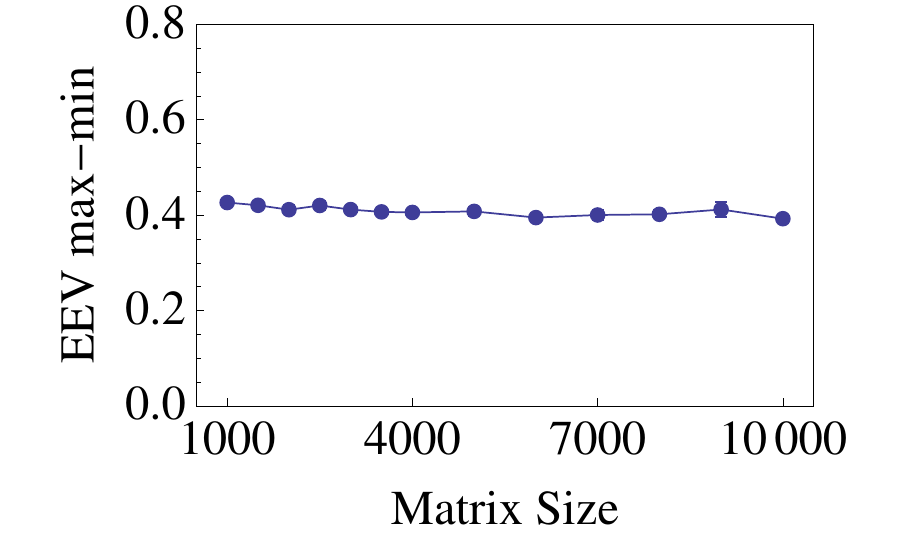}\\
\end{array}$
\caption{Sparse random matrices. Max-min EEV vs matrix size for the full spectrum. Left: even observable.
Right: odd observable.}
\label{minmaxsparse}
\end{figure} 

The equivalence of these two different PRs is not surprising.
If we pick an eigenstate of the pre-quench Hamiltonian, $|\psi_0\rangle$, that has a small $\PR{2}$ relative to the
local basis, it will necessarily be only weakly coupled to the off-diagonal blocks, particularly as the Hamiltonian
matrices are sparse.  $|\psi_0\rangle$ will then be closely related to some post-quench eigenstate.
This implies in turn that $|\psi_0\rangle$ will have a small $\PR{2}$ relative to the post-quench eigenbasis.
Similarly a pre-quench eigenstate $|\psi_0\rangle$ with a large $\PR{2}$ in terms of the local basis will be strongly
affected by the quench in the sense that it is unrelated to any eigenstate of the post-quench eigenbasis,
and so will have a large $\PR{2}$ in terms of this basis.  It is this that lies behind the similar shapes
of Fig. \ref{sampleiprsparsewrtlocal} and Fig. \ref{sampleiprsparse}.

The behavior of the $\PR{2}$ relative to the local basis (and, by this equivalence, the $\PR{2}$ relative to the post-quench eigenbasis)
can be understood directly in the framework of AL.  Our sparse Hamiltonian can be seen as akin to the dynamics of a non-interacting particle hopping
on a Bethe-lattice of fixed 
connectivity, $\ln \HS$, that we introduced in \ref{bethelattice}. Each site has a random potential (the diagonal part of the random Hamiltonian).
We already pointed out that for this model the Anderson localization transition occurs 
with the presence of a \textit{mobility edge} which separates the delocalized states (in the middle 
of the band) from the localized states (in the tails of the energy spectrum).
In our case localization occurs 
when the $\PR{2}$ is $O(1)$, while delocalization is seen for $\PR{2} = O(\HS)$.

\begin{figure}[t]
\centering
$\begin{array}{cc}
\includegraphics[width=0.4\textwidth]{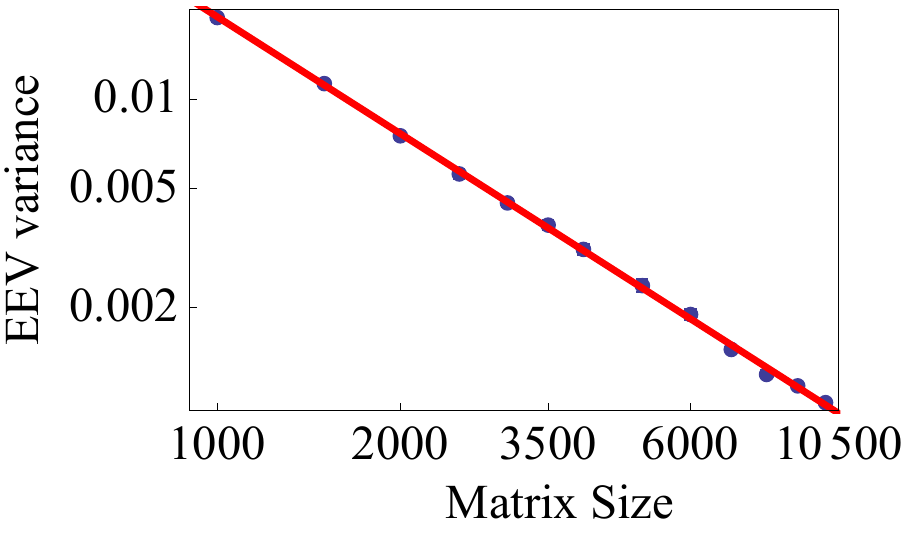} &
\includegraphics[width=0.4\textwidth]{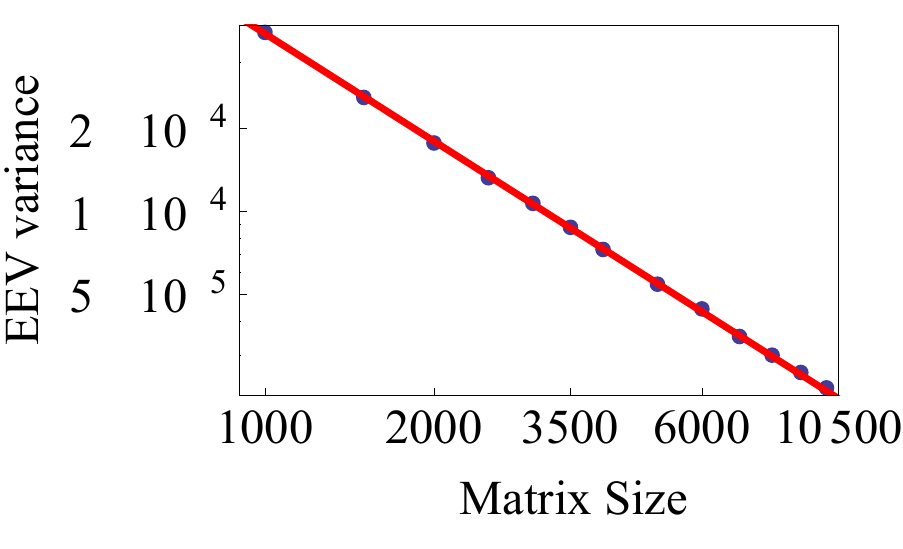}\\
\end{array}$
\caption{Sparse random matrices. EEV variance vs. matrix size for a small energy window around $e=0$. 
The continuous line is the fit $a/x^b$. Left: even observable $b=1.2(9)$.  Right: odd observable $b=1.(2)$.}
\label{EEVvariancesparsecenter}
\end{figure}
\begin{figure}[b]
\centering
$\begin{array}{cc}
\includegraphics[width=0.4\textwidth]{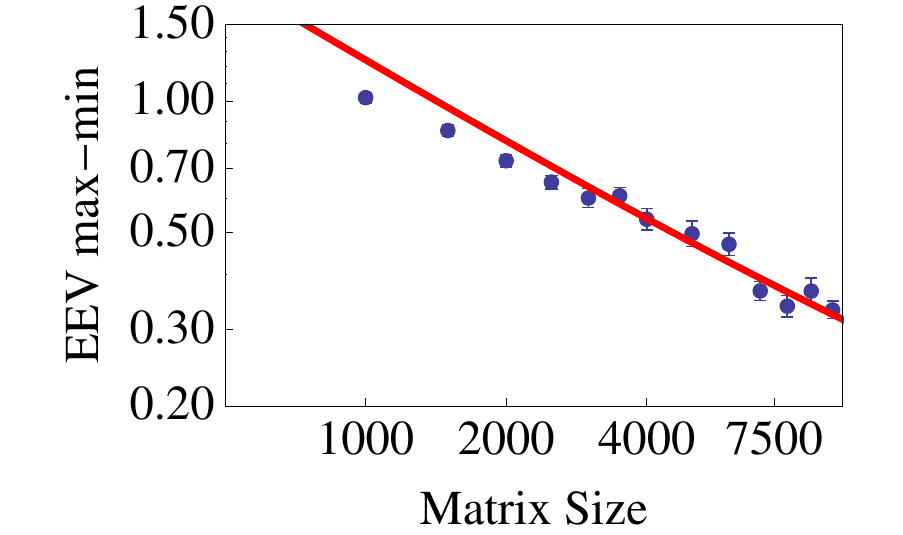} &
\includegraphics[width=0.4\textwidth]{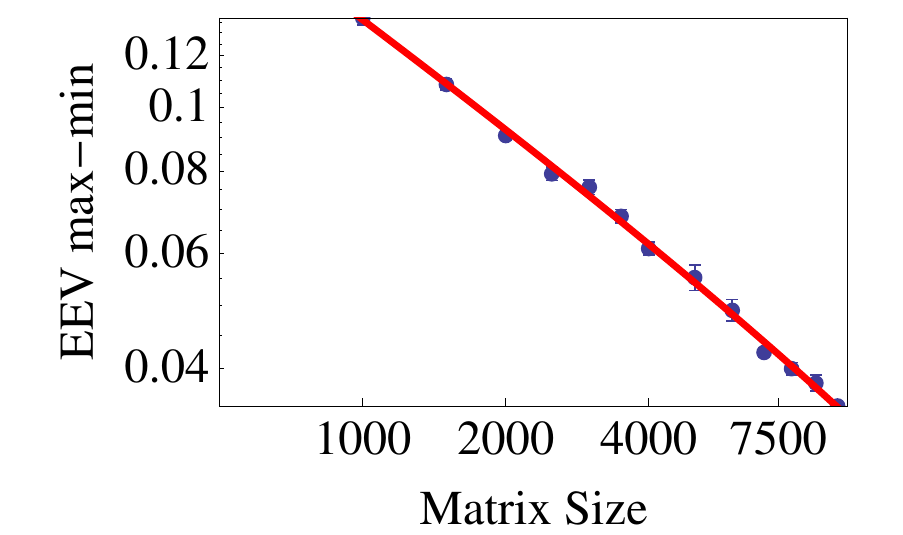}\\
\end{array}$
\caption{Sparse random matrices. Max-min EEV vs. matrix size  for a small energy window around $e=0$ in log-scale. The continuous line is the fit $a/x^b+c$. Left: even observable $e=0$, $b=0.6(5)$, $c=0.06(7)$. Right: odd observable $e=0$, $b=0.5(0)$, $c=0.0(1)$.}
\label{minmaxsparsecenter}
\end{figure}

\begin{figure}[b]
\centering
$\begin{array}{cc}
\includegraphics[width=0.4\textwidth]{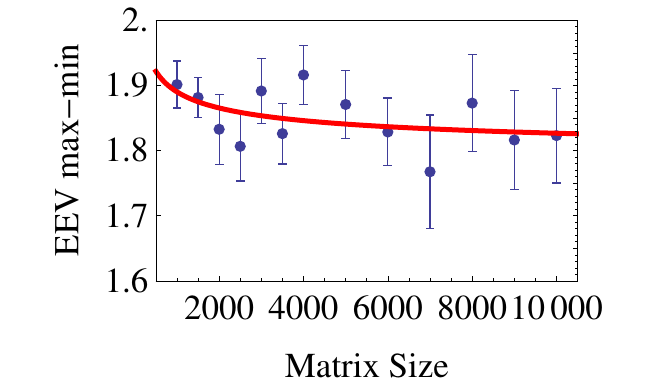} &
\includegraphics[width=0.4\textwidth]{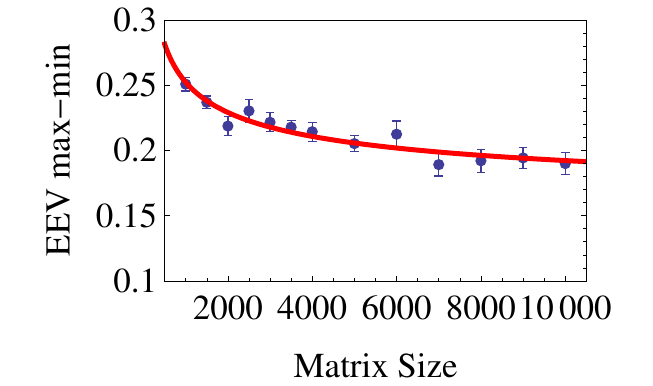}\\
\end{array}$
\caption{Sparse random matrices. Max-min EEV vs. matrix size  for a small energy window around $e=1$. The continuous line is the fit $a/x^b+c$. Left: even observable $e=1$, $b=0.3(5)$, $c=1.(7)$. Right: odd observable $e=1$, $b=0.3(3)$, $c=0.1(4)$.}
\label{minmaxsparseone}
\end{figure}

The position, $E_m$, of the mobility edge can be 
determined by the following equation, derived along the same lines as \cite{abou1973selfconsistent}
\be
\label{aboumobility}
2\ln \HS \int_0^\infty \{ p'(x-E_m) - p'(E_m-x) \} \ln x \,dx= 1 \,\,\,, 
\ee
where $p(x)$ is the probability density of the diagonal terms in our Hamiltonian
\be
\label{gaussianDistr}
 p(x) = \frac{1}{\sqrt{2\pi \ln \HS}} \exp \left(- \frac{x^2}{2 \ln \HS} \right) \,\,\,. 
\ee
States with an $|E| > |E_m|$ such that the right hand side of Eq. (\ref{aboumobility}) is less than $1$ 
are localized.  Otherwise they are delocalized.
The integral in Eq. (\ref{aboumobility}) can be estimated at the leading order in the large $\HS$ limit, and
one obtains
\be
\label{mobility}
1 =  \exp\left(-\frac{E_m^2}{2\ln \HS}\right) \ln\ln \HS \sqrt{\frac{2\ln \HS}{\pi}}
\quad\Rightarrow \quad E_m \simeq \pm \sqrt{\ln \HS \ln \ln \HS} \,\,\,. 
\ee
Thus in the large $\HS$ limit all the states with a non-zero intensive energy $E/\ln \HS$ behave 
as localized. Nonetheless the majority of the states, as they are concentrated in a window
of width $\sigma_H$ (given in Eq. \ref{sparsevariance}) and as  $E_m/\sigma_H \gg 1$, will be delocalized with an $\PR{2} = O(\HS)$.


\begin{figure}[t]
\centering
$\begin{array}{cc}
\includegraphics[width=0.4\textwidth]{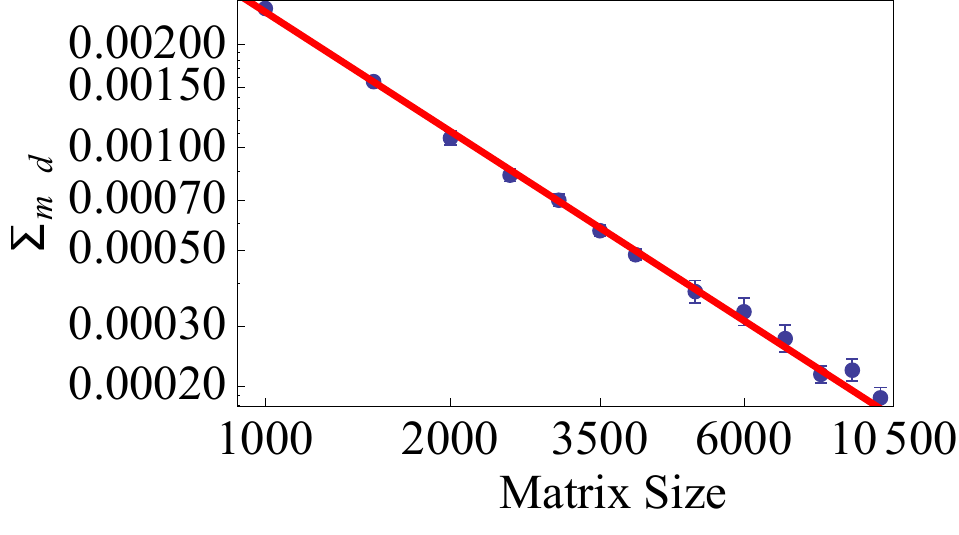} &
\includegraphics[width=0.4\textwidth]{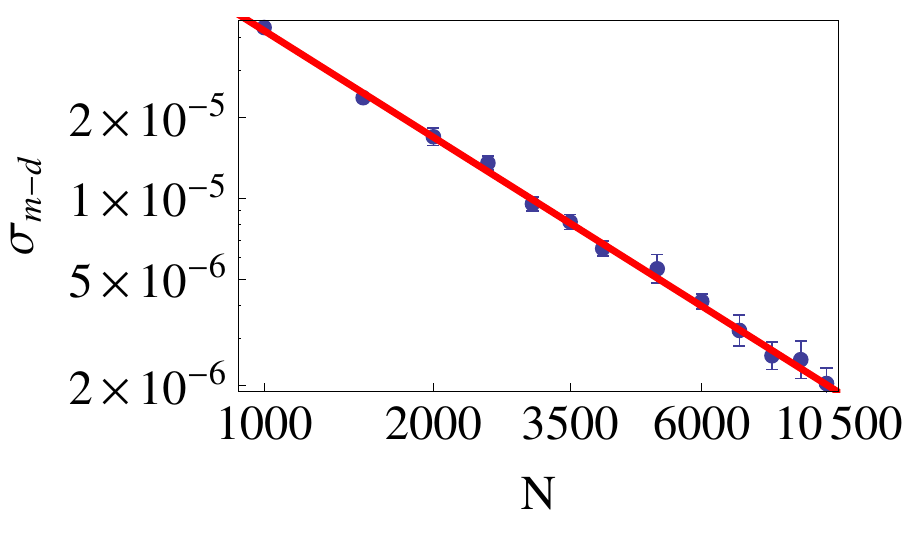}\\
\end{array}$
\caption{Sparse random matrices. $\sigma=\sqrt{({\obs}_\text{micro} - 
{\obs}_\text{diag})^2}$ vs. matrix size for initial states laying in the central part of the 
spectrum $\overline{e}\approx 0$. The continuous lines are the fits $a/x^b$. Left: even observable $b=1.1(5)$.
Right: odd observable $b=1.3(0)$.}
\label{centerscalingsparse}
\end{figure}

\begin{figure}[t]
\centering
$\begin{array}{cc}
\includegraphics[width=0.4\textwidth]{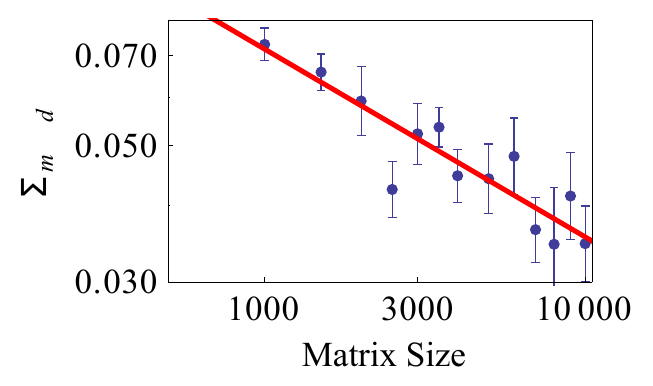} &
\includegraphics[width=0.4\textwidth]{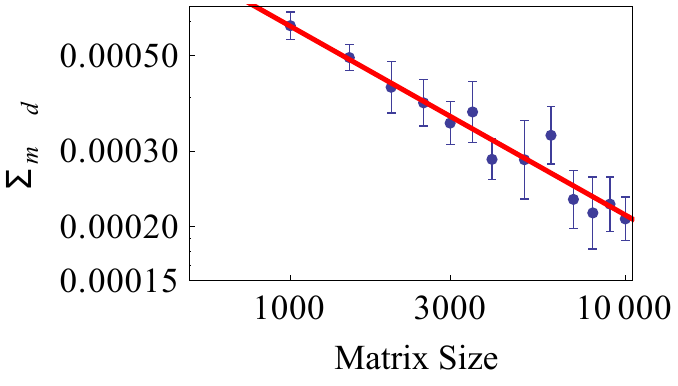}\\
\end{array}$
\caption{Sparse random matrices. $\sigma=\sqrt{({\obs}_\text{micro} - 
{\obs}_\text{diag})^2}$ vs. matrix size for initial states laying in a small window around $e=1$. The continuous lines are the fits $a/x^b$. Left: even observable $b=0.3(1)$.
Right: odd observable $b=0.4(4)$.}
\label{centerscalingsparsee1}
\end{figure}

\begin{figure}[b]
\centering
\includegraphics[width=0.4\textwidth]{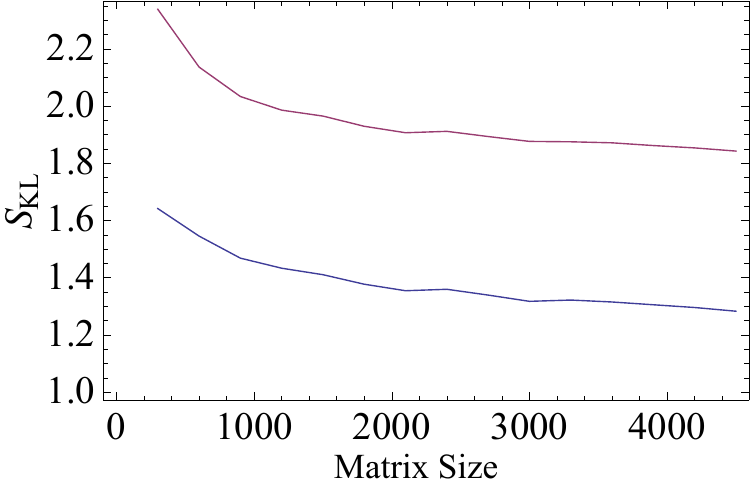}
\caption{Sparse random matrices. Kullback-Leibler entropy vs matrix size for the center of the 
spectrum for the uniform (squares) and the Gaussian distribution (circles). }
\label{KLscaling}
\end{figure}
The localization of states with finite intensive energies has implications for the EEV distribution
vs. $e_{\eig}$.  As the observables have a matrix structure closely related to that of the Hamiltonians,
we expect that localized states to be close to eigenstates of the observables itself, in contrast 
with Eq. (\ref{overlapIPR}) in the dense case.
On such states, the observables will have expectation values far from their zero
average.  On the contrary, the delocalized states mid-spectrum will have EEV values closer to the mean of zero.
In Fig. \ref{EEVsamplesparse} we see numerical verification of this.

This result marks a strong difference with respect to the case of dense matrices.  We also see 
marked differences between the sparse and dense cases with both the variance and the support of 
the EEVs distribution of the observable, 
as shown in Fig. \ref{EEVvariancesparse} and \ref{minmaxsparse}: while the variance 
approaches zero as $\HS$ grows, the support does not, instead tending to a non-zero constant. 
Therefore the scaling Eq. (\ref{maxminscaling}) is no longer applicable, most likely
as the overlaps, $A_{i\beta}$,
and the eigenvalues of the observable, $O_{\eigo}$, in Eq. (\ref{observableexpansion}) can no longer
be considered as independent.

Since the distribution of the overlap coefficients, $c_{\eig}$, are peaked around the energy 
$\overline{e}=\langle \psi_0 |H(h)|\psi_0\rangle$, it is worthwhile
to analyze the scaling
behavior of the distribution of the EEVs in the vicinity of a specific $\bar e$. 
Here we choose two different energy windows, one centered around $\bar e = 0$, lying exactly mid-spectrum and one around $\bar e = 1$. For  $\bar e = 0$ 
the variance and the max-min spread
as functions of the size $\HS$ are plotted in Fig. (\ref{EEVvariancesparsecenter}) 
and Fig. (\ref{minmaxsparsecenter}).  It is this distribution
that is going to determine whether with respect to a particular observable we see thermalization.
We again see that the variance is going to zero, 
while in contrast to the full spectrum, the max-min spread seems to tend, asymptotically, 
to either a very small constant value or to zero.  The errors in our numerics are then not small enough to tell us at $\bar e = 0$
whether there is a complete absence of rare states.
However we can be more definitive for the energy window centered at $\bar e = 1$.
In this window we see (as evidenced in Fig. \ref{minmaxsparseone})
that the max-min spread of the EEV's in the large $\HS$ limit tends to a finite constant for both the even and 
odd observables.

Our numerical data then suggests that there are rare states where the observable remains far from its average value
in each microcanonical
energy window corresponding to non-zero intensive energy, 
differently for what has been observed in the $t-J$ model \cite{santos2010onset,rigol2010quantum,santos2010localization}. 
In these studies of the t-J model, rare states were argued to be absent 
for a range of strengths of the next nearest neighbor interaction, $V'$, and for an energy window centered mid-spectrum.  
In particular with $V'$ small the t-J model is effectively
integrable and rare states were found to be present and with $V'$ large, the model develops energy bands, also compatible with the
existence of rare states.  It is for a middle range of $V'$ that rare states were then found to be 
absent.  Our own study sees some agreement with these results.  And it is natural to think
there would be some agreement at least.  Our study of random matrix model should correspond to the $t-J$ model with intermediate
values of $V'$:  our matrix models are neither integrable nor do they have any notion of energy bands.
In our case, rare states (at least for what we call the odd observable) seem to be vanishing
as system size increases exactly in the center of the spectrum.  However away from this midpoint of
the spectrum, we find that rare states do exist, even in the thermodynamic limit.  It would thus be interesting
to extend the work of \cite{santos2010onset,rigol2010quantum,santos2010localization} to additional energy windows.

Following \cite{biroli2009effect}, we then conclude in the case of sparse matrices
that thermalization may depend on the particular nature of the
initial state and will not occur when such rare states are given a proportionally
large weight in the decomposition of the initial state.
We, however, do not find for the particular initial conditions specified by our quench protocol that 
the rare states are given disproportional weight such that thermalization does not occur.
For both energy windows $\bar e =0$ (see Fig. (\ref{centerscalingsparse})) and $\bar e=1$ (Fig. (\ref{centerscalingsparsee1})),
we see that with increasing matrix size the difference 
between the diagonal and microcanonical ensembles averaged over all initial conditions tends to zero.
This implies that the weighting of rare states in our initial states is not preponderant.  We do note however that 
the vanishing of the difference between ensembles decreases considerably more slowly with system size for the energy
window, $\bar e =1$, than for $\bar e =0$.  We might ascribe this to the presence of rare states at this energy -- even though these states
do not lead to non-thermalization in the thermodynamic limit, they may slow the approach to a thermalized state
as the system size is increased.

We verify this by computing the Kullback-Leibler entropy.  This entropy
is an information theoretic tool used to estimate how close two distributions are. It is
defined as
\be
S_{KL}=\sum_{\eig} P(a)\ln{\frac{P(a)}{Q(a)}} \,\,\,, 
\ee
where $P(a)$ is the expected distribution and $Q(a)$ is the distribution to be
compared. $S_{KL}$ is zero if the two distributions coincide except for sets of
zero measures. In our case we choose $P(a)=|c_{\eig}|^2$ and $Q(a)$ to belong to either 
a uniform or a Gaussian distribution centered about the energy $\overline{e}$. The 
range over which $Q(a)$ is defined has been taken such that the variances of $Q$ and $P$ coincides. 
Fig. \ref{KLscaling} shows the average KL-entropy vs. matrix size for the central
part of the spectrum which indicates in both cases a slow decay
as the system size is increased.  
The slow approach of the distribution $P(a)$ to a distribution $Q(a)$ 
that is both smooth and symmetric about $\bar e$
suggests that rare states are not weighted in a peculiar way, thus permitting thermalization.

In conclusion, for sparse random matrices our numerical data is compatible with the existence of rare states. 
However, the initial states selected by the quench protocol do not seem to have large overlaps with these
rare states and so we typically find thermalization as the end result of our quench process. 

\section{Time Scales of Thermalization}
\label{timescales}
\begin{figure}[t]
\centering
$\begin{array}{cc}
\includegraphics[width=0.4\textwidth]{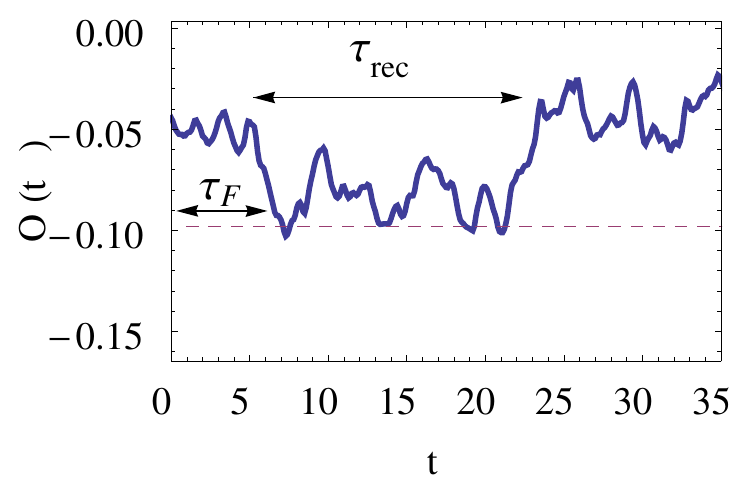} &
\includegraphics[width=0.4\textwidth]{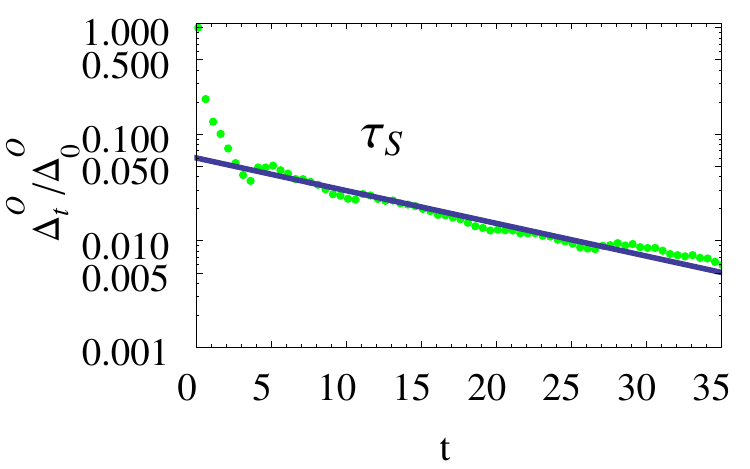}\\
\end{array}$
\caption{Left: time evolution of the observable $\mathcal{O}_t$ versus time $t$. 
Right: time evolution of the fluctuations $\Delta_t^{\mathcal{O}}/\Delta_0^{\mathcal{O}}$ versus time $t$, log-scale on $y$ axis. The straight line is the exponential fit whose slope defines $
\tau_S$.} 
\label{timeEvol}
\end{figure}
The sparse random ensemble, inasmuch as it mimics some characteristic properties of thermalization in systems with
local Hamiltonians, is the right framework to address the study of the thermalization time. 
Interest in this quantity can be traced back to the seminal paper by Von Neumann \cite{neumann1929beweis} regarding 
the quantum ergodic theorem (QET).  The statement made in Von Neumann's paper is that, under suitable 
assumptions (the Hamiltonian has no resonances - meaning that the energy level differences are non degenerate),
any state $\ket{\psi_0}$ in the energy shell $[e-\Delta, e+\Delta]$, will thermalize for most choices of the 
observable and most times $t$, i.e.
$$ 
\Ot \,=\, \bra{\psi_0(t)} \mathcal{O} \ket{\psi_0 (t)} = \Tr{\mathcal{O} \rho_{\text{mc}}} 
\qquad \text{for almost all } t, \mathcal{O} \,\,\,. 
$$
To make the notion of (macroscopic) observable and ``most'' used here precise would require the
development of an involved technical apparatus and so instead
we refer the reader to the existing literature \cite{neumann1929beweis, goldstein2010long}.
Nonetheless we can say there are important differences between this quantum thermalization and 
the classical notion of ergodicity where a time-average is involved. 
It is therefore of interest to give a precise estimate of the time needed for thermalization.
There have been different approaches which have tried to clarify this question
\cite{vznidarivc2012subsystem,masanes2011complexity,brandao2011convergence,short2012quantum}. 
In Fig. \ref{timeEvol}, we examine the 
typical behavior of a realization of a random observable, $\Ot$.  
It decays towards the average value given by the diagonal ensemble:
$$ 
\mathcal{O}_\infty = \Tr{\mathcal{O} \rho_{\text{diag}}} \,\,\,,
$$
and we define the time $\tau_F$ as the first time at which $\mathcal{O}_t$ meets $\mathcal{O}_\infty$.
\begin{figure}[b]
\centering
$\begin{array}{cc}
\includegraphics[width=0.4\textwidth]{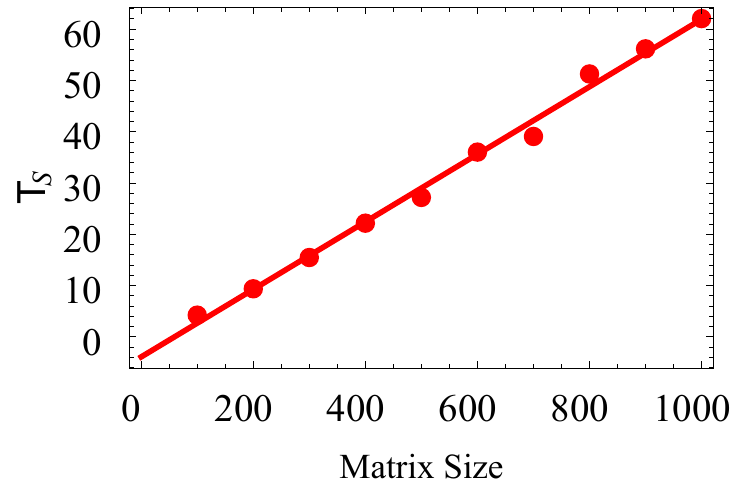} &
\includegraphics[width=0.4\textwidth]{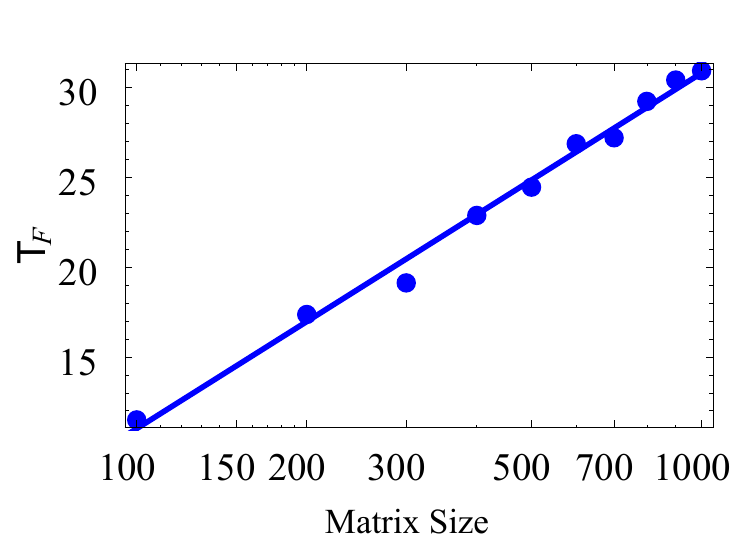}\\
\end{array}$
\caption{Scaling with the matrix size of the two time scale $\tau_S$ (left) with a linear fit ($-3.94 + 0.066 \HS$) and $\tau_F$ (right) with a logarithmic fit ($-28.61 + 8.60 \ln \HS$).}  
\label{timeScaling}
\end{figure} 
We stress here, that even if this quantity has not a direct physical interpretation, it can be considered as a lower bound for the thermalization time.
The time evolution of the observable however keeps fluctuating around the average, due to finite
system size, with it coming close to its initial value after a time, $\tau_{rec}$, the recurrence time.
To quantify these fluctuations we define
$$ 
\Delta^\mathcal{O}_t \equiv \frac{1}{t}\int_0^t ( \Ot - \mathcal{O}_\infty)^2 d\tau  \xrightarrow{t\to\infty} \sum_{\eig,\eigg} c_{\eig}^2 c_{\eigg}^2 O_{\eig\eigg}^2 \equiv \Delta^\mathcal{O}_\infty \,\,\,, 
$$
where we have assumed the absence of energy degeneracies and resonances. 
The quantity $\Delta^\mathcal{O}_t$ can be considered as the variance in the time interval $[0,t]$ of the observable 
and its behavior is plotted in Fig. \ref{timeEvol}: relaxation to the infinite time value is found as expected.
We can fit this curve supposing an exponential relaxation, $e^{- t/\tau_{S}}$, 
defining in this way another time scale, $\tau_{S}$, the time interval needed for the 
relaxation of the fluctuations. Notice that this quantity is the one closer to the Von Neumann formulation: 
indeed, from the Chebyshev inequality, one has a bound on the fraction, $\mu/t$,
of times where the observable has an 
expectation value far from its average
$$ 
\frac{\mu (\tau \in [0,t] \; {\rm ~and~} \; |{\obs}_\tau - \mathcal{O}_\infty| > a)}{t} < 
\frac{\Delta^\mathcal{O}_t}{a^2} \,\,\, .
$$  
As in the thermodynamic limit $ \Delta^\mathcal{O}_\infty \to 0 $, this fraction must also go to zero
in the long time limit. In Fig. \ref{timeScaling}, we see the comparison between the 
two timescales $\tau_{S}$ and $\tau_{F}$ versus the system size. From this plot one can see clearly 
that $\tau_S$ is a long time-scale, with a behavior proportional to $\HS$, i.e.
the size of the Hilbert space.
Notice that this can also be interpreted as the
minimum spectral gap and at the leading order in $\HS$:
$$ 
\min_{\eig\neq\eigg} |\eig - E_{\eigg}| \propto \frac{1}{\HS}. 
$$
In contrast, $\tau_F$ is characterized by a much slower scaling with the size of the system and is 
therefore a fast time-scale. Although it is not easy to 
extract the precise scaling law from the available data, we have fit this data with the form
$$ 
\tau_F(\HS) = a \ln \HS \,\,\, ,
$$
and so taking the scaling of this time scale to go as the volume.
In contrast, the time scale for dense matrices was recently argued to go as the {\it inverse} of the volume\cite{brandao2011convergence}.

\section{Conclusions}
\label{Conclusions}

In this chapter we have addressed the issues of thermalization and the Eigenstate Thermalization Hypothesis in the framework of 
random matrices, aiming to identify  
certain statistical properties of quantum extended systems subjected to a quench 
process. For this purpose we focused our attention on  $Z_2$ breaking quantum 
Hamiltonians, among the simplest theoretical quench protocols. In an attempt to 
encode in our analysis the property of locality, we have considered the ensemble of 
sparse random matrices and we have compared the data coming from this ensemble with 
similar data extracted from the ensemble of dense random matrices. We have found reliable 
evidence of different behavior in the two ensembles.  These differences show up both in the 
$\PR{2}$ of the quench states and in the distribution of the expectation values of the observables on 
post-quench energy eigenstates. In particular, while in the dense random matrix ensemble both the 
variance and the support of the observables vanish with increasing system size $\SN$, 
the sparse random matrix ensemble sees instead strong indications that the variance of EEVs goes to zero while 
the support remains finite as $\SN \rightarrow \infty$. The different 
behavior of the two ensembles can be traced back to the different density of states exhibited by the 
two sets of matrices: while in the dense matrices all states are delocalized in the Hilbert space, 
with almost equal overlap on all energy eigenstates, in the sparse matrices there are instead both 
delocalized and localized states. Localized states give rise to rare values of the expectation values 
of the observables, i.e. values which differ from the typical ones sampled by the micro-canonical ensemble. 
If properly weighted, such localized states may give rise to a breaking of thermalization. In the absence 
of such weighting, as seems to be the case in the initial conditions chosen by our quench protocol,
one instead observes relaxation to the thermal value of the local observables. 

In the framework of sparse random matrices, we have also provided numerical
estimates of the different time scales of thermalization. We have found that it is
possible to identify two time scales: a fast one $\tau_F$ and a slow one $\tau_S$, and that they depend 
differently upon the size of the system.



%% file: Conclusions/conclusions.tex
\def\baselinestretch{1}
\chapter{Final remarks}
\ifpdf
    \graphicspath{{Conclusions/ConclusionsFigs/PNG/}{Conclusions/ConclusionsFigs/PDF/}{Conclusions/ConclusionsFigs/}}
\else
    \graphicspath{{Conclusions/ConclusionsFigs/EPS/}{Conclusions/ConclusionsFigs/}}
\fi

\def\baselinestretch{1.66}

In this work of thesis, we delineated a path in between two important topics such as the characterization of the steady state in closed quantum systems and the existence of a phase transition in disordered and interacting quantum systems. We tried to show that important connections, between the two, exist and that it is fundamental to investigate the ergodicity of the wave function inside the Hilbert space.
In chapter \ref{chapt:mbl}, we presented our work trying to characterize the two phases in the MBL transition. We showed how indicators coming from the AL approach, can be conveniently reformulated in the many-body cases providing good indication for the position of the critical points. The comparison with the single particle case allowed us to identify some important similarities and differencies. The localized phase looks ``less localized'' and the real transition appears as a change of scaling in the wave-function distributions, corresponding to the breaking of ergodicity, instead that by concrete localization in the many-body Hilbert space. This suggested a profitable identification with a Bethe-lattice for large connectivity that remains an open topic for future investigations, both numerical and analytical. Moreover, once the Richardson model has been recognized as toy model for the localized phase, it can be used to compute almost exactly important indicators as the anomalous dimensions, in order to investigate the presence of multi-fractality in this problem.
In chapter \ref{chapt:quench}, we studied the problem of thermalization in closed quantum systems. Instead of focusing on a specific models, we tried to understand what happens in ``typical'' cases. To really define typicality in this framework, we elaborated on the main features that characterize the long time dynamics of a quantum physical system and how they affect the long time behavior. This allowed us to define an interesting ensemble of random matrixes, close friend of the random energy model on the Bethe lattice, already known to the spin glass community. The resulting class of models gave strong indications of the existence of rare states, thus providing an example for a particular mechanism of thermalization, related to a weak version of the ETH, that had been recently conjectured. These numerical results lack of an analytical confirm that could be important in the future. 
We are also interested in the definition of an ensemble of random Hamiltonian that reproduces the features of integrable models. Indeed, it is still open the debate regarding the thermalization or the lack thereof in integrable models. Due to the existence of a set of conserved charges, one could expect that their dynamics is strongly constrained and hardly reaches a thermal steady state. However, a clear answer is difficult in the general case, both for the difficulties in the exact computations of long time expectation values and for the lack of universality of any specific example. Therefore, it would be important to be able to give an answer for the ``typical'' integrable case. This is part of our current research.

We stress once more that many-body localized systems represent an example of interacting, non-integrable systems, that do not reach any thermal equilibrium. So a general understanding of this transition would represent a decisive result for both these important questions.

